\makeatletter
\declare@file@substitution{revtex4-1.cls}{revtex4-2.cls}
\let\old@LaTeX\LaTeX
\documentclass[trackchanges,twocolumn,twocolappendix]{aastex631}
\ifpdf\pdfoutput=1\fi
\let\LaTeX\old@LaTeX
\makeatother

\usepackage{amsmath,xspace,tabularx,textalpha}
\usepackage[all]{hypcap}
\usepackage{xparse}
\newif\iffast\fastfalse
\ifxetex% time consuming font configurations
  \usepackage{xeCJK}
  \iffast\else
    \usepackage{unicode-math}
  \fi
\else
  \usepackage{mathptmx,courier}
  \usepackage[scaled=.92]{helvet}
  \usepackage{CJK}
  \DeclareMathAlphabet{\mathcal}{OMS}{cmsy}{m}{n}
  \AtBeginDocument{\DeclareSIUnit[quantity-product={}]\degree{\text{\textdegree}}}
\fi
\ProvideDocumentEnvironment{CJK*}{m m}{}{}

\usepackage[angle-symbol-over-decimal,unit-mode=text,table-alignment-mode=format,table-number-alignment=center,uncertainty-mode=separate,print-unity-mantissa=false,list-final-separator={, and },list-units=single]{siunitx}

\hypersetup{bookmarksnumbered=true,bookmarksopen=true,pdfauthor=Xuchen Lin}

\input{hack.tex}
\newcommand{\numScritid}{\num{-1.87(51)}\xspace}
\newcommand{\numScrirps}{\num{-0.91(55)}\xspace}
\newcommand{\numScrirpsshort}{{-0.91}\xspace}
\newcommand{\Pmidthres}{{-14.18}\xspace}
\newcommand{\numalpha}{\ensuremath{-0.88^{+0.54}_{-0.25}}\xspace}
\newcommand{\numn}{\ensuremath{4.1^{+5.0}_{-2.8}}\xspace}

\ifxetex
  \newcommand*{\Hi}{H\texorpdfstring{\,}{ }\textsc{i}\xspace}
  \newcommand*{\tHi}{\text{\Hi}\xspace}
  \newcommand*{\sbHi}[1][]{\tsb{\Hi\if\relax\detokenize{#1}\relax\else,#1\fi}\xspace}
\else
  \newcommand*{\Hi}{\texorpdfstring{\ion{H}{1}}{H i}\xspace}
  \newcommand*{\tHi}{\text{\scriptsize\Hi}\xspace}
  \newcommand*{\sbHi}[1][]{\tsb{\scriptsize\Hi\if\relax\detokenize{#1}\relax\else,#1\fi}\xspace}
\fi
\newcommand*{\tsb}[1]{\ensuremath{_{\text{#1}}}}

\newcommand*{\rs}{\ensuremath{r\tsb{s}}\xspace}
\newcommand*{\bmaj}{\ensuremath{b\tsb{maj}}\xspace}
\newcommand*{\Rtid}{\ensuremath{R\tsb{tid}}\xspace}
\newcommand*{\Rso}{\ensuremath{R_{50}}\xspace}
\newcommand*{\Rqo}{\ensuremath{R_{90}}\xspace}
\newcommand*{\Rzsg}{\ensuremath{R_{25,g}}\xspace}
\newcommand*{\Rzoo}{\ensuremath{R_{200}}\xspace}
\newcommand*{\Rsoo}{\ensuremath{R_{500}}\xspace}
\newcommand*{\RHi}[1][]{\ensuremath{R\sbHi[#1]}\xspace}

\newcommand*{\Mzoo}{\ensuremath{M_{200}}\xspace}
\newcommand*{\MHi}{\ensuremath{M\sbHi}\xspace}
\newcommand*{\Mgsoo}{\ensuremath{M_{\text{gas},500}}\xspace}
\newcommand*{\gr}[1][]{\ensuremath{\if\relax\detokenize{#1}\relax\else(\fi g-r\if\relax\detokenize{#1}\relax\else{)_{#1}}\fi}\xspace}
\newcommand*{\Sig}[2][]{\ensuremath{\Sigma_{#2\if\relax\detokenize{#1}\relax\else{,#1}\fi}}\xspace}
\newcommand*{\Stid}{\ensuremath{S\tsb{tid}}\xspace}
\newcommand*{\Stidp}[1][\perp]{\ensuremath{S_{\text{tid},#1}}\xspace}
\newcommand*{\Scri}[1]{\ensuremath{S\tsb{#1,crit}}\xspace}
\newcommand*{\Srps}{\ensuremath{S_\rps}\xspace}
\newcommand*{\Fanc}{\ensuremath{F\tsb{anch}}\xspace}
\newcommand*{\Pram}{\ensuremath{P\tsb{ram}}\xspace}
\newcommand*{\Pmid}[1][]{\ensuremath{P_{\text{mid}\if\relax\detokenize{#1}\relax\else{,#1}\fi}}\xspace}
\newcommand*{\frps}[1][]{\ensuremath{f\tsb{\rps\if\relax\detokenize{#1}\relax\else,#1\fi}}\xspace}
\newcommand*{\ftid}{\ensuremath{f\tsb{tid}}\xspace}
\newcommand*{\fstr}{\ensuremath{f\tsb{str}}\xspace}
\newcommand*{\dprj}{\ensuremath{d\tsb{proj}}\xspace}
\newcommand*{\vepj}{\ensuremath{v\tsb{esc,proj}}\xspace}
\newcommand*{\vesd}{\ensuremath{v\tsb{esc,3D}}\xspace}
\newcommand*{\desd}{\ensuremath{d\tsb{3D}}\xspace}
\newcommand*{\rps}{\ensuremath{\text{RPS}}\xspace}
\newcommand*{\sfr}{\ensuremath{\text{SFR}}\xspace}
\newcommand*{\ssfr}{\ensuremath{\text{sSFR}}\xspace}
\newcommand*{\delSFR}{\ensuremath{\inc\log\sfr}\xspace}
\newcommand*{\delMHi}{\ensuremath{\inc\log\MHi}\xspace}
\newcommand*{\uMsun}{\unit{\Msun}\xspace}
\newcommand*{\tstr}{\ensuremath{t\tsb{str}}\xspace}
\newcommand*{\tdpl}{\ensuremath{t\tsb{dpl}}\xspace}

\newcommand*{\dd}{\mathrm{d}}

\newlength{\myhalfimgsize}
\newcommand*\myplotone[1]{%
  \centering
  \leavevmode
  \includegraphics[width={%
    \ifdim\textwidth=\linewidth%
      2\myhalfimgsize%
    \else\ifdim\linewidth>\myhalfimgsize%
      \myhalfimgsize%
    \else%
      \linewidth
    \fi\fi}]{#1}%
}%
\newcommand*{\autoeqref}[1]{\hyperref[#1]{Equation~(\ref*{#1})}}

\newcommand*{\id}[1]{\##1}

\makeatletter
\@ifpackageloaded{unicode-math}{%
  \newcommand{\inc}{\increment}
}{%
  \newcommand{\inc}{\Delta}
  \newcommand{\coloneq}{:=}
}\makeatother

\DeclareSIUnit\yr{yr}
\DeclareSIUnit\Gyr{\giga\yr}
\DeclareSIUnit\Myr{\mega\yr}
\DeclareSIUnit\pc{pc}
\DeclareSIUnit\Msun{M\ensuremath{_\odot}}
\DeclareSIUnit\kpc{\kilo\pc}
\DeclareSIUnit\Mpc{\mega\pc}
\DeclareSIUnit\dex{dex}
\DeclareSIUnit\deg{deg}
\DeclareSIUnit\mag{mag}

\definecolor{trackchange}{named}{red}
\definecolor{explain}{named}{violet}

\graphicspath{{figures/}}

\defcitealias{2017ApJ...843...16K}{K\&T17}
\defcitealias{2016AJ....152...50T}{CF3}

\shorttitle{Strippings in N4636G}
\shortauthors{Lin et al.}

\begin{document}
\begin{CJK*}{UTF8}{gbsn}

  \title{FAST--ASKAP Synergy: Quantifying Coexistent Tidal and Ram Pressure Strippings in the NGC~4636 Group}

  \correspondingauthor{\cjkname{Jing Wang}{王菁}}
  \email{jwang\_astro@pku.edu.cn}

  \author[0000-0002-4250-2709]{\cjkname{Xuchen Lin}{林旭辰}}
  \affiliation{Department of Astronomy, School of Physics, Peking University, Beijing 100871, People's Republic of China}

  \author[0000-0002-6593-8820]{\cjkname{Jing Wang}{王菁}}
  \affiliation{Kavli Institute for Astronomy and Astrophysics, Peking University, Beijing 100871, People's Republic of China}

  \author[0000-0003-3636-4474]{Virginia Kilborn}
  \affiliation{Centre for Astrophysics and Supercomputing, Swinburne University of Technology, P.O. Box 218, Hawthorn, VIC 3122, Australia}
  \affiliation{ARC Centre of Excellence for All-Sky Astrophysics in 3 Dimensions (ASTRO 3D), Australia}

  \author[0000-0002-2073-2781]{Eric W. Peng}
  \affiliation{Kavli Institute for Astronomy and Astrophysics, Peking University, Beijing 100871, People's Republic of China}

  \author[0000-0002-7422-9823]{Luca Cortese}
  \affiliation{ARC Centre of Excellence for All-Sky Astrophysics in 3 Dimensions (ASTRO 3D), Australia}
  \affiliation{International Centre for Radio Astronomy Research (ICRAR), The University of Western Australia, 35 Stirling Highway, Crawley, WA 6009, Australia}

  \author[0000-0002-9795-6433]{Alessandro Boselli}
  \affiliation{Aix Marseille Univ, CNRS, CNES, LAM, Marseille, France}
  % \affiliation{Aix-Marseille Universit\'{e}, CNRS, CNES, LAM, 13013, Marseille, France}

  \author{\cjkname{Ze-Zhong Liang}{梁泽众}}
  \affiliation{Department of Astronomy, School of Physics, Peking University, Beijing 100871, People's Republic of China}

  \author[0000-0002-3810-1806]{Bumhyun Lee}
  \affiliation{Korea Astronomy and Space Science Institute, 776 Daedeokdae-ro, Daejeon 34055, Republic of Korea}
  % \affiliation{Department of Astronomy, Yonsei University, 50 Yonsei-ro, Seodaemun-gu, Seoul 03722, Republic of Korea}

  \author{\cjkname{Dong Yang}{杨冬}}
  \affiliation{Department of Astronomy, School of Physics, Peking University, Beijing 100871, People's Republic of China}

  \author[0000-0002-7625-562X]{Barbara Catinella}
  \affiliation{ARC Centre of Excellence for All-Sky Astrophysics in 3 Dimensions (ASTRO 3D), Australia}
  \affiliation{International Centre for Radio Astronomy Research (ICRAR), The University of Western Australia, 35 Stirling Highway, Crawley, WA 6009, Australia}

  \author[0000-0003-3523-7633]{N. Deg}
  \affiliation{Department of Physics, Engineering Physics, and Astronomy, Queen's University, Kingston, ON, K7L 3N6, Canada}

  \author[0000-0002-9214-8613]{H. D\'{e}nes}
  \affiliation{School of Physical Sciences and Nanotechnology, Yachay Tech University, Hacienda San Jos\'{e} S/N, 100119, Urcuqu\'{i}, Ecuador}

  \author{Ahmed Elagali}
  \affiliation{Minderoo Foundation, 171--173 Mounts Bay Road, Perth, WA 6000, Australia}
  \affiliation{School of Biological Sciences, The University of Western Australia, 35 Stirling Highway, Crawley, WA 6009, Australia}

  \author[0000-0002-5425-6074]{P. Kamphuis}
  \affiliation{Ruhr University Bochum, Faculty of Physics and Astronomy, Astronomical Institute, D-44780 Bochum, Germany}

  \author[0000-0003-4351-993X]{B. S. Koribalski}
  \affiliation{CSIRO Astronomy and Space Science, Australia Telescope National Facility, P.O. Box 76, NSW 1710, Australia}
  \affiliation{School of Science, Western Sydney University, Locked Bag 1797, Penrith, NSW 2751, Australia}

  \author[0000-0003-4844-8659]{K. Lee-Waddell}
  \affiliation{International Centre for Radio Astronomy Research (ICRAR), The University of Western Australia, 35 Stirling Highway, Crawley, WA 6009, Australia}
  \affiliation{CSIRO Space and Astronomy, P.O. Box 1130, Bentley, WA 6102, Australia}
  \affiliation{International Centre for Radio Astronomy Research (ICRAR), Curtin University, GPO Box U1987, Perth, WA 6845, Australia}

  \author[0000-0001-8496-4306]{Jonghwan Rhee}
  \affiliation{ARC Centre of Excellence for All-Sky Astrophysics in 3 Dimensions (ASTRO 3D), Australia}
  \affiliation{International Centre for Radio Astronomy Research (ICRAR), The University of Western Australia, 35 Stirling Highway, Crawley, WA 6009, Australia}

  \author[0000-0003-2015-777X]{\cjkname{Li Shao}{邵立}}
  \affiliation{National Astronomical Observatories, Chinese Academy of Sciences, 20A Datun Road, Chaoyang District, Beijing 100012, People's Republic of China}

  \author[0000-0002-0956-7949]{Kristine Spekkens}
  \affiliation{Department of Physics and Space Science, Royal Military College of Canada, P.O. Box 17000, Station Forces, Kingston, Ontario, K7K 7B4, Canada}

  \author[0000-0002-8057-0294]{Lister Staveley-Smith}
  \affiliation{International Centre for Radio Astronomy Research (ICRAR), The University of Western Australia, 35 Stirling Highway, Crawley, WA 6009, Australia}

  \author[0000-0002-5300-2486]{T. Westmeier}
  \affiliation{International Centre for Radio Astronomy Research (ICRAR), The University of Western Australia, 35 Stirling Highway, Crawley, WA 6009, Australia}

  \author[0000-0003-4264-3509]{O. Ivy Wong}
  \affiliation{CSIRO Space and Astronomy, P.O. Box 1130, Bentley, WA 6102, Australia}
  \affiliation{International Centre for Radio Astronomy Research (ICRAR), The University of Western Australia, 35 Stirling Highway, Crawley, WA 6009, Australia}
  \affiliation{ARC Centre of Excellence for All-Sky Astrophysics in 3 Dimensions (ASTRO 3D), Australia}

  \author[0000-0001-6163-4726]{Kenji Bekki}
  \affiliation{International Centre for Radio Astronomy Research (ICRAR), The University of Western Australia, 35 Stirling Highway, Crawley, WA 6009, Australia}

  \author[0000-0002-1128-6089]{Albert Bosma}
  \affiliation{Aix Marseille Univ, CNRS, CNES, LAM, Marseille, France}

  \author[0000-0001-9953-0359]{\cjkname{Min Du}{杜敏}}
  \affiliation{Department of Astronomy, Xiamen University, Xiamen, Fujian 361005, China}

  \author[0000-0001-6947-5846]{Luis C. Ho}
  \affiliation{Kavli Institute for Astronomy and Astrophysics, Peking University, Beijing 100871, People's Republic of China}

  \author{Juan P. Madrid}
  \affiliation{Department of Physics and Astronomy, The University of Texas Rio Grande Valley, Brownsville, TX 78520, USA}

  \author[0000-0003-0156-6180]{Lourdes Verdes-Montenegro}
  \affiliation{Instituto de Astrof\'isica de Andaluc\'ia (CSIC), Glorieta de la Astronom\'ia, E-18008, Granada, Spain}

  \author[0000-0002-4911-6990]{\cjkname{Huiyuan Wang}{王慧元}}
  \affiliation{Key Laboratory for Research in Galaxies and Cosmology, Department of Astronomy, University of Science and Technology of China, Hefei, Anhui 230026, People's Republic of China}
  \affiliation{School of Astronomy and Space Science, University of Science and Technology of China, Hefei, Anhui 230026, People's Republic of China}

  \author[0000-0002-9663-3384]{\cjkname{Shun Wang}{王舜}}
  \affiliation{Department of Astronomy, School of Physics, Peking University, Beijing 100871, People's Republic of China}

  \received{2023 January 16}
  \revised{2023 March 30}
  \revised{2023 May 31}
  \accepted{2023 June 18}
  % \published{published date}
  \submitjournal{\emph{The Astrophysical Journal}}

  \begin{abstract}
    Combining new \Hi data from a synergetic survey of ASKAP WALLABY and FAST with the ALFALFA data, we study the effect of ram pressure and tidal interactions in the NGC~4636 group.
    We develop two parameters to quantify and disentangle these two effects on gas stripping in \Hi-bearing galaxies: the strength of external forces at the optical-disk edge, and the outside-in extents of \Hi-disk stripping.
    We find that gas stripping is widespread in this group, affecting 80\% of \Hi-detected nonmerging galaxies, and that 41\% are experiencing both types of stripping.
    Among the galaxies experiencing both effects, the two types of strengths are independent, while two \Hi-stripping extents moderately anticorrelate with each other.
    Both strengths are correlated with \Hi-disk shrinkage.
    The tidal strength is related to a rather uniform reddening of low-mass galaxies ($M_*<\qty{1e9}{\Msun}$) when tidal stripping is the dominating effect.
    In contrast, ram pressure is not clearly linked to the color-changing patterns of galaxies in the group.
    Combining these two stripping extents, we estimate the total stripping extent, and put forward an empirical model that can describe the decrease of \Hi richness as galaxies fall toward the group center.
    The stripping timescale we derived decreases with distance to the center, from \qty{\sim 1}{\Gyr} beyond \Rzoo to $\mathord{\lesssim}\qty{10}{\Myr}$ near the center.
    Gas-depletion happens \qty{\sim 3}{\Gyr} since crossing $2\Rzoo$ for \Hi-rich galaxies, but much quicker for \Hi-poor ones.
    Our results quantify in a physically motivated way the details and processes of environmental-effects-driven galaxy evolution, and might assist in analyzing hydrodynamic simulations in an observational way.
  \end{abstract}

  \section{Introduction}
  \label{sec:intro}
  It is well established that environmental effects are an essential part of galaxy evolution \citep[e.g.,][]{1972ApJ...176....1G,1980ApJ...236..351D,1993ApJ...407..489W,1999MNRAS.308..947A,1999ApJ...518..576P,2009ARA&A..47..159B}.
  They play a dominating role in driving the evolution of low-mass satellite galaxies in general \citep{2014A&ARv..22...74B}, while they also vigorously transform high-mass galaxies under proper conditions \citep{2009AJ....138.1741C}.
  Because environmental effects work through several different physical mechanisms and each of them depends on a number of parameters \citep{2006PASP..118..517B}, it has been challenging to disentangle how they exactly work on galaxies in addition to stellar-mass-related galactic internal effects \citep{2021PASA...38...35C}.

  The primary physical processes producing the environmental effects have been largely identified.
  They can be divided into gravitational and hydrodynamic types, with tidal interaction\footnote{%
    In this paper, we mainly refer to the satellite--satellite interactions, but we do not specifically distinguish between harassment, low-speed interaction, and merger.%
  } and ram pressure stripping (\rps) being probably the most prevalent mechanism in each type \citep{2006PASP..118..517B}.
  With the aid of analytical models and results from controlled simulations that are designed to isolate a given mechanism, it is possible to identify unique conditions or features to select representative samples or prototypes in the real universe that are experiencing strong tidal forces \citep[e.g.,][]{2019ApJ...881..119P,2019MNRAS.482L..55T} or ram pressure \citep[e.g.,][]{2017ApJ...844...48P,2022ApJ...925....4M}, but details are still uncertain due to the lack of constraints from observation \citep[e.g.,][]{2008MNRAS.389.1619F,2015MNRAS.451.2663H}.
  In targeted observations conducted with a similar idea, the \Hi gas has been a popular tracer for environmental effects, because when it is abundant it is easily perturbed by those effects \citep{2009AJ....138.1741C}, while when it is poor it signals the beginning of star formation quenching \citep{2016A&A...596A..11B}.
  These previous studies found that the \Hi masses quickly decrease (within tens to hundreds of megayears) once the galaxies reach the ``stripping zone'' of ram pressure and show one-sided tails \citep{2015MNRAS.448.1715J,2017ApJ...838...81Y}, while the effect of tidal interactions may be more complex and can be conflicting \citep{2018MNRAS.478.3447E,2022ApJ...934..114Y}, but doubts remain whether these galaxies are representative enough.

  The targeted studies enable us to explore the parameter space of each physical mechanism, but they are just the first step toward understanding the effects and roles of each mechanism in galaxy evolution in a cosmological context.
  The cosmological context sets the actual occupation distribution of local galaxies in the parameter space.
  More specifically, it determines the typical properties of galaxies when they fell into their current cluster, as well as the dynamic condition of the clusters when these galaxies travel through them.
  In this context, preprocessing can start far before satellites reach their current cluster \citep{2015ApJ...806..101H,2021MNRAS.507.2949M}, and less massive satellites tend to be satellites for a longer time \citep{2012MNRAS.423.1277D}.
  More massive satellites deplete the \Hi and quench faster both since they enter the first cluster and since they enter the current cluster, because they are closer to being depleted or quiescent before becoming satellites \citep{2018ApJ...865..156J,2021MNRAS.501.5073O}.
  The cold dark matter (CDM) hierarchical assembly paradigm also produces substructures in clusters, which provide unique local environmental conditions like enhanced galaxy number densities and shocked intracluster medium \citep[ICM;][]{2019MNRAS.484..906R}.

  Despite these above, it is challenging to pinpoint the effect of each physical mechanism when we look into the general population of satellite galaxies.
  Indeed, the effects of different environmental processes often mix in less prototypical galaxies \citep{2016MNRAS.461.2630M}.
  Determined by their physical nature, tidal interaction and ram pressure tend to both strengthen in higher-density environment and when they act on low-mass galaxies \citep{2006PASP..118..517B}.
  In addition to statistical arguments \citep{2016MNRAS.461.2630M}, there are observational examples where the two effects coexist in the same galaxy, and resolved \Hi images often provide the crucial supporting evidence \citep{2003A&A...398..525V,2007ApJ...659L.115C}.
  Tidal interaction and ram pressure also interplay with each other.
  For example, tidal interactions can assist \rps by distributing gas to the outer stellar disk where the restoring force is weaker \citep{2016MNRAS.455.2994M}, while ram pressure may prevent tidal stripping (TS) on the leading side of the motion of galaxies \citep{2018A&A...620A.164B}.
  To make things more complex, the galactic internal effects, including the stellar feedback, can interfere with the two environmental processes, by pushing the gas to a location or kinematic status more prone to stripping \citep{2017ApJ...836L..13K,2022A&ARv..30....3B}.

  Possibly because of these difficulties, the role of the environment is much less established in groups compared to massive clusters: groups have less extreme environments and thus more complex combinations of environmental mechanisms.
  Although based on detailed analysis of high-resolution images of the neutral or ionized gas, individual satellite galaxies undergoing \rps or TS have been identified in the group environment \citep[e.g.,][]{2018MNRAS.480.3152V}, statistically establishing the importance of these effects is still difficult.
  Most observational statistical studies tend to go around these complexities, use general parameters such as local densities to describe the environment, and focus on the consequence of the different effects combined together \citep{2017MNRAS.466.1275B,2017MNRAS.469.3670S,2021PASA...38...35C}.
  Motivated by the observed relatively long quenching time for star formation in satellite galaxies, it has been suggested that starvation following \rps of the hot-gas halo, as opposed to \rps of the neutral gas, is the primary mechanism for quenching in groups \citep[e.g.,][]{2015ApJ...806..101H,2017MNRAS.469.3670S}.
  Similar conclusions have been reached by cosmological semianalytical simulations \citep{2008MNRAS.389.1619F,2020MNRAS.498.4327X}.
  On the other hand, many simulations also suggest that \rps of the cold gas might be necessary to reproduce the observed distribution of \Hi deficiency for satellites \citep{2020MNRAS.498.4327X}, particularly those in massive groups \citep{2021MNRAS.505..492A}.
  In cosmological simulations, the tidal interaction on gas is less discussed, which can be more complex than \rps as it can induce both the removal and accretion of gas as suggested by recent zoom-in hydrodynamic simulations \citep{2022MNRAS.509.2720S}.
  Recent theoretical studies have more clearly pointed out the degeneracy between the internal feedback and environmental effects in cosmological semianalytical models \citep{2017MNRAS.471..447S} and the lasting discrepancy between the theoretical prediction and observation of \Hi gas in cosmological hydrodynamic simulations of group environmental effects \citep{2019MNRAS.483.5334S}.

  It is now the time when we need to disentangle the effects of different environmental processes, in order to move forward on the topic of environment-driven galaxy evolution.
  To solve the same problem from the theoretical side with cosmological simulations, a standard approach is to implement analytical recipes or prescriptions directly deduced from physical models or empirically summarized from controlled simulations \citep[e.g.,][]{2007A&A...468...61D} to track the effects of different environmental mechanisms \citep{2015ARA&A..53...51S}.
  These recipes typically depict the contrast between restoring and perturbing forces on the interstellar medium.
  A similar approach can be applied to the observational data that deeply detect and spatially resolve the distribution of gas in galaxies for the whole clusters.
  Contiguous large maps that cover the infalling and virialized regions of clusters provide both a relatively cosmologically representative view (in contrast to targeted, small-field observations) and a large number of galaxies.
  The latter could assist in statistically reducing the observational uncertainties (of relative distance, velocity, density, etc.)\ due to projections.
  Subgalactic spatial resolution is necessary to derive the localized forces, while high sensitivity is essential for tracing the late stage of environmental processing.

  The aforementioned approach of using analytical prescriptions to separate environmental effects has been tested with data from the prepilot and pilot surveys of Widefield ASKAP $L$-band Legacy All-sky Blind surveY \citep[WALLABY;][]{2022PASA...39...58W}, which has a field of view of \qty{30}{\deg^2} per pointing.
  As a conservative preparation, these experiments dealt with prototypical clusters and/or groups where either the tidal or ram pressure effects tend to dominate.
  Based on the empirical model of \citet{1991A&A...244...52E}, \citet{2022ApJ...927...66W} derived the tidal strength parameter \Stid to quantify the instantaneous effect of tidal interaction on the optical disk.
  They showed that \Stid reasonably traced the tidally driven effects of \Hi-disk shrinking and optical-disk reddening for dwarf galaxies in the Eridanus supergroup.
  Based on a revised form of the analytical model from \citet{1972ApJ...176....1G}, \citet{2021ApJ...915...70W} derived the amount of strippable \Hi due to RPS, \frps.
  With \frps, they were able to characterize the diversity of \rps effects on reducing the galactic \Hi mass in the Hydra cluster.
  The next step is to combine these two prescriptions for galaxies in less prototypical clusters and/or groups, and to test how far one may reach in disentangling the \rps or TS effects of \Hi-removal.

  In this work, we analyze the evolution of galaxies under the RPS and TS effects in the NGC~4636 group (N4636G)\@.
  The paper is organized as the following.
  The main information of the group is described in \autoref{sec:sample} and basic statistical properties of members are reported in \autoref{sec:popu}.
  The \Hi data, which are described in more details in Sections \ref{sec:sample} and~\ref{sec:measurement} together with the multiwavelength data, come from a Five-hundred-meter Aperture Spherical radio Telescope (FAST) and WALLABY synergetic observational program.
  We are therefore able to reach a low \Hi-mass limit, and to spatially resolve the large \Hi disks and deblend close pairs at the same time.
  In \autoref{sec:stid_srps}, based on results and experience from the previous studies \citep{2021ApJ...915...70W,2022ApJ...927...66W}, we develop a new parameter \ftid to quantify the amount of \Hi strippable due to TS, and another new parameter \Srps to quantify the instantaneous effect of \rps on the \Hi disk at the edge of stellar disk.
  In \autoref{sec:env}, we study how galaxy properties vary in response to changes in these parameters (\ftid, \frps, \Stid, and \Srps).
  In \autoref{sec:strip}, we put these results together and use a highly simplified model to describe the \Hi stripping history of galaxies when they fall into N4636G\@.
  We derive the \Hi stripping timescales, and predict the \Hi depleting timescales based on this model.
  The results are not very different from those previously obtained from hydrodynamical simulations, semianalytical models, or statistically inferred results from the \Hi mass or star-formation rate (\sfr) distribution of galaxies in clusters and/or groups.

  Throughout the paper, we assume a Lambda CDM cosmology, with $\Omega\tsb{m}=0.3$, $\Omega_\Lambda=0.7$, $h=0.7$, and use the \citet{2003PASP..115..763C} initial mass function when estimating the stellar mass or \sfr.
  We only consider neutral atomic gas in this paper, since \Hi disks are usually much more extended than molecular gas and thus more likely to be stripped.
  In addition, it is probably the major component of neutral gas in low-mass galaxies \citep{2017ApJS..233...22S}.
  All measurements obtained in this paper can be found in \autoref{sec:app:catalogue}.
\end{CJK*}

\section{Sample and Data}
\label{sec:sample}
\begin{figure*}
  \myplotone{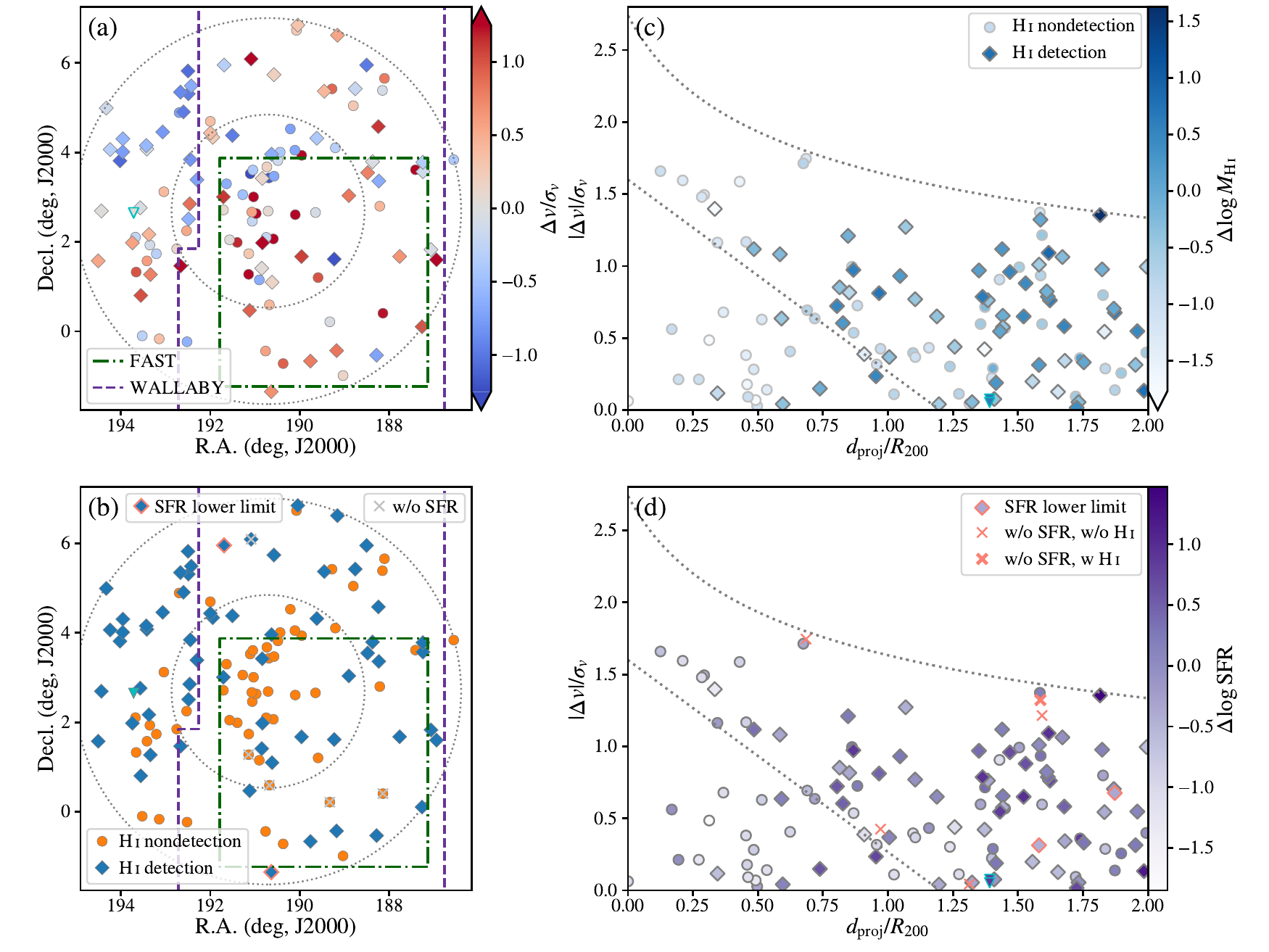}
  \caption{Overview of the galaxy samples.
    The galaxy pair (\id{110} and \id{111}) are labeled as triangles with cyan edge.
    (a)~The spatial distribution of galaxies, color-coded by radial velocities~$\inc v$ relative to the group center.
    The observation fields of FAST (dashed--dotted green line) and WALLABY (dashed purple line) are drawn.
    Two dotted gray circles of radii one and two times \Rzoo indicate the range of N4636G\@.
    (b)~The spatial distribution of \Hi~detections (blue diamonds) and nondetections (orange circles).
    Galaxies without \sfr measurements are overlaid with gray crosses, and those with only lower-limit estimations of \sfr are edged with salmon pink.
    (c)~Galaxies plotted on projected PSD, color-coded by their deviations from the mean \MHi relation of local LTGs (see \autoref{fig:ms}).
    The abscissa is the projected distance~\dprj from the center of N4636G normalized by~\Rzoo;
    the ordinate is $|\inc v|$ normalized by N4636G's 1D velocity dispersion~$\sigma_v$.
    \Hi~nondetections (light-gray edged circles) are color-coded according to their upper limits of~\MHi (see \autoref{sec:popu}).
    The projected escape velocity~\vepj profile and the virialized region are plotted as dotted gray lines.
    (d)~Projected PSD color-coded by deviations from SFMS (see \autoref{fig:ms}).
    \Hi~detections and nondetections without \sfr measurements are represented as bold and regular salmon crosses respectively.
    Salmon-edged diamonds are galaxies with only lower-limits of \sfr.}
  \label{fig:sample}
\end{figure*}

\subsection{The NGC~4636 Group}
\label{ssec:sample_select}
The N4636G is centered at $\alpha=\ang{190.7084}$, $\delta=\ang{2.6880}$ \citep{2002ApJ...567..716R}, at a distance of \qty{16.2}{\Mpc} and a heliocentric velocity of \qty{919}{\km\per\s} \citep[PGC1--42734 in][hereafter \citetalias{2017ApJ...843...16K}]{2017ApJ...843...16K}.
X-ray analysis suggests that the group has a characteristic radius of \qty{0.61}{\Mpc}, with an enclosed total mass of $\Mzoo\sim \qty{2.5e13}{\Msun}$ \citep{2002ApJ...567..716R}.
Here the characteristic radius is taken as~\Rzoo, the radius within which the average density is 200~times the critical density of the universe.
The 1D velocity dispersion, $\sigma_v\sim \qty{278}{\km\per\s}$, is deduced from \Mzoo following the relation by \citet{2008ApJ...672..122E}.
We note that, although this relation is based on simulations of halos more massive than $\num{1e14}\qty[parse-numbers=false]{h^{-1}}{\Msun}$, it is compatible with the preliminary simulation results with a lower mass limit reported by \citet{2022EPJWC.25700018F}.

We compile a catalog of galaxies around the N4636G center with  redshift measurements from several \Hi and optical catalogs, which will be introduced later in Sections \ref{ssec:hi_sample} and~\ref{ssec:optical_sample}.
We define the members of N4636G as galaxies having both a small projected distance (\dprj) and a small relative radial velocity ($\inc v$) with respect to the group's center.
We require $\dprj<2\Rzoo$ and $|\inc v|<\vepj$, where \vepj is the projected escape velocity at~\dprj.
We have removed possible interlopers and finally identified 119~galaxies potentially belonging to the group.
Please refer to \autoref{sec:app:vesc} for more details on escape velocity, interlopers, and possible influence of the nearby Virgo cluster.

\autoref{fig:sample}(a) shows the spatial distribution and redshifts of galaxies in N4636G\@.
\autoref{fig:sample}(c) and~(d) are the projected phase space diagrams \citep[PSDs; e.g.,][]{2015MNRAS.448.1715J}, which are able to qualitatively show the infalling status of galaxy members by plotting radial velocity and projected distance relative to the group center \citep[e.g.,][]{2013MNRAS.431.2307O}.
The color-coding of these projected PSDs reflects gas-richness and star formation and will be discussed in \autoref{sec:popu}.

\subsection{\Hi Samples}
\label{ssec:hi_sample}
Among 119~galaxies belonging to the group, 63 have \Hi~detections.
One pair of \Hi-detected galaxies (\id{110} and \id{111}) are undergoing a merger.
Their evolution is thus dominated by merging, instead of \rps or premerger tidal interaction.
We thus exclude this pair from the following examination and call the remaining 61~galaxies the \emph{\Hi~sample}.
The other 56 \Hi-nondetected galaxies are among the \emph{optical-only sample}.
The spatial distributions of \Hi and optical-only samples can be found at \autoref{fig:sample}(b).

Most of \Hi sample is detected by either FAST, Arecibo Legacy Fast ALFA \citep[ALFALFA,][]{2005AJ....130.2598G,2018ApJ...861...49H}, or WALLABY, all of which will be introduced below.
Besides, one \Hi-detected galaxy, \id{49}, belongs to N4636G and is observed by \Hi Parkes All Sky Survey (HIPASS) only \citep{2004AJ....128...16K}.

\subsubsection{FAST Data}
Using the single-dish radio telescope FAST (channel width \qty{\sim4}{\km\per\s}, beam size \ang{;\sim2.9;}), \citet{2022RAA....22i5016Z} observed a $\ang{\sim 5}\times\ang{5}$ square region covering N4636G, which is indicated in \autoref{fig:sample}(a).
With the high sensitivity of FAST, they pushed the detection limit in \Hi mass by \qty{\sim 0.4}{\dex} deeper than that of the previous ALFALFA data.
Specially, FAST provides four new \Hi-detections (\id{5}, \id{24}, \id{33}, and \id{58}).
When the \Hi flux is measured with more than one telescope, the FAST flux is preferred, except for two galaxies near the edge of the FAST field (\id{4} and \id{77}).
In the end, the \Hi fluxes of 17~galaxies are obtained with FAST\@.

\subsubsection{ALFALFA Data}
The ALFALFA survey observed the northern sky and thus covers a large fraction of the N4636G \citep{2018ApJ...861...49H}.
It detects 57~\Hi sources belonging to the group.
\citet{2022RAA....22i5016Z} has shown the high level of consistency between the FAST and ALFALFA fluxes for overlapping sources in the N4636G\@.
After excluding FAST-detected sources and merger systems, ALFALFA provides fluxes for 42 galaxies in the \Hi~sample.

\subsubsection{WALLABY Data}
The WALLABY pilot survey \citep{2022PASA...39...58W} observed the NGC~4636 field, and the footprint is plotted in \autoref{fig:sample}(a).
As an interferometric \Hi survey, WALLABY has a beam size of \ang{;;\sim 30} and can provide moderate-resolution \Hi maps \citep[e.g.,][]{2021ApJ...915...70W}.
Among 19~galaxies detected by WALLABY belonging to this group, five galaxies have data with enough resolution to study their moment~(0) maps (see also the discussion in \autoref{ssec:hi_prop} and \autoref{sec:app:hi_mock}).
They also enable checking of the ALFALFA and FAST fluxes for possible contamination from neighbors.
One such contamination is identified (\id{79} detected by ALFALFA), and we adopt the \Hi flux measured with WALLABY\@.
A comparison between the ASKAP (Australian Square Kilometre Array Pathfinder) and ALFALFA fluxes for galaxies in N4636G can be found in \citet{2022PASA...39...58W}.

\subsection{The Super Catalog of Redshifts}
\label{ssec:optical_sample}
We compile the super catalog of these 119~group member galaxies with redshift measurements by combining the \citetalias{2017ApJ...843...16K} galaxy catalog (91 galaxies; 45 \Hi-detections), the Sloan Digital Sky Surveys (SDSS) Data Release (DR) 16 \texttt{SpecObj} table \citep[79; 32,][]{2016AJ....151...44D,2020ApJS..249....3A}, and four new \Hi-detections by FAST\@.
These catalogs are crossmatched with each other to remove the duplicates.
The redshift values from \Hi surveys are preferred to those from optical catalogs, and those by the SDSS DR16 are preferred to the other optical ones.
The final super catalog is presented in \autoref{sec:app:catalogue}.

\subsection{The SDSS and DECam Images}
We use $g$-~and $r$-band images from SDSS DR12 \citep{2000AJ....120.1579Y,2015ApJS..219...12A} and the DECam Legacy Survey (DECaLS) DR9 \citep{2019AJ....157..168D} for optical photometric measurements.
SDSS has typical seeings (as FWHM) of \ang{;;1.44} and~\ang{;;1.32} in both bands respectively, and the corresponding depths are \qtylist{21.84;20.84}{\mag}.
For DECaLS, these parameters are \ang{;;1.29},~\ang{;;1.18}; \qtylist{23.72;23.27}{\mag} respectively.

Since the resolution and depth of DECaLS are slightly better than those of SDSS, DECaLS images are preferred to study dwarf galaxies.
The background removal pipeline of DECaLS, however, is not optimized for extended sources, and large galaxies suffer from flux loss on the periphery \citep{2019AJ....157..168D}.
Therefore, we use SDSS photometry for our \emph{high-mass galaxies} ($M_*\ge\qty{1e9}{\Msun}$) and DECaLS photometry for \emph{low-mass galaxies} ($M_*<\qty{1e9}{\Msun}$).
Since some of our discussions below divide the sample into low- and high-mass subsets, such a choice of optical survey provides consistency.
More details on combining these two types of fluxes are presented in \autoref{sec:app:sdss_ls}.

\section{Measurements}
\label{sec:measurement}
\subsection{Optical Photometry}

We make cutouts of images centered on each galaxy.
Since the sample galaxies typically have a spatial extent comparable to the field of view of SDSS, SDSS image frames are mosaicked using \emph{SWarp}%
\footnote{\url{https://github.com/astromatic/swarp/releases/tag/2.41.4}}
\citep{2002ASPC..281..228B}, with overlapping regions averaged.
Because the downloaded images have gone through initial background subtraction, no additional background removal is conducted during mosaicking.
A few frames have problematic background removal by the SDSS pipeline, and they are discarded.
DECaLS images are processed with the same procedure if the images that we downloaded are not large enough to cover the whole galaxy.

The general procedure of the photometric pipeline is the same as that of \citet{2017MNRAS.472.3029W}.
The following are the main steps.
\begin{enumerate}
  \item \emph{Masks.}
        We deblend and mask contaminating light from neighboring sources.
        We combined \emph{SExtractor}%
        \footnote{\url{https://github.com/astromatic/sextractor/releases/tag/2.25.0}}
        \citep{1996A&AS..117..393B} and \emph{photutils} \citep{2021zndo...5525286B} to generate masks, based on the $r$-band image from SDSS or DECaLS\@.
        A ``cold and hot'' method \citep[e.g.,][]{2004ApJS..152..163R} is applied.
        In the ``cold'' mode, we run SExtractor twice.
        The first run, with detecting threshold set to $1.5\sigma$, detects all clearly separated neighboring sources and masks them.
        The second run, however, unmasks faint clumpy features possibly belonging to the peripheral parts of the target by setting a threshold of $1.0\sigma$ and a deblending parameter of $0.005$.
        In the ``hot'' mode, foreground stars are detected using \texttt{DAOStarFinder} in photutils, with box size of $10\times 10$\,pixels for background estimation, detect threshold as~$5\sigma$, and roundness threshold as~$0.7$.
        A star candidate is masked only if its peak is larger than 5~times the local background level and is not around the galaxy center.
        All masks are dilated by $1.5$~times the seeing size, checked with $g$-band images, and manually adjusted if necessary.
  \item \emph{Background subtraction and galaxy geometry.}
        After we flag the pixels belonging to the sample galaxy and those masked in the last step, the background of the image is modeled using a 2D linear equation and removed.
        We replace the masked pixels with the centrosymmetric or the azimuthally averaged value around the galaxy center.
        Function \texttt{detect\_sources} of photutils is used to measure the center, position angle, and ellipticity with a threshold of $1.5\sigma$.
  \item \emph{Surface brightness (SB) profile.}
        The geometry parameters of $r$-band images are used for generating photometric annuli of all bands.
        We measured the $3\sigma$ clipped average value in each elliptical ring as the SB\@.
        At the outer region, annuli enlarge with a geometric step of $1.05$ times.
        When 15 contiguous annuli's SB values are linearly uncorrelated with their sizes by a $p$-value of $0.05$ (using $F$-test), we consider the profile flattens hereafter.
        The size ratio between the largest and the smallest of these 15 annuli is typically around 2.
        The $3\sigma$ clipped average of these 15~values is subtracted as residue background, and 10~times their clipped standard deviation is used as the threshold for final profile.
  \item \emph{Total fluxes.}
        We then measure the growth curve (GC) of the flux within each elliptical aperture as a function of aperture size.
        Again, the GC is considered flattened with 15~uncorrelated contiguous flux values.
        The $3\sigma$ clipped average of these 15~values is then considered to be the total flux of the galaxy, and their scatter is taken as the uncertainty.
        Some of the GCs rise or fall slightly beyond the flat region and then flatten again, and we attribute it to variations in the local background.
        We take the extent of rise or fall as the extra uncertainty of background estimation and propagate this into the uncertainty of the flux.
\end{enumerate}

\citet{2019AJ....157..168D} reported that the difference between fluxes measured by DECaLS and SDSS can be modeled as a cubic function of $g-i$ color in SDSS\@.
We use the SDSS color measurements and this model to remove the systematic difference of DECaLS fluxes from the SDSS ones.
For consistency, the $g$-~and $i$-bands' SDSS fluxes used here are measured with the GC method using apertures based on the DECaLS $r$-band geometric parameters.

For all fluxes, the Galactic extinction is corrected based on the dust map of \citet{1998ApJ...500..525S} and the extinction curve of \citet{1989ApJ...345..245C}.

\subsection{Size}
We characterize the size of stellar disk with~\Rzsg, the radius at which the SB drops to \qty{25}{\mag.arcsec^{-2}}.
The~\Rzsg is derived through linearly interpolating the $g$-band SB profile.
We want to use this radius to trace the edge of disk growth, and thus it is defined in $g$ band here, which is relatively blue.

By cubically interpolating the GC, we also get the radii enclosing $50\%$ and $90\%$ of fluxes, \Rso and \Rqo, in $r$ band.
The $r$ band is used here instead, in order to track the radial structure of stellar mass dominated by old stars.

We note that the so-called \emph{radius} here is always the semimajor axis of the corresponding elliptical annulus, accounting for projection effects.

\subsection{Color}
\label{ssec:color}
We calculate for each galaxy the total \gr color and \gr color profile, the latter of which is truncated when the uncertainty reaches \qty{0.1}{\mag}.
We further define three localized color parameters of each galaxy at the center, at~\Rso, and at~$2\Rso$, denoted as \gr[0], \gr[\Rso], and \gr[2\Rso] respectively.
They are derived from regions within $0.2\sqrt{2}\Rso$, between $0.8$~and $1.25\Rso$, and between $1.6$~and $2.5\Rso$, based on the GC\@.

\subsection{Stellar Mass and \sfr}
\label{ssec:mass}

The total \gr is converted to the $r$-band stellar mass-to-light ratio following \citet{2009MNRAS.400.1181Z}.
We then determine the stellar mass with the $r$-band luminosity.
Stellar mass profiles are obtained similarly using the $r$-band SB profiles and \gr color profiles after correction for the inclination, with the assumption of infinitely thin disks.

We obtained \sfr{}s following the methods of \citet{2017MNRAS.472.3029W}.
The same photometric pipeline of \citet{2017MNRAS.472.3029W} is used to derive fluxes from the Galaxy Evolution Explorer \citep[GALEX;][]{2005ApJ...619L...1M} and the Wide-field Infrared Survey Explorer \citep[WISE;][]{2010AJ....140.1868W} images.

W4-band images from the WISE are used for an estimate of the dust-attenuated part of the \sfr, with far-ultraviolet (FUV) images from the GALEX for the unattenuated part.
Near-ultraviolet (NUV) data are used when FUV images are unavailable.
The luminosity in each band is converted to \sfr using the equations of \citet{2013seg..book..419C}.
There are 58~galaxies out of~119 having no FUV images, and three of them are also missing NUV data (\id{49},~\id{65}, and \id{75}).
W4-band luminosities provide the lower-limit estimation of \sfr for \id{49}~and \id{75}, leaving the \Hi-detected galaxy \id{65} without \sfr measurement.
Besides, \Hi-nondetected \id{28},~\id{50}, and~\id{69}, although covered by GALEX, have neither ultraviolet nor W4 detection and as a result have no \sfr measurement either.

\subsection{\Hi Properties}
\label{ssec:hi_prop}

\Hi~mass values are calculated using \Hi~fluxes from the catalog compiled in \autoref{ssec:hi_sample}.
We derive the size of \Hi~disk, \RHi, from \Hi~mass, using the relation given by \citet[with a \qty{\sim 0.06}{\dex} scatter]{2016MNRAS.460.2143W}, where \RHi is the radius where \Hi~surface density reaches \qty{1}{\Msun\per\square\pc}.
For eleven galaxies with a derived \RHi larger than $\bmaj\sqrt{a/b}$, we also measured their \RHi from their WALLABY moment~(0) map.
Here, \bmaj is the WALLABY beam FWHM (\ang{;;\sim 30} or \qty{2.3}{\kpc}), and $a$ and $b$ are the major-axis and minor-axis of the galaxy measured from optical images.
We confirmed that they indeed follow the relation after correcting the smearing effect by adopting $\RHi = \sqrt{\RHi[obs]^2-(\bmaj/2)^2}$ \citep{2016MNRAS.460.2143W}, where \RHi[obs] is the radius directly derived from the deprojected surface density profile.
The ellipticity of the \Hi disk is assumed to be the same as the one of the stellar disk.
More discussion on $\bmaj\sqrt{a/b}$ as the criterion of selecting a reliable \RHi directly measured from images can be found at \autoref{sec:app:hi_mock}.

\subsection{ICM Properties}

\begin{figure}
  \myplotone{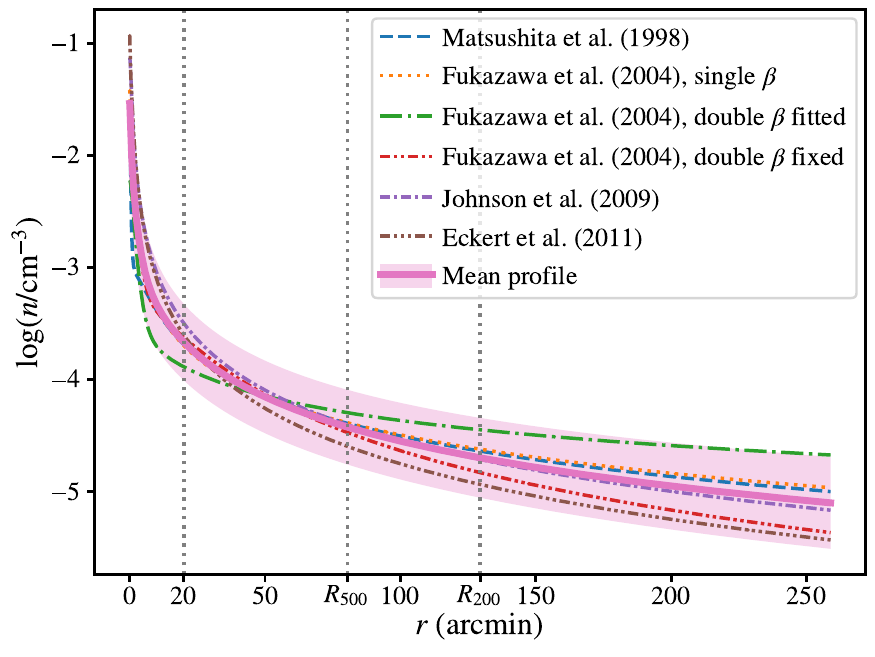}
  \caption{Different ICM density models for N4636G from the literature (see the legend) and the averaged ICM density profile (bold pink line).
    The pink shading indicates the uncertainty, considering the scatter of different models and the uncertainty of~\Mgsoo estimation.
    Historical X-ray surveys covered less than \ang{;20;} (vertical dotted gray line) from the group center.
    The~\Rsoo and~\Rzoo of N4636G are also drawn as vertical lines.
    \nocite{1998ApJ...499L..13M,2004PASJ...56..965F,2009ApJ...706..980J,2011A&A...526A..79E}}
  \label{fig:icm}
\end{figure}

We use the findings of earlier publications to create a reasonable model of the ICM density profile $\rho(r)$ of N4636G\@.
The final ICM number density profile~$n(r)$, shown in \autoref{fig:icm}, is the geometric average of six different models of this group from the literature \citep{1998ApJ...499L..13M,2004PASJ...56..965F,2009ApJ...706..980J,2011A&A...526A..79E}.
These models are normalized to a uniform ICM mass within~\Rsoo, $\log\left(\Mgsoo/\uMsun\right)=\num{11.72(32)}$.
Beyond $r=\ang{;20;}$, our averaged model can be well fitted as
\begin{equation}
  n(r) = \qty{11.7e-3}{\cm^{-3}}\left(r/\unit{arcmin}\right)^{-1.32},
\end{equation}
with a scatter of \qty{\sim 0.3}{\dex}.

We caution that the actual measurements of X-ray brightness, from which these models were derived, are limited to a small radial range of \ang{;20;} ($\mathord{\sim}\Rzoo/6$) from the group center.
The ICM mass density is $\rho(r)=1.4m\tsb{p}n(r)$, where $m\tsb{p}$ is the mass of the proton, and where the factor~$1.4$ accounts for the presence of helium.
Similar to~\Rzoo, \Rsoo is the radius within which the average density is 500~times the critical density.
The value of~\Mgsoo is calculated from~\Mzoo following \citet{2020MNRAS.492.4528S}.

\section{Overview of the Galaxy Population}
\label{sec:popu}
\begin{figure*}
  \myplotone{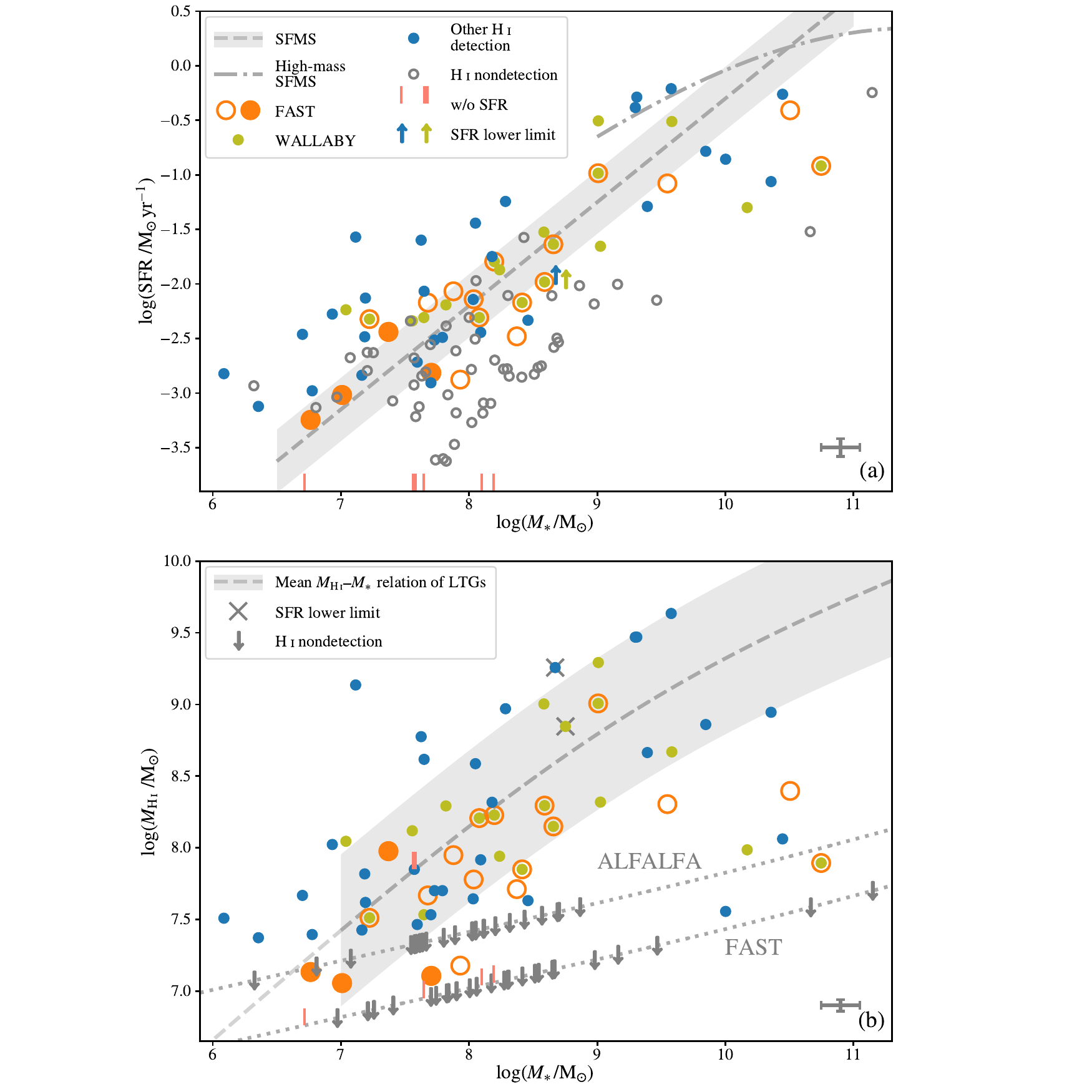}
  \caption{The \sfr and \Hi~mass (\MHi) of our sample galaxies, compared with the average relation from star-forming nearby galaxies.
    (a)~\sfr vs.~stellar mass $M_*$.
    \Hi~nondetections are indicated by empty gray dots; \Hi~detections by WALLABY, FAST, and previous \Hi~surveys are plotted as olive dots, orange circles, and blue dots respectively.
    Estimations of lower limits of \sfr are plotted as upward arrows in the corresponding color.
    All WALLABY detections have been covered by previous ALFALFA or HIPASS surveys, while FAST provides four new \Hi~sources (orange filled circle).
    The stellar masses of \Hi-detected and -nondetected galaxies without \sfr measurements are pointed out with bold and regular vertical salmon ticks respectively.
    As a reference, the fitted local SFMS from \citet{2014MNRAS.445..899C} is given as dashed gray line, and the scatter of \qty{0.29}{\dex} is given as shading.
    The local SFMS of high-mass galaxies from \citet{2016MNRAS.462.1749S} is also plotted as dashed--dotted gray line.
    The typical uncertainties of data points are plotted at the lower right corner.
    (b)~\MHi vs.~$M_*$.
    Most of the symbols are the same as those in~(a).
    The upper limits of~\MHi (dotted gray line) of \Hi~nondetections (gray arrow) are provided, calculated with FAST or ALFALFA parameters according to their coordinates.
    Galaxies that only have lower limits of \sfr are overlaid with gray crosses.
    The mean \MHi relation of local LTGs from \citet{2018RMxAA..54..443C} is plotted as dashed gray line, with the extrapolated part of $M_*<\qty{1e7}{\Msun}$ translucent.
    Its scatter of \qty{0.53}{\dex} is given as shading.
    The typical uncertainties are also given.}
  \label{fig:ms}
\end{figure*}

We present our measurements of \sfr and \Hi~mass \MHi in \autoref{fig:ms} against the stellar mass $M_*$.
We also plotted the star-forming main sequence \citep[SFMS; from][]{2014MNRAS.445..899C} and the mean \MHi relation of local late-type galaxies \citep[LTGs;][]{2018RMxAA..54..443C} for comparison.

It is clear that most of \Hi~detections lie very close to the SFMS (\qty{1.5}{\dex} below it at most), which is consistent with the view that the galaxies need neutral gas to sustain the star formation \citep{2020ApJ...890...63W,2021ApJ...918...53G}.

Our \Hi~sample generally follows and scatters around the mean \MHi relation.
At the high-mass end, one can find \Hi~detections nearly \qty{2}{\dex} below the relation;
at the low-mass end, however, the detection limit meets the relation, which results in a bias toward gas-rich galaxies.
Notwithstanding, FAST lowers the limit by \qty{\sim 0.4}{\dex} compared with ALFALFA \citep{2022RAA....22i5016Z}.
For the optical-only sample, we estimated the upper limit of~\MHi mainly following \citet{2021ApJ...915...70W}, which is based on the rms of data and the line widths predicted from the baryonic Tully--Fisher relation \citep{2000ApJ...533L..99M}.
We confirm that the predicted line widths are all larger than the velocity resolution.

The offset relative to the SFMS or the mean \MHi relation can be seen as an indicator of star-formation strength or \Hi richness \citep[e.g.,][]{2010MNRAS.408..919S}.
We denote them as $\delSFR\coloneq\log\left(\sfr/{\sfr\tsb{MS}}\right)$ and $\delMHi\coloneq\log\left(\MHi/{M\sbHi[MS]}\right)$.
By doing so, the underlying dependence on~$M_*$ in galaxy evolution (i.e., the secular evolution) is removed in a first-order approximation.
In this study, we refer to the galaxies above (or below) the mean \MHi relation as \emph{gas-rich} (or \emph{gas-poor}) ones.

\autoref{fig:sample}(c) and~(d) show how \delMHi and \delSFR change as a function of position in the projected PSD, which roughly indicates the infalling stage.
There is a general trend that both offsets drop toward the group center.
In particular, they decrease significantly inside the virialized region (triangle bordered by gray dashed lines), suggesting that galaxies there have been strongly processed by environmental effects recently or in the long past.

\section{Quantifying the Tidal and Ram Pressure Effects}
\label{sec:stid_srps}

We consider two environmental effects, the gravitational tidal interaction (with an effect of either stripping or perturbation), and the hydrodynamic ram pressure (with an effect of either stripping or compression).
We use two types of parameters to describe the effect of stripping: the strength of stripping at the optical-disk edge, and the fraction of strippable \Hi.

We note that we derive these two parameters from only the total \Hi flux;
though, \citet{2021ApJ...915...70W} have shown that it is possible to calculate them from \Hi moment~(0) map.
The reason is that only five galaxies are resolved enough for this calculation (see Appendices \ref{sec:app:hi_mock} and~\ref{sec:app:fstr} for more information).
Still, we provide the atlas of these five galaxies and brief discussion in \autoref{sec:app:overlay}.

\subsection{The Strength of Stripping at the Optical-disk Edge}

\citet{1991A&A...244...52E} described the strength of tidal interaction in simulations using
\begin{equation}\label{eq:stid_orig}
  \frac{M\tsb{p}}{M\tsb{g}}\left(\frac{R\tsb{g}}{\delta}\right)^3
  \frac{\inc t}{T},
\end{equation}
This quantity compares the impulse transferred by tidal force and the inherent momentum of self-revolution.
Here ``p'' and ``g'' denoted the perturber and the target galaxy respectively; $M$ is the total mass of galaxy, $R\tsb{g}$ is the characteristic radius of ``g,'' and $\delta$ is the minimum distance between these two galaxies during the whole encounter.
The value $\inc t$ is the interaction time, measured as the time needed for ``p'' to rotate \qty{1}{rad} with respect to ``g'' when they are nearest, while $T$ is the time for ``g'' to self-rotate \qty{1}{rad}.

\citet{2022ApJ...927...66W} modified \autoeqref{eq:stid_orig}, made it suitable for observation, and explored the effects of tidal interaction on the optical disk in the Eridanus supergroup.
They replaced the minimum distance $\delta$ with the current projected distance $\delta\tsb{proj}$, and calculate $\inc t$ using the difference of heliocentric radial velocity $\inc v\tsb{rad}$.
They defined
\begin{align}
  \Stid & = \sum\tsb{all p} \frac{M\tsb{p}}{M\tsb{g}}\left(\frac{R\tsb{g}}{\delta\tsb{proj}}\right)^3
  \frac{\inc t}{T}\notag                                                                              \\
        & = \sum\tsb{all p} \frac{M\tsb{p}}{M\tsb{g}}\left(\frac{R\tsb{g}}{\delta\tsb{proj}}\right)^3
  \frac{\delta\tsb{proj}/\sqrt{(\inc v\tsb{rad})^2 + V\tsb{circ}^2}}{R\tsb{g}/V\tsb{circ}},\label{eq:stid}
\end{align}
which is summed over all possible perturbers ``p.''
In \autoeqref{eq:stid}, the ratio $\inc t/T$ is transformed into the ratio of angular velocity.
The $V\tsb{circ}$ is the circular velocity of ``g'' calculated from the baryonic mass of the galaxy with the Tully--Fisher relation \citep{2000ApJ...533L..99M}, using the sum of stellar mass and the helium-corrected \Hi mass.
If the galaxy is \Hi-nondetected, the upper limit of \MHi is used.
The $V\tsb{circ}$ term in $\sqrt{(\inc v\tsb{rad})^2 + V\tsb{circ}^2}$ is a smooth parameter introduced by \citet{2022ApJ...927...66W} to avoid zero divides, but it also reflects that when two galaxies have a low relative velocity, the circular velocity determines the timescale of tidal interaction.

We follow the procedure of \citet{2022ApJ...927...66W} to derive \Stid.
The total mass $M$ of the galaxy is estimated as $V\tsb{circ}^2\Rzsg/G$, where $G$ is the gravitational constant.
The characteristic radius $R\tsb{g}$ is chosen as the \Rzsg.
We note that $M\tsb{p}$ should have been the total mass within the virial radius of the perturber's dark matter halo, instead of just within \Rzsg.
Theoretically and in a first-order approximation, however, the optical disk size linearly scales with the virial radius \citep{1998MNRAS.295..319M}, and thus our $M\tsb{p}$ is roughly a uniformly scaled estimate.
Since in this study only the relative values of \Stid are important, this systematic offset of total mass is to some extent properly accounted for.
The major uncertainty is from the projection effects, which we estimate using simulation data as \qty{0.41}{\dex} (see \autoref{sec:app:proj}).

Massive galaxies surrounding the group could add to the tidal interaction.
Using the coordinate and redshift-independent distance in the Cosmicflows-3 catalog \citep[hereafter \citetalias{2016AJ....152...50T}]{2016AJ....152...50T}, we found 13~galaxies within $3\Rzoo$ from the center of N4636G that are not in our sample.
They are included as potential perturbers.
Their stellar masses are measured using SDSS images, and ALFALFA \Hi~fluxes are adopted when possible.

We generalize the idea of \Stid to express the strength of \rps as
\begin{equation}
  \Srps = \frac{\Pram}{\Fanc(\Rzsg)},
\end{equation}
where \Pram is the ram pressure, estimated as $\rho(\dprj)(\inc v)^2$ \citep{1972ApJ...176....1G}, where $\rho(r)$ is the volumetric mass density of ICM at a distance~$r$ from the group center.
The anchoring force $\Fanc$ describes the gravity exerted on gas from the stellar disk and the gas disk itself, estimated as \citep{2021ApJ...915...70W}
\begin{equation}\label{eq:fanc}
  \Fanc = 2\pi G\left(\Sig{*}+\Sig{\text{gas}}\right)\Sig{\text{gas}},
\end{equation}
where $G$ is the gravitational constant, and \Sig{*} and~\Sig{\text{gas}} are deprojected stellar and cold-gas surface density respectively.
The value at~\Rzsg is used.
We mostly follow the method of \citet{2021ApJ...915...70W} for the determination of \rps parameters.
More details of \Fanc are given in \autoref{sec:app:fstr}.
The major source of uncertainty for \Srps is \Pram, the projective random uncertainty of which is estimated as \qty{0.48}{\dex} using simulation data (see \autoref{sec:app:proj} for more discussion).

We note that, due to the way that \Pram is derived, \Srps can practically extend to physically meaningless infinitely low values when the galaxy evolution is almost unaffected by \rps.
Relatedly, \Srps has a much larger dynamical range (\qty{\sim 6}{\dex}) than \Stid has (\qty{\sim 2}{\dex}).
We thus manually set \Srps for galaxies with $\log\Srps<\num{-3.2}$, which are 10\% of the sample, as invalid, assuming the \rps effect on these galaxies to be insignificant in galaxy evolution.
As shown later in \autoref{ssec:disk_size}, this does not change our classification of \rps- or TS-dominant galaxies (introduced in \autoref{ssec:ts_rps_comparison}).

\subsection{The Extent of Stripping: Fraction of Strippable \Hi}
\label{ssec:frps_ftid}
We follow the method of \citet{2021ApJ...915...70W} to derive \frps that represents the extent of \rps within the galaxy.
This parameter measures the fraction of \Hi gas that is subjected to \rps, i.e., with $\Pram>\Fanc$.
The method makes use of the \Hi size--mass relation, assuming a characteristic \Hi-surface-density profile \citep{2020ApJ...890...63W} that we truncate at $1.5\RHi$.
We summarize the technical details in \autoref{sec:app:fstr}.

We further define a similar parameter for TS, \ftid.
The main idea is to generalize~\Stid to each radius~$r$ in the galaxy and to use a critical value \Scri{tid} to determine the stripping radius.
The conventional definition of tidal truncation radius \citep{1984ApJ...276...26M,2003MNRAS.341..434T} is not adopted here because it is purely gravitational and does not account for hydrodynamic effects.

The tidal strength as a function of the galactocentric distance~$r$ is derived as
\begin{align}
  \Stid(r) & = \sum\tsb{all p} \frac{M\tsb{p}}{V\tsb{g}^2(r)r/G}
  \left(\frac{r}{\delta\tsb{proj}}\right)^3 \frac{\inc t}{T}\notag \\
           & = \frac{V\tsb{circ}^2}{V\tsb{g}^2(r)}
  \frac{r^2}{\Rzsg^2}\Stid,
\end{align}
where the timescale ratio $\inc t/T$ is fixed.
The total mass within $r$ is calculated with $V\tsb{g}(r)$, the rotational velocity of ``g'' at $r$.
\citet{2017ApJ...836..152L} gave an observational relation between the localized radial acceleration $g(r)=V\tsb{g}^2(r)/r$ and the stellar gravitational field profile, which we calculate using the stellar mass~$M_*(r)$ within~$r$ as $g_* = GM_*(r)/r^2$.
Therefore we can derive
\begin{equation}\label{eq:stidr}
  \Stid(r) = \frac{g(\Rzsg)r}{g(r)\Rzsg}\Stid.
\end{equation}
We assume that the tidal interaction affects all mass outside of tidal radius \Rtid, which satisfies $\Stid(\Rtid)=\Scri{tid}$.
Here \Scri{tid} is the critical value that will be introduced in \autoref{ssec:s_crit}.
Then \ftid is calculated as the fraction of gas out side of \Rtid.
Similar to \frps, those \ftid's smaller than $1\%$ are set to $0$.

One would assume that the influence of tidal interaction would also reflect in the stellar population.
We confirm that six galaxies have $\Rtid<\Rzsg$, all of which are dwarf galaxies (\id{5}, \id{57}, \id{58}, \id{79}, \id{107}, and \id{112}).
We find obvious irregularity and asymmetric structures in three by visual inspection (\id{5}, \id{58}, and \id{79}).
However, since these dwarf galaxies are faint and it is hard to estimate the intrinsic irregularity of them, we thus did not quantify this and consider this question beyond the scope of this paper.

\subsection{The Difference between \texorpdfstring{$S$}{S} and \texorpdfstring{$f$}{f}}
Both $f$ and $S$ quantify the significance of environmental effects but describe different aspects.
The major cause is that \Hi and stellar (or star-forming) disks have distinct radial profiles and extensions.
For example, an \Hi-rich, massive galaxy may simultaneously have a high $f$ and a low $S$: a significant portion of the \Hi could extend beyond the stellar disk, while the environmental forces at \Rzsg are not high enough compared to the restoring ones.
Thus, $f$ reflects the frontier of gas stripping while $S$ describes the environmental influence at the edge of the stellar disk.
The latter will be used to study the dependence of the consequent or instantaneous optically related properties of the galaxy on environmental effects (\autoref{sec:env}), and the former will be used to construct continuity equations to trace the process of gas stripping (\autoref{sec:strip}).

\subsection{The (Lack of) Correlation between \rps and TS Instantaneous Strengths}
\label{ssec:ts_rps_comparison}

\begin{figure*}
  \myplotone{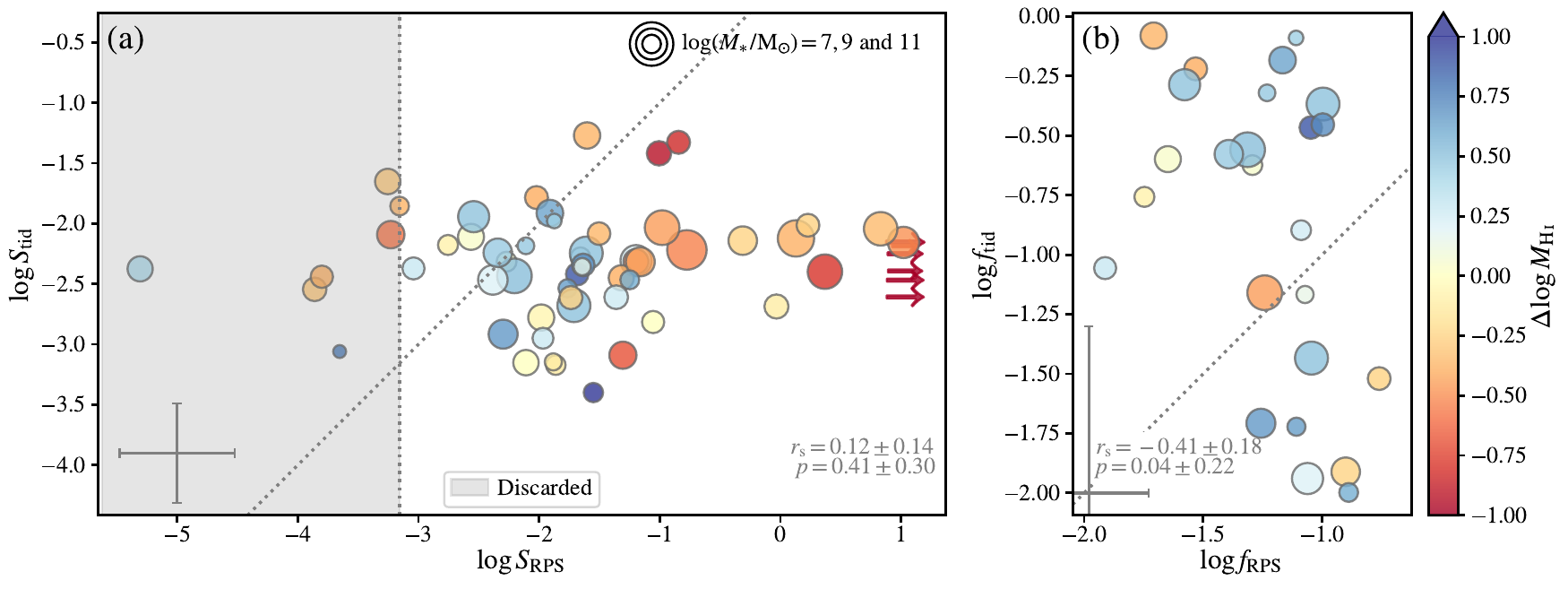}
  \caption{The relation between \rps and tidal interaction among \Hi-detected galaxies.
    Circles represent individual galaxies, color-coded by the deviation from the mean \MHi relation.
    Their radii are proportional to the stellar mass.
    (a)~Environmental strength \Srps vs.~\Stid.
    The Spearman rank correlation~\rs and the $p$-value are given at the lower right corner, the uncertainties of which are obtained with bootstrap.
    The last 10\% of \Srps's are not included in correlation calculation and is covered with gray shading.
    The line of $\Srps=\Stid$ is drawn as the dotted gray line.
    Five galaxies have so little \Hi~gas that $1.5\RHi$ is smaller than \Rzsg.
    As a result, its~\Srps is infinite, and it is plotted as an arrow pointing rightward.
    The typical uncertainties of \Stid and \Srps are given at the lower left corner.
    (b)~Strippable gas fraction \frps vs.~\ftid.
    Only galaxies with both $f$ nonzero are plotted.
    The line of $\frps=\ftid$ is drawn.
    The correlation between \frps and \ftid is given.
    The typical uncertainties of \ftid and \frps are given at the lower left corner (only one side of the error bar is plotted due to limited space).}
  \label{fig:env_compare}
\end{figure*}

We examine the correlation of strengths between RPS and TS with the two types of parameters defined above.

We note that, while \Srps and \Stid are defined in a similar way, they should not be directly compared.
Indeed, their respective roles as \emph{perturbation} and \emph{stabilizer} are quite different.
However, it is reasonable to compare the rank values.
In \autoref{fig:env_compare}(a), we plot two $S$'s against each other.
The \Srps and \Stid show no significant rank correlation, measured by the Spearman rank correlation coefficient ($\rs=0.12$).
This indicates that the instantaneous strengths of \rps and tidal interaction are independent.

In \autoref{fig:env_compare}(b) we plot the two $f$'s, and they show a moderate rank anticorrelation ($\rs=-0.41$), implying that the two effects prefer different working conditions.
Thus for most galaxies, one type of stripping clearly overrides the other one at a time.
We define our \emph{\rps sample} (27 galaxies) and \emph{TS sample} (22) as those with $\frps>\ftid$ and $0\neq\ftid\ge\frps$ respectively.
Among these 49 galaxies, 25 have both \frps and \ftid larger than $0$, indicating that they are undergoing \rps and TS simultaneously.
Like \Stid and \Srps, the uncertainties of \ftid and \frps are dominantly caused by observational projection.
We plot the typical values of uncertainties in the figure, and the estimation procedure is given in \autoref{sec:app:proj}.

\begin{figure*}
  \myplotone{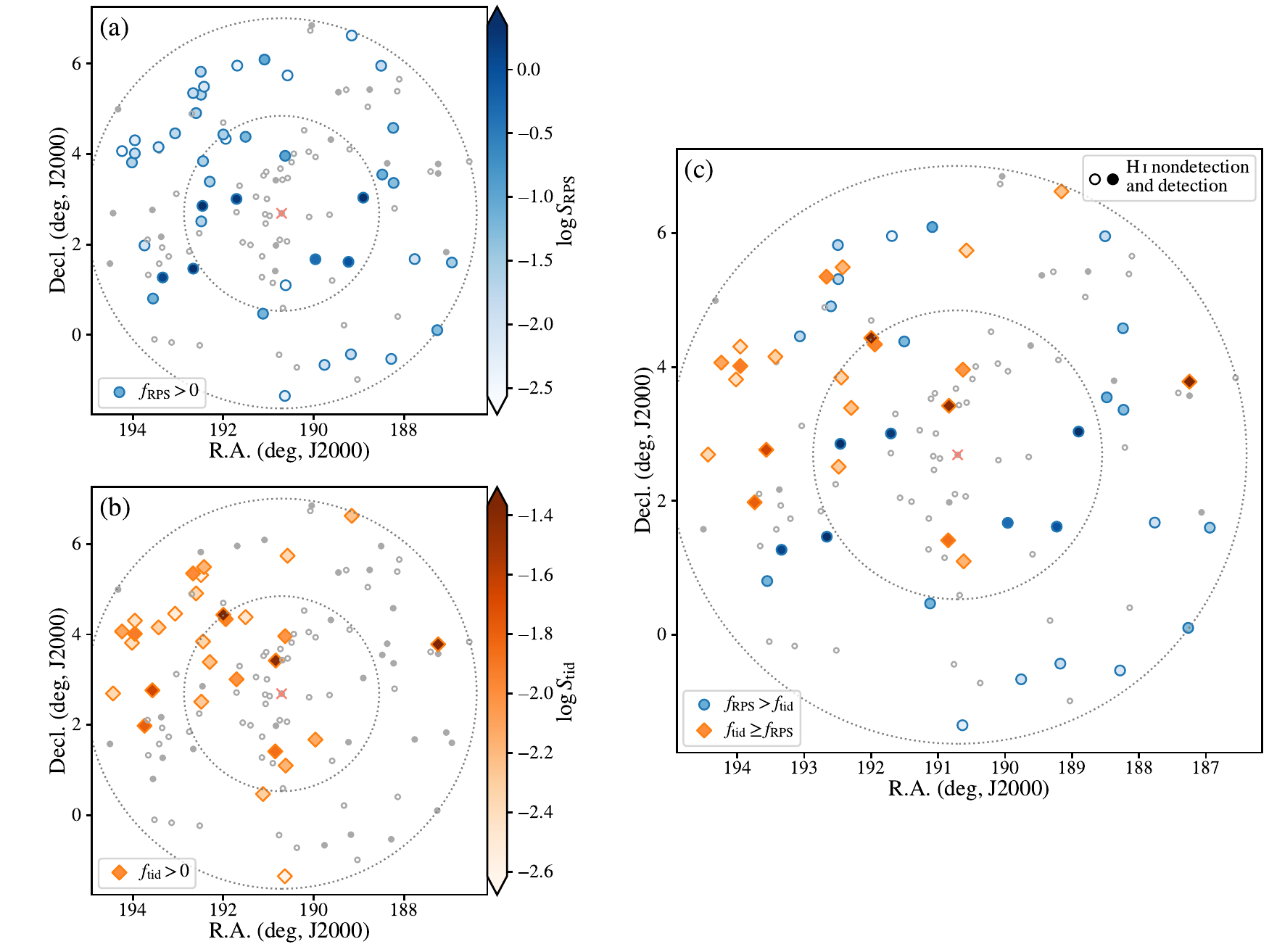}
  \caption{The spatial distribution of gas stripping.
    \Hi~detections and nondetections are plotted as filled and empty symbols respectively.
    Irrelevant galaxies (i.e., their corresponding $f=0$) in each panel are plotted as gray dots.
    Two dotted gray circles of radii 1~and 2~times \Rzoo indicate the range of N4636G, and the group center is labeled as a red cross.
    (a)~The \rps strength map.
    Circles represent galaxies having gas strippable by ram pressure ($\frps>0$), color-coded by~\Srps.
    The limits of the color bar are the 5th and 95th percentiles among all color-coded galaxies.
    (b)~The tidal effect strength map.
    Galaxies having gas tidally strippable ($\ftid>0$) are plotted as diamonds, color-coded by~\Stid.
    The limits of the color bar are selected alike.
    (c)~The stripping status map.
    Large colored symbols are those having strippable gas, blue circles being the \rps sample and orange diamonds the TS sample.
    They are color-coded exactly as those symbols in (a) and~(b) are.
  }
  \label{fig:spatial_dist}
\end{figure*}

We further investigate how the effects of RPS and TS processes are spatially distributed in N4636G\@.
\autoref{fig:spatial_dist}(a) and~(b) are maps of galaxies undergoing gas stripping by ram pressure and tidal interaction respectively, color-coded by the corresponding $S$.
A galaxy is plotted as a filled circle (RPS) or diamond (TS) as long as it has a nonzero value of the corresponding~$f$, no matter how strong the other kind of stripping is.

One remarkable trend is that most galaxies undergoing TS are located in the northeast of the group while the southwestern part is almost void of such galaxies.
In contrast, the galaxies undergoing \rps are spread over the entire volume of the group.
Besides, the galaxies with the largest \Srps's are within the~\Rzoo or the east of N4636G\@.
The value of~\Stid shows no obvious relation with the position in the group.
The trend corroborates the theoretical prediction of RPS and tidal interaction.
At the group center, infalling galaxies get the largest relative velocity, and the ICM is densest, making RPS effective.
While the galaxy number density is also the largest at the group center, which is favorable to TS, the high relative velocity shortens the impact time of possible galaxy--galaxy interactions.

\autoref{fig:spatial_dist}(c) gives the spatial distribution of the TS and \rps sample.
In the southwest of N4636G, \rps galaxies are present, but no TS galaxies are found there.
In the northeast, where the number density is higher, TS galaxies outnumber \rps ones.
As we show in \autoref{sec:app:substructure}, there seems to be abundant substructures in the eastern part of the group.

In summary, the TS and RPS instantaneous strengths and extents seem highly independent in N4636G with our models.
The TS and RPS samples are distributed within a similar projected distance from the group center, but the TS is more sensitive to presence of localized substructures.
This may partly explain the relative independence of their strengths.

\section{Dependence of Galactic Properties on Current Environmental Effects}
\label{sec:env}

\begin{figure*}
  \myplotone{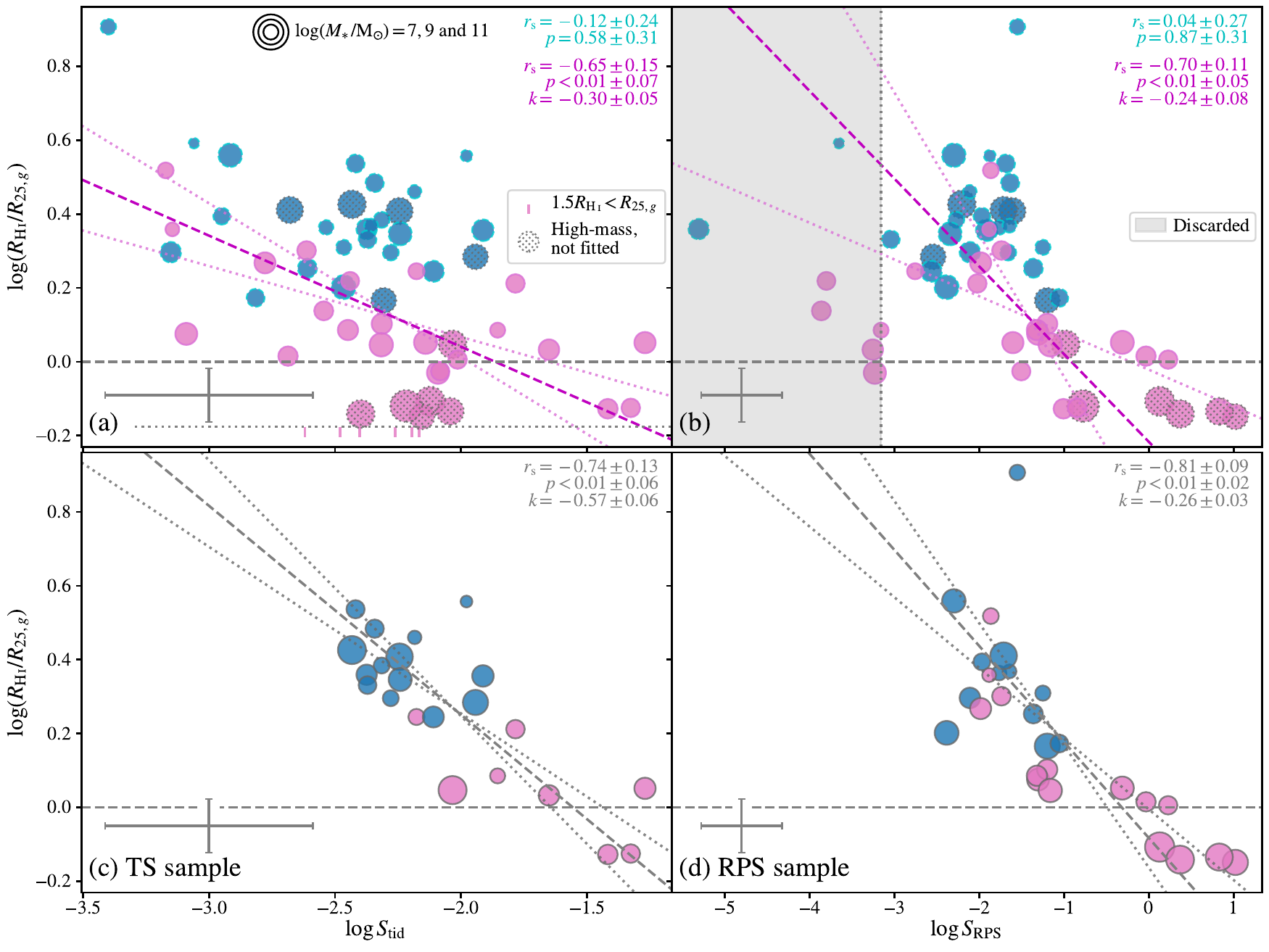}
  \caption{The relation between \Hi-to-optical disk size ratio ($\RHi/\Rzsg$) and strengths of environmental effects in galaxies.
    The size of the symbols have the same meanings as those in \autoref{fig:env_compare}, and gas-rich and -poor galaxies are plotted as blue and pink circles respectively.
    In panels (a) and~(b), high-mass galaxies are hatched with gray lines and are \emph{not} included in the linear fitting or correlation analysis.
    The typical uncertainties are given at the lower left corner of each panel.
    (a)~Size ratio plotted against tidal strength \Stid.
    For \emph{low-mass} gas-poor and -rich galaxies, the Spearman rank correlation~\rs and the $p$-value are given at the upper right corner in respective colors, the uncertainties of which are obtained with bootstrap.
    The bisector linear fitting of the low-mass gas-poor sample is plotted as dashed magenta line, with two orthogonal fittings as dotted lines.
    The bisector fitting line crosses the $\RHi=\Rzsg$ line (dashed gray), at $\log\Stid=\numScritid$.
    The slope $k$ of the line is also reported with the fitting uncertainty.
    (b)~Size ratio plotted against \rps strength~\Srps.
    Galaxies with the lowest 10\% of~\Srps's are not used for correlation calculation or fitting, and they are covered in shading.
    The intercept of the bisector fitting line is $\log\Srps=\numScrirps$.
    Panels (c)~and (d) are the same as (a) and~(b), except that only the TS sample or the RPS sample is plotted, that high-mass galaxies are included, and that the gas-rich and gas-poor samples are not distinguished.
    The intercepts are \numlist{-1.52(27);-0.34(14)}, overlapping the $1\sigma$ confidence intervals of those in (a) and~(b).}
  \label{fig:size_ratio}
\end{figure*}

\begin{figure*}
  \myplotone{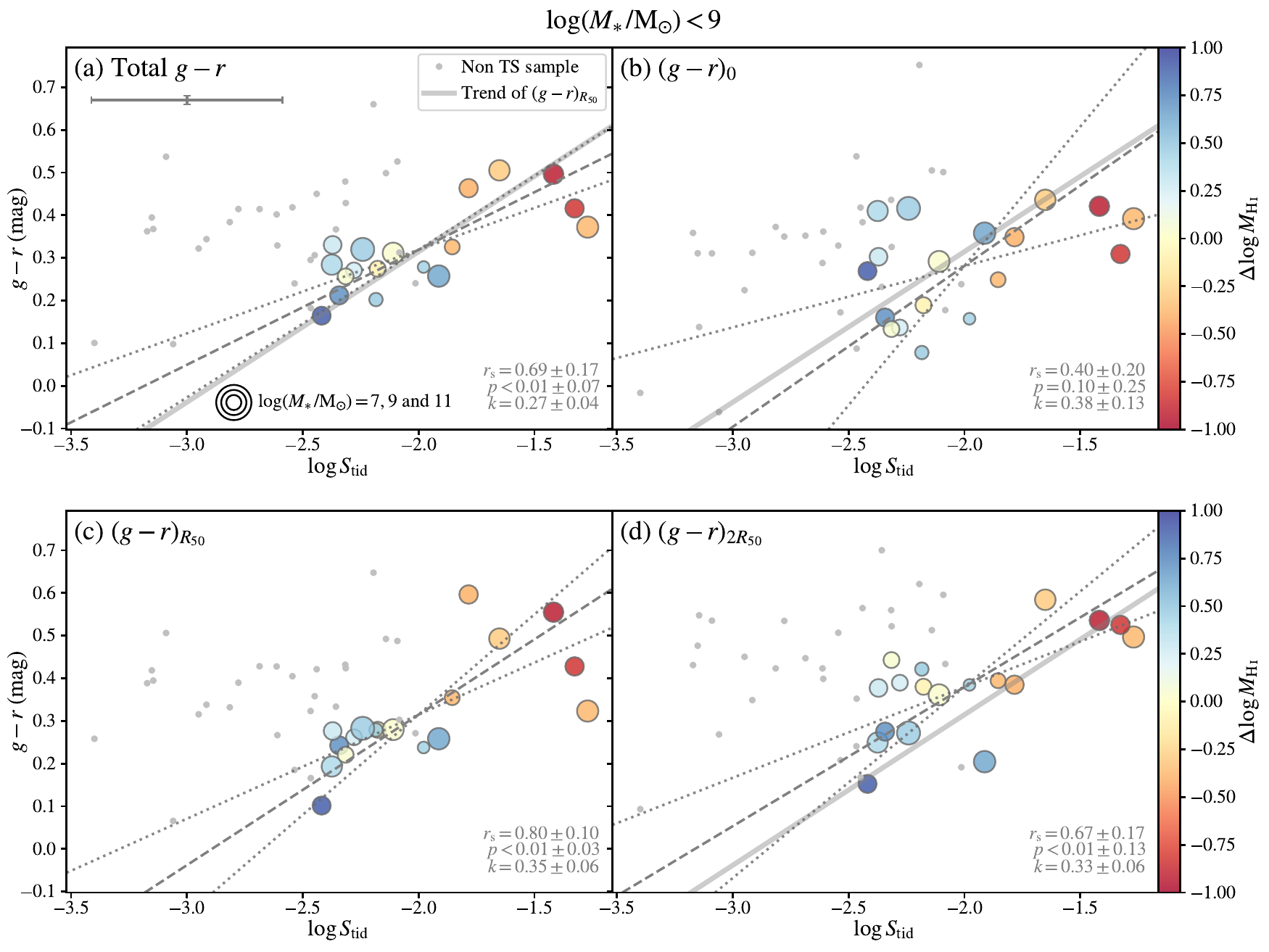}
  \caption{The \gr color of \Hi-detected low-mass galaxies plotted against tidal strength~\Stid.
    The color is measured (a)~using total fluxes, (b)~at the galactic center, (c)~at \Rso, or (d)~at $2\Rso$ (see \autoref{ssec:color} for details).
    Galaxies in the TS sample ($\ftid\ge\frps$) are plotted as circles, color-coded as those in \autoref{fig:size_ratio}.
    The size of circles represents stellar mass.
    For these galaxies, the Spearman correlation coefficient and the bisector fitting are given.
    The fitted line at \Rso is plotted in all other panels as bold gray lines for comparison.
    Remaining \Hi~detections are plotted as gray dots, all of which are not involved in correlation calculation or fitting.
    The typical uncertainties for low-mass TS galaxies are given in panel~(a).}
  \label{fig:s_tid}
\end{figure*}

\begin{figure}
  \myplotone{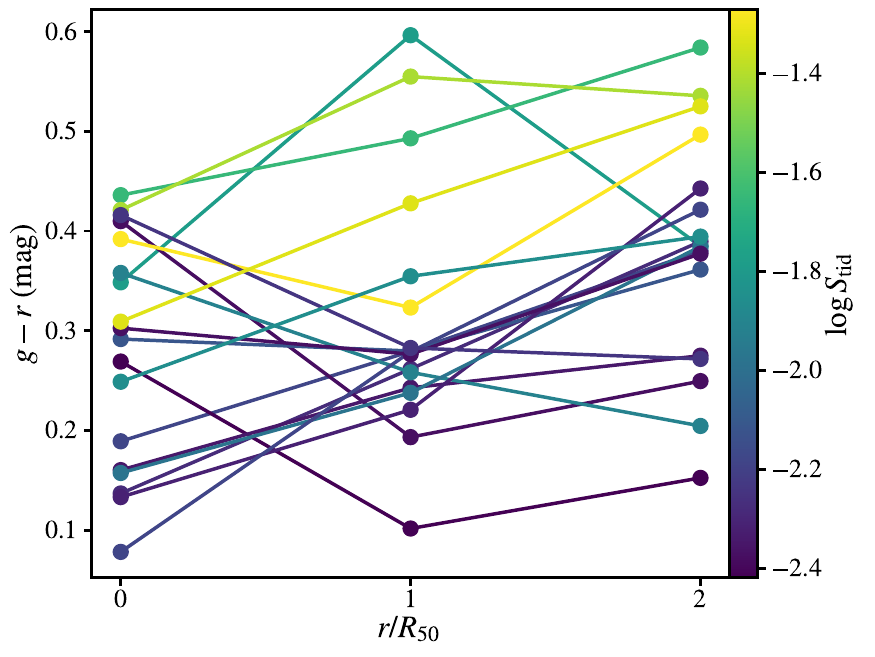}
  \caption{A reillustration of \autoref{fig:s_tid}(b)--(d).
    The \gr color of \Hi-detected low-mass galaxies is plotted against the position of color measurement, color-coded by the tidal strength~\Stid.}
  \label{fig:s_tid_link}
\end{figure}

\begin{figure*}
  \myplotone{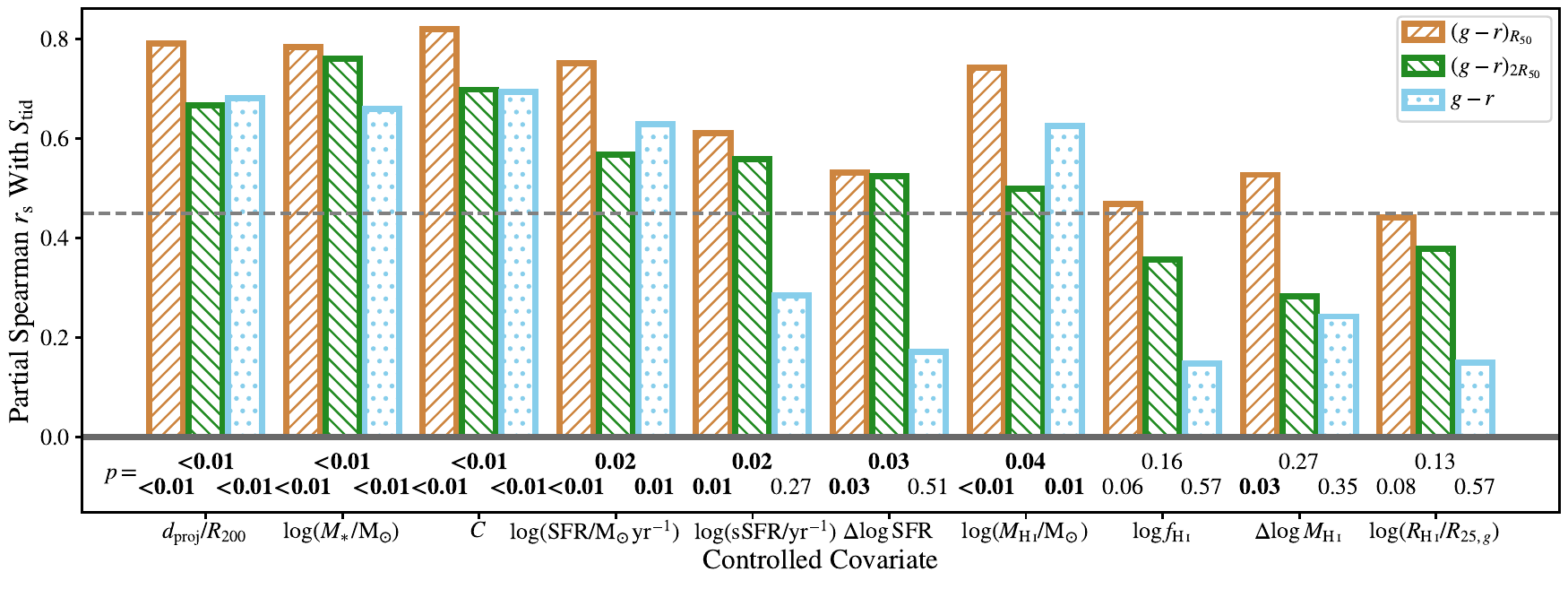}
  \caption{Partial Spearman correlation coefficients~\rs for three $(\gr)$--\Stid relations given in \autoref{fig:s_tid}.
    Orange slashed, green backslashed, and blue dotted bars correspond to the relation at \Rso, at $2\Rso$, and of the whole galaxy respectively.
    The controlled covariates are labeled along the $x$-axis.
    Only low-mass TS galaxies, i.e., colored symbols in \autoref{fig:s_tid}, are used.
    All of them have confident \MHi and \sfr measurements.
    The $p$-values of these partial correlations are listed at the bottom of the figure, boldfaced ones smaller than~$0.05$.
    The threshold for strong correlation, $|\rs| = 0.45$, is drawn as a gray dashed line.}
  \label{fig:s_tid_partial}
\end{figure*}

\subsection{The Change in \Hi~Disk Size with Environmental Strengths}
\label{ssec:disk_size}

\autoref{fig:size_ratio}(a) and~(b) show how the \Hi~disk size~\RHi (deduced from~\MHi, see \autoref{ssec:hi_prop}) changes with \Stid or~\Srps.
We focus on low-mass galaxies (nonhatched circles) since they are thought to be more susceptible to environmental effects \citep{2014A&A...570A..69B}.
Some galaxies have their \Rzsg larger than $1.5\RHi$, the largest radius of our model \Hi profile.
Their gas disks are heavily truncated, indicating that the gas depletion has almost finished.
They will also have an infinite \Srps with our methods.
Thus, we do not include them in our analysis of this section and simply indicate them as short pink bars in \autoref{fig:size_ratio}(a).

In \autoref{fig:size_ratio}(a), no correlation is found between the size ratio and \Stid for \Hi-rich, low-mass galaxies.
However, strong anticorrelations are found for \Hi-poor, low-mass galaxies, as in \citet{2022ApJ...927...66W}.

Similarly, in \autoref{fig:size_ratio}(b), \Hi-rich galaxies show a better correlation between the size ratio and \Srps than \Hi-poor galaxies do.
Also, galaxies that we discarded (with the lowest 10\% of \Srps, under gray shading) do not follow the relation, corroborating the idea that, with such a low \Srps, other mechanisms, such as tidal interaction, play a larger role.

These anticorrelations indicate ongoing gas-stripping by ram pressure or tidal effects, which are likely to happen independently since no correlation is found between \Stid and~\Srps (see \autoref{fig:env_compare}).
The drop of gas-richness (reflected by the size ratio here) should be the consequence of continuous stripping, while $S$'s measure a more instantaneous process.
Therefore, their correlation is likely due to the corresponding environmental process progressively strengthening during the infall of a galaxy.
The stripping can also be self-enhancing:
the removal of gas at the outskirts reduces the local anchoring force, which makes further stripping easier and which enlarges the corresponding $S$, possibly also giving rise to the anticorrelation.

The lack of correlation of the gas-rich sample, however, does not necessarily mean the absence of gas-stripping.
It is possible that, for many of these gas-rich galaxies, neither TS nor RPS is well established, since the gas-rich sample also contains more newcomers to the group.
This would be more significant if the gas-stripping is a cumulative process.
It is also possible that weaker effects, like starvation, are more significant than direct stripping for gas-rich galaxies at this stage.

\subsubsection{The Deviation of Critical Stripping Strength}
\label{ssec:s_crit}
The fitted relation [\autoref{fig:size_ratio}(a)] for low-mass, gas-poor galaxies between the size ratio and~\Stid intersects the line of $\RHi=\Rzsg$.
The intercept marks a critical value $\log\Scri{tid}=\numScritid$,%
\footnote{The uncertainty is derived from bisector fitting results, and consistent with the uncertainty from projection.}
where the \Hi~disk \emph{has been} stripped close to the radius \Rzsg while $\Stid(\Rzsg)=\Scri{tid}$.
It naturally measures the level of~\Stid needed for the TS to have obvious cumulative consequence, and we adopt it in the calculation of~\ftid.
Due to the aforementioned chronology, the \ftid values might be slightly overestimated, since there is a delay between the onset and the completion of stripping.
However, the overestimation is mitigated by the fact that \Hi~disks can actually extend farther out than \RHi.

In principle, we may calibrate a similar $\log\Scri{\rps}$ of \numScrirps, and define \frps in a similar way to \ftid.
However, we notice the relatively large scatter and sparsity of data points near the intercept of \autoref{fig:size_ratio}(b), and we thus prefer to stick to the well-established model of \citet{1972ApJ...176....1G}.
Nevertheless, we discuss how our major results may change if we use \Scri{\rps} to derive \frps in \autoref{ssec:app:strip_model}.

With the aforementioned intercepts, we could calculate the strippable gas fraction and select the RPS and TS samples as mentioned in \autoref{ssec:frps_ftid}.
We replotted \autoref{fig:size_ratio}(a) and~(b) as \autoref{fig:size_ratio}(c) and~(d) with galaxies in the corresponding sample only.
Galaxies in both samples show significant anticorrelations of the two parameters regardless of the gas-richness and stellar mass.
This implies that the noncorrelation of the gas-rich sample in \autoref{fig:size_ratio}(a) and~(b) could be due to the fact that the relevant stripping process needs more time to show its effects.
Besides, the trends in \autoref{fig:size_ratio}(c) and~(d) are much more tight.
It is possibly because, by selecting the TS and \rps sample, we removed galaxies with lower signal-to-noise ratio in the $S$ measurements, and simplified a more complex mixture of physical processes.
The scatter is thus lowered.

We note that relations in \autoref{fig:size_ratio}(a) and~(b) are flatter than those in (c) and~(d), probably due to the coexistence of two stripping mechanisms among many galaxies.
The relations in \autoref{fig:size_ratio}(c) and~(d) give compatible intercepts with those given in (a) and~(b), suggesting that our derivation of critical stripping strength is self-consistent.

\subsection{The Radial Change of Color Due to Tidal Interaction}
\label{ssec:color_tidal}

Tidal forces could induce the redistribution of cold gas, either by stripping or by inflow.
We examine the relation between \Stid and both the localized and global colors for the N4636G galaxies.
When optimizing the photometric pipeline, we have separated the galaxies into two subsets, the low- and high-mass ones, with a dividing line at $M_*=\qty{1e9}{\Msun}$.
Such a division is also supported by different patterns of reddening found in \citet{2022ApJ...927...66W} for the Eridanus group galaxies.
Based on past ALFALFA observations, this line also divided galaxies into two groups exhibiting different properties \citep{2012ApJ...756..113H}, and in theory, it separates the regimes of galaxies being less and more prone to stellar feedback \citep{2005ARA&A..43..769V}.
We focus on the low-mass galaxies in the following because the high-mass TS subset is small (three galaxies).
We notice that adding the high-mass galaxies to the low-mass subset would simply add noise to our calculations (shown in \autoref{sec:app:highmass_ts}).

\subsubsection{The Trends}
We show in \autoref{fig:s_tid}(a) that, for low-mass galaxies in the TS sample, the Spearman test indicates significant correlations of \gr with~\Stid.
In contrast, no significant correlations are found for the low-mass galaxies in the RPS sample [gray dots in \autoref{fig:s_tid}(a)], as TS is not the dominating mechanism for these galaxies.
Similar correlations of low-mass galaxies in the TS sample are found for \gr[\Rso] and \gr[2\Rso] with~\Stid, but only tentative for \gr[0], as other panels of \autoref{fig:s_tid} show.

We fitted the color with \Stid at three radii, and the bisector-fitted lines are shown in \autoref{fig:s_tid}.
Compared with the relation at \Rso (shown as the bold gray line), the line fitted at $2\Rso$ (or galactic center) is redder (or bluer) than it.
Besides, three fitted lines are parallel within the uncertainties.
This indicates that, under tidal interaction, these galaxies statistically undergo a rather uniform reddening while retaining a blue core.

We exclude the possibility that the reddening is the result of dust attenuation, instead of tidal interaction, using the method by \citet[their Section~5.2.1 and Figure~12]{2022ApJ...927...66W}.
The global level of attenuation, $A_g-A_r$, is estimated from the W4-band and total \sfr measurements \citep{2000ApJ...533..682C,2007ApJS..173..293W}.
A tight Spearman \emph{anticorrelation} is found between $A_g-A_r$ and \Stid with $p<0.05$, which strengthens the trends in \autoref{fig:s_tid}.

To better illustrate this uniform reddening, we replot \autoref{fig:s_tid} as \autoref{fig:s_tid_link}, where the localized color is displayed as a function of galactocentric radius.
The lines are color-coded by \Stid.
Among these 18 galaxies, 12 have $\gr[\Rso]-\gr[0]>\qty{-0.03}{\mag}$ and $\gr[2\Rso]-\gr[\Rso]>\qty{-0.03}{\mag}$, where \qty{-0.03}{\mag} accounts for uncertainty.
If we do not consider the uncertainty, nine galaxies are strictly blue-cored.

We note that both \gr color (global, at \Rso, or at $2\Rso$; see Figures \ref{fig:s_tid} and~\ref{fig:s_tid_link}) and \Hi richness [\autoref{fig:size_ratio}(c)] are found to depend on \Stid.
They are known to correlate with each other, and both of them are known to correlate with other galactic properties.
We thus use the partial Spearman rank correlation to find out how significant ($p$-value) and strong (\rs-value) \Stid directly influences the colors.
We controlled for 10 possible parameters and calculated the respective partial rank coefficients between three different colors and \Stid, shown in \autoref{fig:s_tid_partial}.
The correlations of all colors are still significant after fixing \dprj, concentration $C$, or aggregate values like stellar mass, \MHi, or \sfr.
But after being controlled for specific values from which the $M_*$ is effectively divided [such as specific \sfr (\ssfr), $f\sbHi$, \delSFR, or \delMHi{}], the color parameters show weaker correlations with \Stid, especially the global \gr.
Nevertheless, \gr[\Rso] still shows significant partial correlation in most cases.

These results indicate that tidal effects possibly affect the global color indirectly, through changing the global \Hi richness ($f\sbHi$ or \delMHi), while the change in global color reflects a change in the integral star-forming status (\delSFR or \ssfr).
It seems, however, that the localized colors, especially at \Rso, are affected in more direct ways, possibly by causing gas-inflows or changing the localized star-formation conditions.

\subsubsection{Comparison with Trends in the Eridanus Group}

\begin{figure*}
  \myplotone{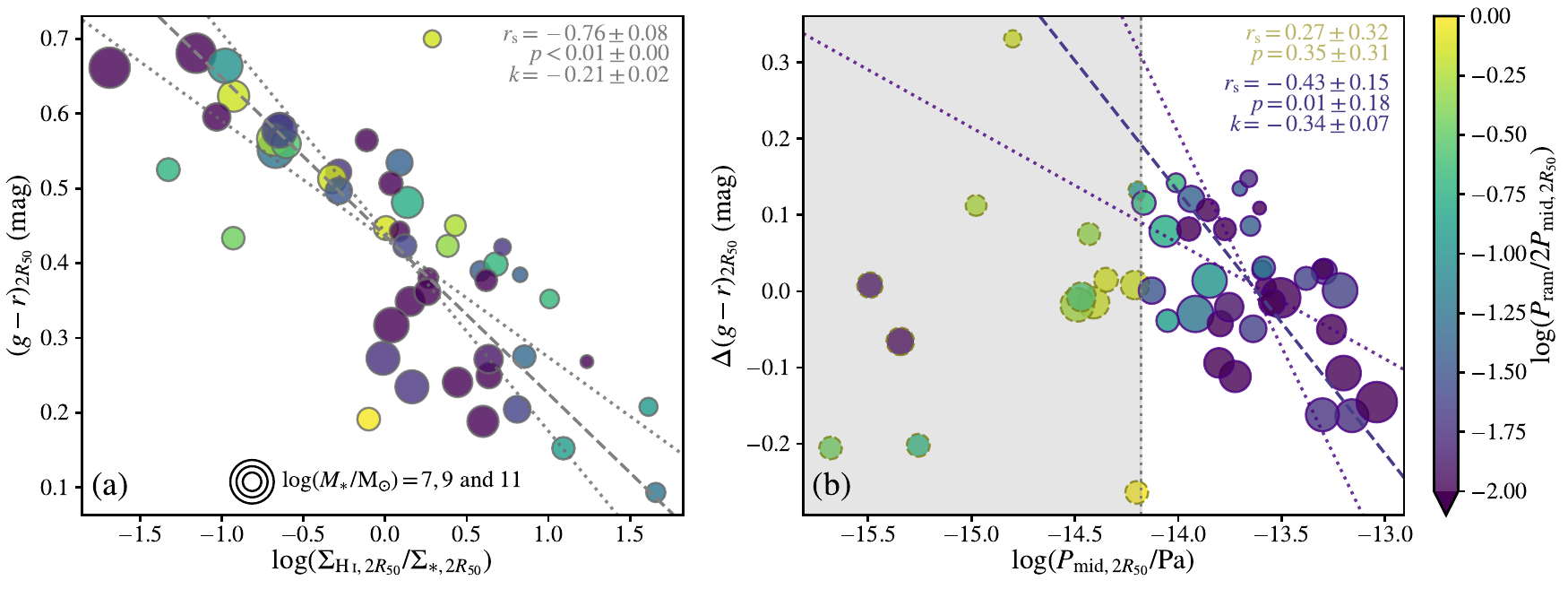}
  \caption{Additional effects of ram pressure~\Pram on the \gr color at $2\Rso$.
    Only galaxies whose predicted \Hi-disk sizes satisfy $1.5R\sbHi>2\Rso$ are plotted, color-coded by the importance of~\Pram to the mid-plane gas pressure at $2\Rso$, i.e., \Pmid[2\Rso].
    The size of circles represents stellar mass.
    (a)~The relation between the \gr color at $2\Rso$ and the ratio of \Hi~mass density to stellar mass density at $2\Rso$.
    The bisector linear fitting result is plotted, and the slope is provided.
    The Spearman rank correlation and $p$-value of the relation are given with bootstrapped uncertainty.
    (b)~The deviation of \gr[2\Rso] from the fitted relation in~(a) vs.~\Pmid[2\Rso].
    A threshold of $\log\left(\Pmid[2\Rso]/\unit{Pa}\right)$ ($\Pmidthres$, vertical dotted gray line) is chosen to let samples above it have the strongest correlation.
    The Spearman correlation coefficient and the bisector fitting are given in dark purple.
    The correlation of galaxies below the threshold is listed in yellow.}
  \label{fig:p_mid}
\end{figure*}

A similar study \citep{2022ApJ...927...66W} conducted for low-mass galaxies in the Eridanus group found the reddening of \gr with \Stid to be most prominent in the central region, and less significant at or beyond \Rso, contrary to our findings in the N4636G\@.
Such a difference between the reddening patterns of these two groups indicates that the replenishment of star-forming gas near galaxy centers should be either suppressed in the Eridanus group or enhanced in the N4636G\@.

It is known that unperturbed dwarf irregular galaxies tend to show positive color gradients \citep[``blue cores,'' e.g.,][]{2012AJ....143...47Z}.
Such a pattern is possibly maintained by fountain-driven gas-inflow \citep{2014ApJ...796..110E}, or the higher star-formation efficiency (SFE) at the galaxy center \citep{2020A&A...644A.125B}.
In groups, fountain-driven gas-inflow may be suppressed by the RPS of the circumgalactic medium \citep[CGM;][]{2018ApJ...863...49B};
also, gas affected by tidal interactions may have a high level of turbulence, which can suppress the star formation \citep{1902RSPTA.199....1J,2011ApJ...741...12B}.

These effects may explain why Eridanus galaxies failed to maintain blue cores when they started to redden.
The question then becomes how N4636G galaxies, while undergoing global reddening, manage to better replenish their inner disks with gas than Eridanus galaxies do.

Closely comparing two groups, we find that their ranges of \Stid are similar, but N4636G galaxies are systematically \Hi-richer.
The median \delMHi of \Hi-detected galaxies in N4636G (\qty{-0.11}{\dex}) is significantly higher than that in Eridanus (\qty{-0.34}{\dex}), and there are more \Hi-detected galaxies in N4636G (43\%) above the \MHi--$M_*$ relation than in Eridanus \citep[17\%;][]{2021MNRAS.507.2300F,2022ApJ...927...66W}.
This difference may explain the different reddening patterns.
Aside from its tidal-stripping effect, galaxy interactions are a major factor responsible for gas-inflows in galaxies at low redshift, which is more efficient in gas-rich environments, i.e., when the neighbors are gas-rich \citep{2018MNRAS.479.3952B,2019ApJ...882...14M}.
This is because the extended \Hi disks are more likely to encounter each other and/or interact with the CGM, which cause shocks and ram pressure compression \citep{2019ApJ...882...14M}.
As a result, massive gaseous clumps are likely to form in the outer disks, which impose torques and drive inflows and/or migrate radially in as a result of dynamic friction \citep{2018MNRAS.479.3952B}.
Thus, the N4636G galaxies, which on average have more \Hi-rich neighbors than the Eridanus ones do, have more efficient tidally driven gas-inflows, being more likely to retain blue cores.

As to why Eridanus galaxies are on average more \Hi-poor than N4636G galaxies, the reason might be in the different dynamic status of the two groups.
In the middle of a major merger with another two groups, the Eridanus group has a spatial extent that is \num{\sim 2.5}~times%
\footnote{Here we compare the maximum radial extents of the Eridanus group and of the NGC~1332 group reported by \citet{2006MNRAS.369.1351B}.
  They have similar \Rsoo deduced from respective velocity dispersion.}
larger than expected for its velocity dispersion \citep{2006MNRAS.369.1351B}.
It indicates that its member galaxies used to be much closer to each other in the past than in the currently observed snapshot.
The tidal interaction in the past might also be much stronger than that indicated by the current \Stid, which has efficiently accelerated the removal and consumption of the \Hi.
This indicates that, aside from the virial mass, the dynamic status of groups also plays a role in determining the current properties of the galaxies within.

\subsection{Additional Influence of Ram Pressure on Color}
\label{ssec:color_rp}

We searched for signatures of color reddening as a result of \rps.
We conducted a similar analysis as in \autoref{ssec:color_tidal} for RPS, but do not find significant trends.
The reason is possibly that removing outlying \Hi by \rps does not significantly affect star formation in the optical disk \citep{2021PASA...38...35C}, at least not as much as by tidal interaction as suggested by our results.

On the other hand, we also searched for signatures of excess blue colors (indicating star-formation enhancement) induced by ram pressure.
The conversion efficiency of \Hi to the molecular gas, thus the SFE for \Hi gas, is reported to be related with the local midplane interstellar pressure \citep{2002ApJ...569..157W,2008AJ....136.2782L,2010ApJ...721..975O}.
It is suggested that ram pressure might increase this midplane pressure and, as a result, boost star formation \citep[e.g.,][]{2016AJ....151...78P,2017MNRAS.467.4282M,2018ApJ...866L..25V,2020ApJ...899...98V}.
Although there are also observations against this \citep[e.g.,][]{2012A&A...543A..33V}, several recent studies found enhanced molecular gas fractions in \rps galaxies in clusters \citep[e.g.,][]{2020ApJ...901...95C,2020ApJ...897L..30M,2021ApJ...921...22C,2022ApJ...941...77R}.

To verify this idea, we use the formula below \citep{2010ApJ...721..975O} to calculate the midplane gas pressure:
\begin{equation}\label{eq:p_mid}
  \Pmid=\frac{1}{2}\Pram+\pi G\left(\Sig{*}+\frac{1}{2}\Sig{\text{gas}}\right)\Sig{\text{gas}},
\end{equation}
assuming that ram pressure is exerted on one side of the disk, which gives rise to the first $1/2$ factor.
Here \Sig{*} and \Sig{\text{gas}} are the deprojected stellar and cold-gas surface density respectively.
Their derivations are described in \autoref{sec:app:fstr}.
\autoeqref{eq:p_mid} implies that \Pmid depends also on the local baryonic surface density, and the increase of~\Pmid caused by~\Pram would therefore be more significant on the periphery.
We thus focus on the~\Pmid and \gr color at~$2\Rso$, the latter as an indicator of local long-term \ssfr.

The localized \gr color as a proxy for \ssfr is naturally correlated with the localized gas fraction \citep{2020ApJ...890...63W}.
It is thus necessary to control for the effect of $\Sig{\tHi}/\Sig{*}$ before investigating any further dependence of \gr on \Pram.
In \autoref{fig:p_mid}(a), we fitted the relation between \gr[2\Rso] and $\log\left(\Sig[2\Rso]{\tHi}/\Sig[2\Rso]{*}\right)$, which is tight.
The role of~\Pram would then reflect in the \emph{color offset} from this relation, $\inc\gr[2\Rso]$.
\autoref{fig:p_mid}(b) shows the relation between it and the midplane pressure.
The points are color-coded by $\log(\Pram/2\Pmid[2\Rso])$, i.e., the fraction of the \Pram~contribution in~\Pmid\ [factor $1/2$ is included due to \autoeqref{eq:p_mid}].

For galaxies with a~\Pmid[2\Rso] larger than the threshold of \qty[parse-numbers=false]{10^\Pmidthres}{\Pa}, the color offset and the midplane pressure show a significant negative Spearman correlation.
This anticorrelation does not strongly depend on the threshold.
Thresholds ranging from \qty[parse-numbers=false]{10^{-14.2}}{\Pa} to \qty[parse-numbers=false]{10^{-13.8}}{\Pa} mostly give $p$-values below $0.05$, and \qty[parse-numbers=false]{10^\Pmidthres}{\Pa} gives the lowest $p$-value.
For these galaxies, $\Pram/2$ comprises at most 25\% of the whole \Pmid.
Therefore, although there truly is a regulation of star formation by the midplane pressure, it is hard to assess the role of~\Pram.

The~\Pram plays a more important role in~\Pmid for galaxies below the threshold.
However, there is no correlation between $\inc\gr[2\Rso]$ and~\Pmid.
This might be so because the pressure-regulating mechanism requires a threshold for star formation to take place, or the star-formation enhancement by ram pressure demands a certain direction of orbit and is thus rare.
We also note that \gr generally traces star formation on a gigayear timescale, and thus the lack of correlation could be due to the \Pram-triggered star-formation enhancement being too short-lived or too weak for \gr to indicate.
The H\textalpha{} emission may provide a better tracer for star-formation enhancement of this kind (X. Lin et al.\ in preparation).
Nevertheless, we find no evidence of a statistical link between \Pram and a significant (or long-term) increase of star formation in these galaxies.

\section{The Process of Gas Stripping}
\label{sec:strip}

\subsection{The Dependence of \texorpdfstring{$f$}{f} on Galactic Properties}
\label{ssec:f_str}

\begin{figure*}
  \myplotone{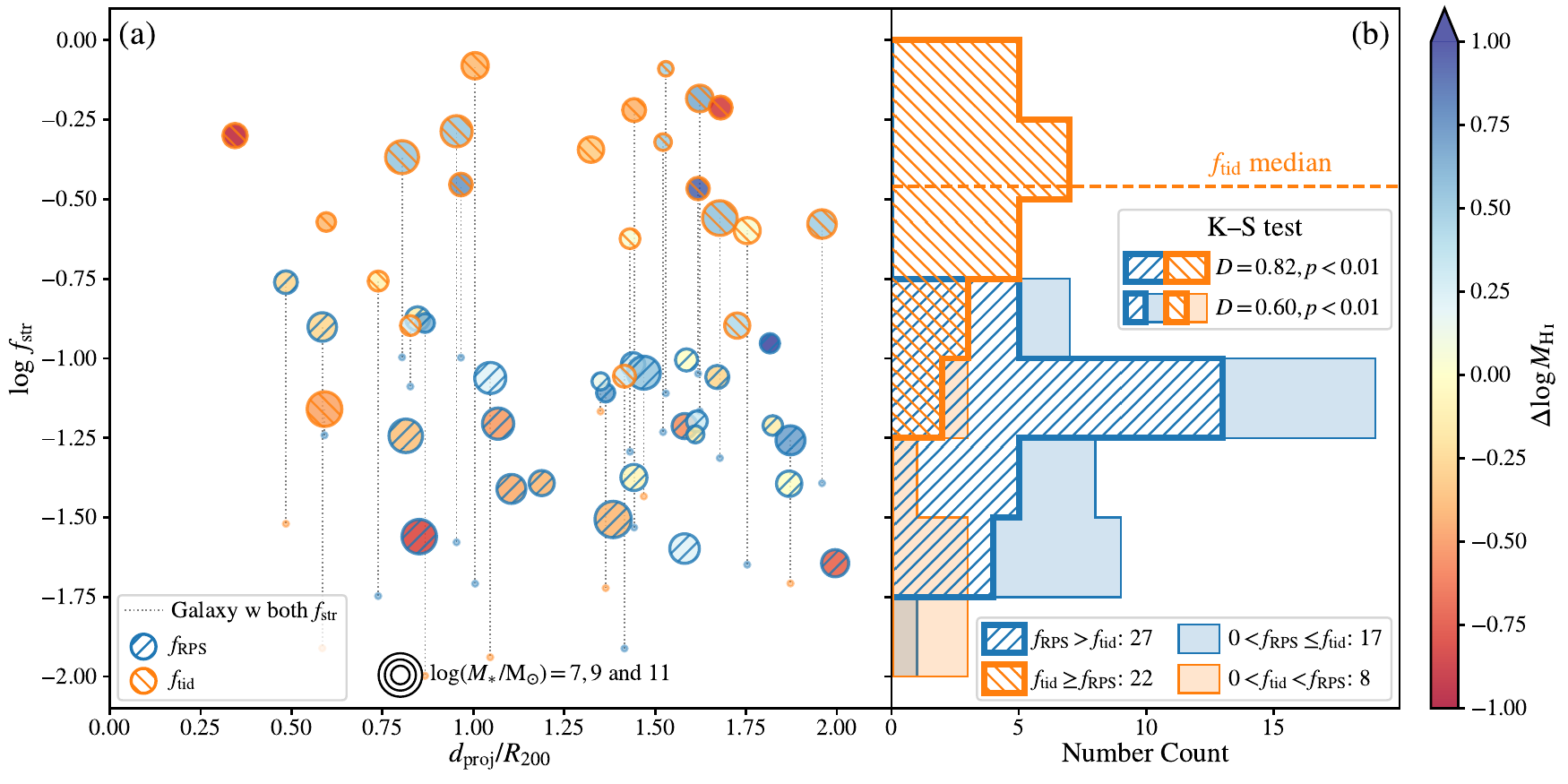}\\
  \myplotone{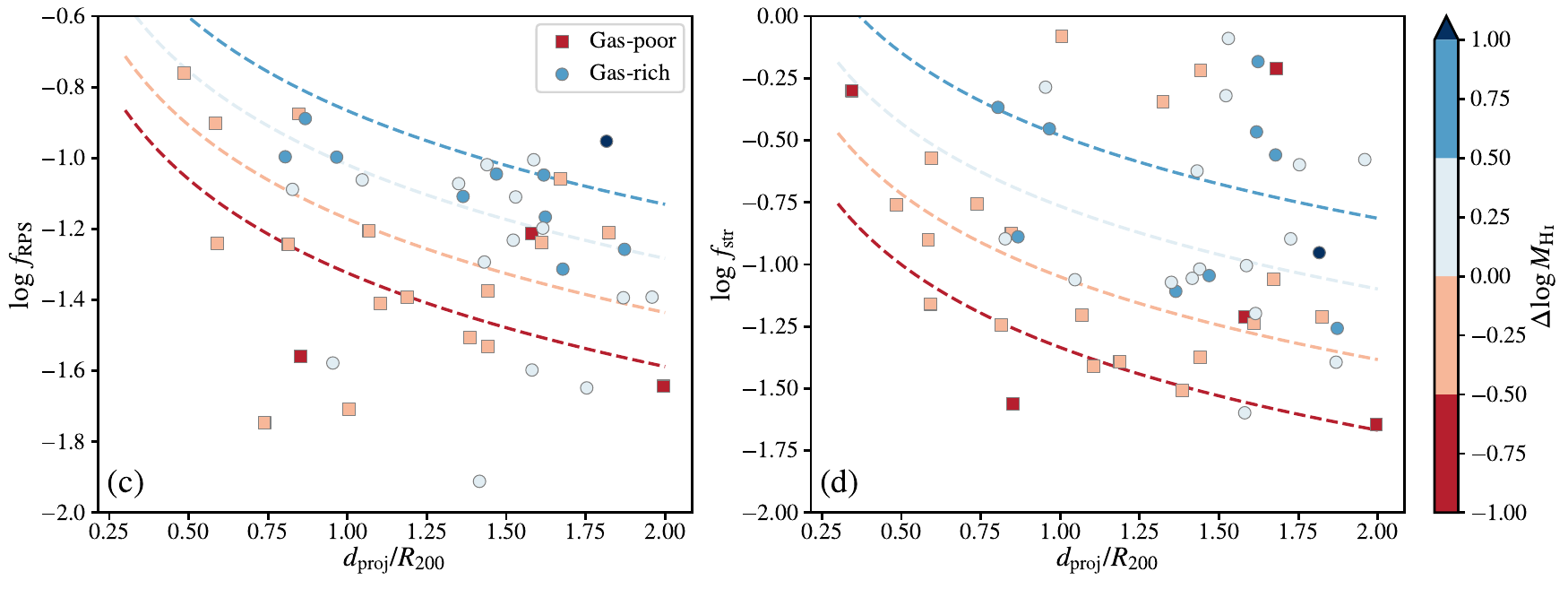}
  \caption{Comparison of the fractions of gas strippable by ram pressure (\frps) and by tidal interaction (\ftid).
    (a)~How the strippable gas fraction~\fstr changes with the projected distance~\dprj from the center of N4636G\@.
    If a galaxy has both \frps (blue-edged slashed) and~\ftid (orange-edged backslashed) positive, the larger one is plotted as a circle, with the other plotted as a translucent dot.
    These two symbols are connected with a gray dotted thin line.
    The circle, color-coded by \delMHi, has a radius related to the stellar mass.
    Thus the blue-edged circles and orange-edged circles correspond to the \rps and TS sample respectively.
    (b)~Distributions of the~\frps of the \rps sample (blue slashes) and the~\ftid of the TS sample (orange backslashes).
    The positive \frps of the TS sample (translucent blue patch) and the positive \ftid of the \rps sample (translucent orange patch) are stacked as well.
    The results of K--S tests between \frps and~\ftid, either limited to the \rps and TS samples or including all nonzero values, are given.
    An orange dashed line is drawn at the median value of~\ftid.
    (c)~Fitting result of \frps using \autoeqref{eq:f_fit}.
    Galaxies are divided (and color-coded) into four bins according to \delMHi, and the four corresponding fitted lines are plotted as dashed lines of respective colors.
    Squares (and circles) are gas-poor (gas-rich) galaxies.
    (d)~Fitting result of \fstr.
    The symbols are the same as (c).}
  \label{fig:f_str}
\end{figure*}

\begin{deluxetable}{c*{3}{R@{; }L}}
  \tablecaption{Partial Rank Correlations of $f$'s.}
  \label{tab:f_str_corr}
  \tabletypesize{\footnotesize}
  \tablehead{
    \colhead{Variable} & \multicolumn2c{\frps}
    & \multicolumn2c{\ftid} & \multicolumn2c{\fstr}
  }
  \startdata
  % \begin{tabular}{c*{3}{S[table-format=+1.2]@{; }S[table-format=1.2]}}
%   \toprule
%   {Variable} & \mc{2}{c}{\frps} & \mc{2}{c}{\ftid} & \mc{2}{c}{\fstr} \\
%   \midrule
  \dprj      & -0.48            & 0.01             & 0.10             & 0.67 & -0.23 & 0.11 \\
  \delMHi    & 0.55             & <0.01            & -0.01            & 0.98 & 0.32  & 0.03
%   \\\bottomrule
% \end{tabular}

  \\\enddata
  \tablecomments{The partial Spearman rank correlation coefficient \rs and the $p$-value are presented side-by-side, separated by a colon.
    The coefficient is calculated between the column and row headings;
    the other parameter in column ``Variable'' is set as the controlled covariate.
    For example, the \rs value of $-0.48$ is calculated between \frps and \dprj after controlling for \delMHi.
    Please refer to \autoref{fig:f_str}.}
\end{deluxetable}

\begin{deluxetable}{cCCCC}
  \tablecaption{Fitted Parameters of \autoeqref{eq:f_fit}.}
  \label{tab:f_str_fit}
  \tabletypesize{\footnotesize}
  \tablehead{
    $f$ & \colhead{$A$} & \colhead{$B$} & \colhead{$C$} & \colhead{$\sigma_{\log f}$}\\
    &&&& \colhead{(\unit{\dex})}
  }
  \startdata
  % \begin{tabular}{cS[table-format=+1.2(1)]S[table-format=1.2(2)]S[table-format=+1.2(2)]S[table-format=1.2]}
%   \toprule
%   $f$   & {$A$}        & {$B$}       & {$C$}        & {$\sigma_{\log f}$} \\
%         & {}           & {}          & {}           & {(\unit{\dex})} \\
%   \midrule
  \frps & -1.10\pm0.04 & 0.31\pm0.09 & -0.88\pm0.22 & 0.16 \\
  \fstr & -0.91\pm0.05 & 0.57\pm0.11 & -1.11\pm0.28 & 0.29
%   \\\bottomrule
% \end{tabular}

  \\\enddata
  \tablecomments{Fitting results of \frps and \fstr are given with uncertainty.
    The scatters around the best fitting relation after excluding outliers, $\sigma_{\log f}$, are also listed.
    Please refer to \autoref{fig:f_str}(c) and~(d).}
\end{deluxetable}

The strippable \Hi fraction $f$ is intended to describe the instantaneous extent of stripping in an \Hi disk.
\autoref{fig:f_str}(b) presents the value distribution of \frps and \ftid.
The Kolmogorov--Smirnov (K--S) tests show that their spatial distribution is different.
Tidal interactions can induce a larger extent of simultaneous stripping than ram pressure does.
Yet \rps is spatially more widespread than TS (see \autoref{fig:spatial_dist}) and as a result should not be neglected in a study of the galactic environment.
Our RPS sample is actually more populated than the TS sample.

\autoref{fig:f_str}(a) shows \ftid (orange-edged backslashed) and \frps (blue-edged slashed) as a function of projected distance~\dprj.
Under the assumption that the outermost gas is most susceptible to stripping, \ftid and \frps cannot be added, and the larger one is dominant.
So if a galaxy has both strippings ongoing, the smaller $f$ is plotted as a dot in \autoref{fig:f_str}(a).
We take the larger one as the strippable gas fraction
\begin{equation}
  \fstr = \max\left\{\frps,\ftid\right\}.
\end{equation}
Thus all filled circles in \autoref{fig:f_str}(a) are \fstr data points.

We find that neither \ftid nor \frps shows a significant correlation with \dprj.
The reason is likely that $f$ further depends on the gas-richness of galaxy:
a galaxy with an extended \Hi~disk is likely to have more gas under stripping than gas-poor ones do under the same conditions.
In \autoref{fig:f_str}(a), the data points are colored by the \Hi richness \delMHi and it shows that gas-rich galaxies are generally indeed located at the region of large $f$ and \dprj.
We further verify this scenario by calculating the partial rank correlations and report the results in \autoref{tab:f_str_corr}.
The \frps is dependent on \delMHi (or \dprj) at a given \dprj (or \delMHi), and \fstr is dependent on \delMHi at a given \dprj.
The dependence of \fstr on \dprj is not that significant due to the inclusion of \ftid, which depends on neither parameter (consistent with the discussion in \autoref{ssec:ts_rps_comparison}).

We thus conclude that with a given \dprj, the extent of stripping, \fstr, increases with an increasing \delMHi, and that \fstr also tentatively increase toward the group center.
The shape of this trend is largely set by the \rps, while the TS mainly contributes by systematically elevating the level of \fstr.
We fit the relation of \fstr and \frps as a function of \dprj and \delMHi,
\begin{equation}
  \label{eq:f_fit}
  \log f = A + B\delMHi + C \log\frac{\dprj}{\Rzoo},
\end{equation}
where $A$,~$B$, and~$C$ are the fitting parameters.
The fitting results for \frps (\fstr) are presented in \autoref{tab:f_str_fit} and are plotted in \autoref{fig:f_str}(c) and (d).
There is only one galaxy with $\delMHi=1.62>1$, and we exclude it from fitting.
During the fitting of \fstr, we further exclude six outliers with $2.25\sigma$ clipping.
The final fitted relations have rather uniform scatters $\sigma_{\log f}$, which are also reported in \autoref{tab:f_str_fit}.

\subsection{Empirically Modeling the Stripping Process}
\subsubsection{The Motivation}

\begin{figure}
  \myplotone{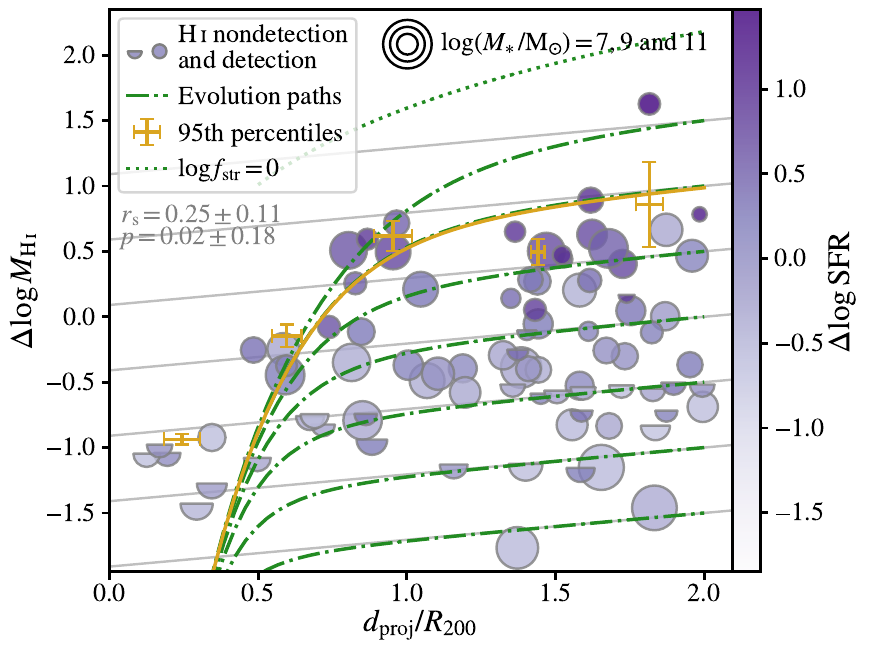}
  \caption{The evolution of galaxies' \delMHi as they infall.
    Only galaxies with an $\ssfr>\qty{1e-11}{\per\yr}$ are plotted in this figure.
    \Hi~detections (circle) and nondetections (semicircle, plotted using the upper limit) are color-coded by \delSFR.
    The radius of these circles and semicircles correlates with the stellar mass.
    The medians of \dprj and the 95th percentiles of \delMHi (including \Hi~nondetections but not low-\ssfr galaxies) in five even \dprj~bins are given in orange, and the uncertainties are estimated with bootstrap.
    Hypothetical evolutionary paths for infalling galaxies with different initial \delMHi are plotted as green dashed--dotted lines.
    The Spearman rank correlation and the $p$-value are given.
    The galaxy pair are indicated by cyan-edged triangles.
    The line of $\fstr=100\%$ is plotted as green dotted lines, above which galaxies should not appear in our model.
    The evolutionary path with star formation as the only gas consumption is plotted as gray lines for reference.}
  \label{fig:logmhi_drop}
\end{figure}

A major goal of quantifying the strengths of \rps and TS is to find out how they together shape the \Hi distribution (and subsequently influence star formation) of satellite galaxies.
In \autoref{sec:env} we have already discussed the respective influences of these two environmental processes on \Hi~gas and optical colors,
while, in \autoref{ssec:f_str} we have shown how gas stripping is prevalent and varies throughout the group.

Correspondingly, the gas content and the star-formation status of infalling star-forming galaxies may change with \dprj as well.
We plot them in \autoref{fig:logmhi_drop}, and try to model them with known information in this section.
We note that we only plotted galaxies with an $\ssfr>\qty{1e-11}{\per\yr}$ in \autoref{fig:logmhi_drop} to better trace the recent infalling galaxies \citep{2021PASA...38...35C}.
There are significant correlations (Spearman rank) between the \delMHi and \dprj ($p=0.02$) and between \delMHi and \delSFR ($p<\num[print-unity-mantissa=true]{1e-13}$).
The latter corroborates the link between the gas-stripping and quenching \citep{2022ARA&A..60..319S}.
If low-\ssfr galaxies are included, the aforementioned correlations remain significant, and \delSFR and \dprj will also show a significant correlation.
For a similar diagram featuring all galaxies in the group, please refer to \autoref{sec:app:evol}.

The pattern of \Hi becoming on average more deficient toward the group center is similar to previous results for other groups and clusters \citep[e.g.,][]{2013AJ....146..124H,2015MNRAS.448.1715J}.
Hereafter, we take an empirical approach that is based on both physical considerations and observations, to model the drop of \delMHi with decreasing \dprj.
Similar approaches have been used by \citet{2003A&A...398..525V} to infer (based on the multiwavelength morphologies) the infall orbits of galaxies near the Virgo cluster center, by \citet{2014A&A...570A..69B} to distinguish the role of \rps and strangulation in the reddening of satellites in massive clusters, and by \citet{2015MNRAS.448.1715J} to explain dramatically rising fraction of \Hi nondetections driven by \rps toward the small-distance and high-velocity region of massive clusters.
Compared to these early works mostly focusing on massive clusters, our major improvements are the inclusion of TS and timescales for stripping the strippable \Hi \citetext{\tstr, also see \citealp{2021ApJ...915...70W} for deviation of RPS time scales in an RPS-dominated environment}.

\subsubsection{The Empirical Model}
\label{ssec:strip_model}

The core part of this empirical, simple model is the decreasing rate of the \Hi amount during the infall process.
We make use of \fstr as a function of \dprj and \delMHi that is fitted in \autoref{ssec:f_str}.
Combining it with \tstr (the time needed for all strippable gas to be removed under the same environment), and adding a persistent gas consumption due to the star formation, leads to a simplified prescription of the decreasing rate of the \delMHi.

We ignore the influence of gas accretion, feedback driven outflows, and return of mass from stellar evolution.
Most of these effects are related to the star formation, and they together can be viewed as a modification to the gas consumption rate due to star formation.
They may be better constrained in the future when more information becomes available as the WALLABY survey progresses.
We also ignore the increase of stellar mass, which will further reduce \delMHi, in the main part of this paper.
The influence of these approximations on the main results is discussed in \autoref{ssec:app:strip_model}.
Besides, previous simulations \citep{2012MNRAS.422.1609T,2013A&A...556A..99J,2017MNRAS.469...80Q,2018MNRAS.479.4367K} and observations \citep[e.g.,][]{2021ApJ...921...22C} found the falling-back of stripped gas, especially when the ram pressure is unsteady or changes abruptly, which could slow down the quenching.
In this model, however, we focus on the first-time infalling of galaxies before they pass the pericenter, and thus falling-back would be less important and is ignored here.

This model additionally has the following components and assumptions.
\begin{enumerate}
  \item \emph{The infall starts at $2\Rzoo$ and the time since then is traced by \dprj.}
        The \dprj of infalling galaxies decreases at a constant velocity of~$\sigma_v$, the 1D velocity dispersion of the group.
        We use the half-crossing time $T_0=\Rzoo/\sigma_v=\qty{2.146}{\Gyr}$ as the time unit.
        By doing so, we have assumed radially infalling orbits for these relatively star-forming galaxies, and thus the calculation is only statistically meaningful.
        We also presumed that stripping is weak enough beyond $2\Rzoo$, which is confirmed later.

  \item \emph{The \tstr depends only on \dprj but not on \delMHi.}
        Once \Hi becomes strippable, the restoring force is fully counteracted by the stripping force, and the acceleration for stripped \Hi to leave the galactic disk is determined by the residual stripping force fully depending on the external environmental conditions.
        This scenario is in concept consistent with the theoretical prediction of TS-assisted RPS and vice versa \citep{2022A&ARv..30....3B}.
        We simplistically parameterize \tstr as $10^\alpha d^nT_0$ without further physical motivation, where $\alpha$ and $n$ are free parameters to be derived, and $d\coloneq\dprj/\Rzoo$.
        We have ignored additional pushes by the stellar feedback and the clumpy or multiphase nature of gas \citep{2017ApJ...836L..13K}.

  \item \emph{We use the most \Hi-rich galaxies as the reference population to fit the model.}
        We do this because they are least affected by the detection limit of \MHi.
        Also, the most \Hi-rich galaxies at the outskirts are very likely to remain the most \Hi-rich after infalling.
        We thus focus on the most gas-rich galaxies at each \dprj, which form the upper envelope of the data points in \autoref{fig:logmhi_drop}.
        This envelope is almost horizontal beyond \Rzoo, but plummets within it.
        We quantitatively describe this envelope with the 95th percentiles of \delMHi in five even bins of $0.4\Rzoo$ width, which are plotted in \autoref{fig:logmhi_drop} in orange.
        The percentiles are calculated without galaxies with $\ssfr\le\qty{1e-11}{\per\yr}$, and the result including them can be found at \autoref{ssec:app:strip_model}.
        The choice of bin edges alleviates the influence of the gap at $\num{\sim 1.25}\Rzoo$, possibly linked with the second turnaround radius of groups \citep{1989RvMP...61..185S,2015AJ....149...54T}.

  \item \textit{Galaxies have a uniform SFE of \qty{0.22}{\per\Gyr} \citep{2017ApJS..233...22S}.}
        This corresponds to an intrinsic slope of $0.205/\Rzoo$ on the \delMHi--\dprj diagram (gray lines in \autoref{fig:logmhi_drop}).
        We note that the SFE of star-forming galaxies significantly decreases toward low $M_*$ \citep{2012ApJ...756..113H}.
        This effect actually cancels out and mitigates the systematic bias caused by the assumption of holding $M_*$ constant (see \autoref{ssec:app:strip_model}).

\end{enumerate}

With all these assumptions, the empirical model is described by the following differential equation,
\begin{equation}\label{eq:ivp}
  \frac{\dd\delMHi}{\dd d} = \frac{\log\left[1/(1-\fstr)\right]}{\tstr/T_0}+0.205,
\end{equation}
where $1-\fstr$ is the fraction of remaining gas and where \fstr is calculated using \autoeqref{eq:f_fit}.

\subsection{The Stripping Timescale}

\begin{figure*}
  \myplotone{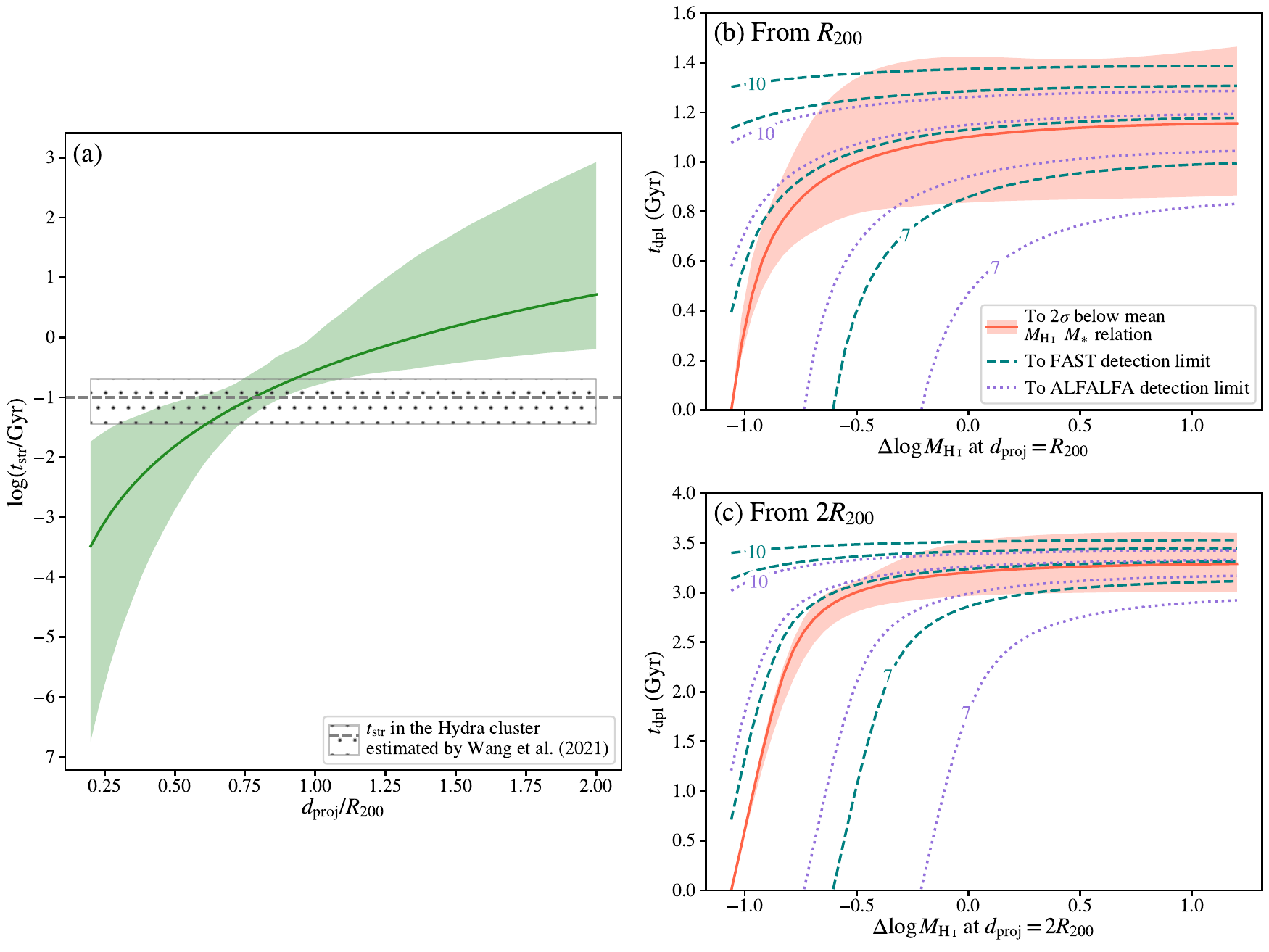}
  \caption{Model timescales of gas stripping and depletion.
    (a)~Time needed for all strippable gas to be stripped, \tstr, as a function of~\dprj.
    To estimate the uncertainty, we use random parameters of $\alpha$ and $n$ from MCMC to construct an ensemble of stripping timescales.
    The shading area is bordered by the 16th and 84th percentiles of the timescale distribution, indicating the $1\sigma$ uncertainty.
    As a reference, the \tstr of RPS estimated by \citet{2021ApJ...915...70W} is plotted.
    The median value is the dashed gray line, and the 16th and 84th percentiles are given as gray dotted shadings.
    (b)~Duration of galaxies becoming gas poor, \tdpl, vs.~the initial \delMHi at $\dprj=\Rzoo$.
    With different definitions of \emph{gas poor} and different stellar mass, the relation is different.
    Red solid, green dashed, and purple dotted lines give the time for galaxies to reach $2\sigma$ (\qty{1.06}{\dex}) below the mean \MHi relation, FAST detection limit, and ALFALFA detection limit respectively.
    For the latter two, the results for galaxies with $\log\left(M_*/\uMsun\right)$ of \numlist{7;8;9;10} are different.
    A galaxy with a higher stellar mass has a longer depletion time.
    The uncertainty of the red solid line is given in shading and is obtained similarly as the uncertainty in~(a).
    Panel~(c) is similar to~(b) but starting time is when $\dprj=2\Rzoo$.
    The abscissa is the value there, too.}
  \label{fig:str_time}
\end{figure*}

\autoeqref{eq:ivp} can be solved as an initial value problem numerically with given parameters $\alpha$, $n$, and $T_0$.
We use the Python package \emph{emcee} \citep{2013PASP..125..306F} to fit these three parameters with Markov Chain Monte Carlo (MCMC) methods, and the details are reported in \autoref{sec:app:evol_fit}.

The median values of two parameters with a $1\sigma$ confidence interval are $\alpha=\numalpha$ and $n=\numn$, and the corresponding fitted evolutionary path is plotted as the orange solid line in \autoref{fig:logmhi_drop}, which reproduces the gradual-then-steep feature and follows five percentile points well.
We additionally plotted several evolutionary paths with different initial \delMHi at $2\Rzoo$ as green dashed--dotted lines.
Galaxies with less gas tend to spend more time losing gas almost solely by star formation (i.e., following gray lines) and to drastically lose gas by stripping later.
Within \Rzoo, the evolutionary paths gradually converge, suggesting the overwhelming effect of stripping there.

\autoref{fig:str_time}(a) plots our model of stripping timescale estimation adopting these two values and shows the corresponding $1\sigma$ scatter.
At the outer region of the group, it takes several gigayears to strip all strippable gas.
The timescale there could even reach the age of the universe given the large error bar.
At \Rzoo, $\tstr$ decreases to hundreds of megayears.
It drops rapidly toward the group center, reaching a submegayear scale.

Interestingly, the range of $\tstr$ is largely consistent with those estimated by \citet{2021ApJ...915...70W} within \Rzoo of the Hydra cluster using a different method.
Different from N4636G, the dominating stripping effect in the Hydra cluster is \rps, which might be assisted by thermal evaporation \citep{2021ApJ...915...70W}.
Compared to the Hydra cluster, N4636G is 6~times lower in \Mzoo, resulting in a twice lower velocity dispersion and on average a 4 times lower ram pressure \citep{1972ApJ...176....1G}.
N4636G also has a $3.3$ times lower ICM temperature and a $4.5$ times less efficient thermal evaporation \citep{1977ApJ...211..135C}.
While instantaneously and for individual galaxies, TS and \rps are possibly highly independent, it seems that, on a longer time-scale during which the satellite infalling and for the whole satellite population in the group, TS and \rps may work together to make stripping efficient in N4636G, as suggested by the simulations of \citet{2016MNRAS.455.2994M}.

\subsection{The Depletion Timescale}
With these deduced evolutionary paths of \delMHi, it is possible to translate them into the timescale of the depletion of gas, \tdpl, since crossing the group border defined as $2\Rzoo$ or as \Rzoo, the radius where stripping becomes dominant.

We first use a more controlled definition of the \tdpl:
the time needed to drop to 2~times the scattering below the mean \MHi--$M_*$ relation, i.e., $2\times\qty{0.53}{\dex}$.
With our simple model of infalling and this definition of~\tdpl, \tdpl is independent of the stellar mass and is only a function of the initial \delMHi.
The red solid lines in \autoref{fig:str_time}(b) and~(c) show these depletion timescales since crossing \Rzoo and $2\Rzoo$ and their scatter respectively.
All galaxies with an initial \delMHi larger than \qty{-0.5}{\dex} have long and similar timescales of \qtylist{\sim 1;\sim 3}{\Gyr}.
We note that this \qty{\sim 2}{\Gyr} difference between the \Rzoo and $2\Rzoo$ values is similar to $T_0$, the crossing time of one \Rzoo, indicating that the stripping is mild beyond \Rzoo.
The \tdpl of galaxies with less gas drops significantly.

If we define depletion from a more observational point of view as being \Hi nondetectable, the story will be quite different.
The \Hi detection limit hinges on the stellar mass, and for galaxies with lower $M_*$, the threshold is higher relative to the mean \MHi relation, resulting in a shorter \tdpl.
\autoref{fig:str_time}(b) gives the \tdpl relations of different stellar masses detected by ALFALFA (purple dotted) and by FAST (green dashed).
Galaxies of low stellar mass become \Hi-nondetected much quicker than those of high $M_*$ do, and the dependence of $M_*$ is more obvious for initially gas-poor galaxies.
Meanwhile, the galaxies with a stellar mass of \qty{1e10}{\Msun} retain a~\tdpl of \qty{\sim 1.2}{\Gyr} even when their initial \delMHi is as low as \num{-1.0}.
A similar discussion applies to $2\Rzoo$ results shown in \autoref{fig:str_time}(c).

In summary, it takes \qty{\sim 1}{\Gyr} to deplete the galactic \Hi (to $2\sigma$ below the mean \MHi--$M_*$ relation) since galaxies enter \Rzoo, within which stripping dominates.
If we start the clock at $2\Rzoo$ instead, the depletion time is largely uniform around \qty{3}{\Gyr} for \Hi-rich galaxies, but for HI poor galaxies, it shortens dramatically depending on the starting \delMHi.
This result is highly consistent with \tdpl of group galaxies in the literature \citep[e.g.,][]{2013MNRAS.432..336W,2015ApJ...806..101H,2019A&A...632A..78J,2020ApJS..247...45R,2021PASA...38...35C,2021A&A...648A..31L,2021MNRAS.501.5073O,2022ApJS..263...40M}.

The new aspect of the analysis here is that we use a relatively observational perspective to separate the effect of different physical mechanisms (RPS, TS, and star-formation), which may lead to gas depletion in group environments.
We quantify the combined stripping effect of \rps and TS (with \fstr and \tstr) by adding them up after characterizing each of them, instead of using the more conventional way of inferring them as a whole from a mixed consequence (e.g., by quantifying \delMHi as a function of local density or dark matter halo mass).
The \tstr we derived and its trend with \dprj provides a new observational constraint, which may be useful to break degeneracies in more theoretically oriented models.
The \tstr combines the effect of \rps and TS, but the relative role of each has been separated and addressed in Sections \ref{ssec:ts_rps_comparison} and~\ref{ssec:f_str}, based on which it is decided that \tstr is sufficient to capture the combined effect.

The experiment above also suggests that studying the environmental effects on galactic \Hi is sensitive to the definition of \Hi depletion other than using \delMHi, and the depth of \Hi data.
A secondary dependence on $M_*$ and other parameters may be introduced by those definitions.
Studies of the comparison of environmentally driven galactic \Hi depletion between studies should be done with caution.
We also again stress that here only the first-time infalling of galaxies into the group is considered.

\section{Summary and Conclusion}

In this paper, we have developed a method to quantify the strength of TS and RPS using two types of parameters $S$ and $f$.
The $S$ describes the instantaneous strength of these effect at the optical-disk edges, while the $f$ describes the radial extent of \Hi affected by each of these effects.
We investigate the response of satellite galaxies to these two effects, make a phenomenon-based simple prescription of gas stripping to trace the change of \MHi, and estimate the timescale of gas stripping and depletion as galaxies infall in the N4636G\@.
We summarize our major results as follows.

First and most importantly, we provide a promising method to separate the effects of RPS and TS on galaxies in N4636G\@.
It is supported by the following results.
\begin{enumerate}
  \item \textbf{RPS and TS coexist in the group but do not grow simultaneously.}
        The RPS and TS affect 72\% and 49\% of \Hi-detected nonmerging satellite population respectively, and 41\% of the satellites are undergoing both strippings.
        Among these satellites, 44\% (36\%) are mainly influenced by \rps (TS)\@.
        The values of $S$ of the two kinds of stripping effects are independent, while the $f$'s of two effects are moderately anticorrelated for the same galaxy.
        The \frps increases toward the group center and in more gas-rich galaxies, while \ftid does not show similar relations.
        In N4636G, TS generally reaches a higher $S$ and $f$ than RPS does.

  \item The RPS (TS) sample shows a clearer correlation between $\RHi/\Rzsg$ and \Srps (\Stid) than the remaining galaxies.

  \item The low-mass ($M_*<\qty{1e9}{\Msun}$) subset of the TS sample shows significant trends of optical color (throughout the disks, at \Rso, and at $2\Rso$) reddening with stronger \Stid, while the remaining low-mass galaxies of the group do not.
\end{enumerate}

On the separate and cooperative role of TS and RPS in driving galaxy evolution, we find the following.
\begin{enumerate}
  \item \textbf{The \Hi disks respond to \rps and TS similarly, by shrinking, but the reddening of the low-mass optical disks does not.}
        For relatively \Hi-poor galaxies, the \Hi-to-optical disk size ratio shows anticorrelations with $S$ of both environmental effects.
        The low-mass TS galaxies show a reddening across the galaxy with the increase of \Stid, but the increase of \Srps has no similar effect on either all galaxies or the \rps ones.

  \item \textbf{The average stripping timescale well characterizes the strengthening process of RPS and TS when galaxies infall from beyond the virial radius to near the group center.}
        It drops from nearly the Hubble time when galaxies are at $2\Rzoo$ to less than \qty{1}{\Gyr} at \Rzoo, and then less than \qty{100}{\Myr} at $0.5\Rzoo$.
        As a result, active stripping and passive strangulation are the more dominant mechanisms to deplete the \Hi when galaxies are within and beyond \Rzoo respectively.
        Galaxies experience a first slow (timescale \qty{\sim 3}{\Gyr}) and then fast \Hi depletion after crossing $2\Rzoo$ and $\num{\sim 0.5}\Rzoo$ respectively, qualitatively consistent with the conclusion of many previous studies \citep[e.g.,][]{2015ApJ...806..101H}.
\end{enumerate}

Comparing the TS and RPS effects in N4636G with those in more extreme environments, we find the following.
\begin{enumerate}
  \item Compared to the more TS-dominated Eridanus supergroup, the trend of \Hi disks shrinking in response to TS is similar, but the pattern of reddening in low-mass optical disks is not.
        While the low-mass optical disks in the Eridanus supergroup redden inside out \citep{2022ApJ...927...66W}, those in the N4636G redden throughout the disks rather uniformly.

  \item Compared to the more RPS-dominated Hydra cluster \citep{2021ApJ...915...70W}, there are less $f$ values reaching unity, but the time needed to strip the strippable \Hi of galaxies within \Rzoo is similar.
        The efficient stripping suggests the cooperative effect of RPS and TS on stripping the galaxies in N4636G\@.

  \item In contrast to some previous findings in jellyfish galaxies under strong RPS \citep[e.g.,][]{2016MNRAS.455.2994M,2018MNRAS.476.4753J,2019MNRAS.484..906R}, we do not find evidence for enhanced star formation (i.e., being bluer than expected) in \rps-affected galaxies using the \gr color as the tracer.
\end{enumerate}

Putting these results together, our efforts to disentangle \rps and TS are reflected in different behaviors of \Hi-disk shrinking and color reddening when either dominates the galaxy or the whole group.
The different effects and weights of \rps and TS in different groups raise caution on the conventional operation to blindly stack satellites by the normalized projected distance or PSD of corresponding groups.
On the other hand, the consistent behavior of \Hi decreasing with both types of $S$ and in different groups enables the possibility of empirically combining these two effects, and supports our derivation of \fstr.

Despite the encouraging results, we need keep in mind the possible uncertainty of using a median \Sig{\tHi} profile when deriving the stripping strengths ($S$) and extents ($f$).
We must also be cautious about the limited statistics (e.g., data points are sparse close to \Scri{tid}) and related large uncertainties throughout the analyses.
The method of deriving \tstr relies on simple assumptions, including effectively assuming that the gas accretion, outflow, and stellar mass loss cancel out and disappear in the term of SFE\@.
These aspects should be improved when more data come in from FAST and the main survey of WALLABY \citep{2020Ap&SS.365..118K} and when comparable hydrodynamic simulation mocks are analyzed in a similar observational way.

Our \Hi sample has combined both interferometric and single-dish surveys to achieve a full group coverage as well as to exploit the advantage of each.
While some results (e.g., the upper envelope of the \delMHi--\dprj relation, \autoref{fig:logmhi_drop}) benefit from the capability of FAST to detect a wide dynamic range in \delMHi, the basis of the analysis (i.e., deriving $f$ and $S$ based on a uniform shape of \Hi profile) is supported by the consistent \Hi size--mass relation of the few galaxies with resolved WALLABY images in this group.
The deviation of \frps and \Stid was also calibrated and tested using resolved images from WALLABY in pilot studies \citep{2021ApJ...915...70W,2022ApJ...927...66W}.
When more complexities of galaxy evolution are studied in the future, the cooperation of the two types of data in a similar manner will continue to be powerful.

\section*{Acknowledgments}
% \begin{acknowledgments}
% 202110001069
We thank the anonymous referee for their constructive and helpful
comments.

J.W. acknowledges research grants from Ministry of Science and Technology of the People's Republic of China (No.\ 2022YFA1602902) and the National Science Foundation of China (No.\ 12073002).

B.L. acknowledges the support from the Korea Astronomy and Space Science Institute grant funded by the Korea government (MSIT; Project No.\ 2022-1-840-05).

P.K. acknowledges financial support by the German Federal Ministry of Education and Research (BMBF) Verbundforschung grant 05A20PC4 (Verbundprojekt D-MeerKAT-II)\@.

A. Bosma acknowledges support from the Centre National d'Etudes Spatiales (CNES), France.

L.C.H. was supported by the National Science Foundation of China (11721303, 11991052, 12011540375, 12233001) and the China Manned Space Project (CMS-CSST-2021-A04, CMS-CSST-2021-A06).

H.Y.W. is supported by NSFC No.\ 12192224.

L.V.M. acknowledges financial support from grants CEX2021-001131-S funded by MCIN/AEI/10.13039/ 501100011033, RTI2018-096228-B-C31 and PID2021-123930OB-C21 by MCIN/AEI/10.13039/ 501100011033, by ``ERDF A way of making Europe'' and by the European Union and from IAA4SKA (R18-RT-3082) funded by the Economic Transformation, Industry, Knowledge and Universities Council of the Regional Government of Andalusia and the European Regional Development Fund from the European Union.

% WM#1
Parts of this research were supported by High-Performance Computing Platform of Peking University.

% FAST

This work has used the data from the Five-hundred-meter Aperture Spherical radio Telescope (FAST)\@.
FAST is a Chinese national mega-science facility, operated by the National Astronomical Observatories of Chinese Academy of Sciences (NAOC)\@.

% WALLABY
The Australian SKA Pathfinder is part of the Australia Telescope National Facility, which is managed by CSIRO\@.
Operation of ASKAP is funded by the Australian Government with support from the National Collaborative Research Infrastructure Strategy.
ASKAP uses the resources of the Pawsey Supercomputing Centre.
Establishment of ASKAP, the Murchison Radio-astronomy Observatory, and the Pawsey Supercomputing Centre are initiatives of the Australian Government, with support from the Government of Western Australia and the Science and Industry Endowment Fund.
We acknowledge the Wajarri Yamatji people as the traditional owners of the observatory site.

% SDSS
Funding for the Sloan Digital Sky Survey IV has been provided by the Alfred P. Sloan Foundation, the U.S. Department of Energy Office of Science, and the participating institutions.

SDSS-IV acknowledges support and resources from the Center for High Performance Computing at the University of Utah.
The SDSS website is \url{www.sdss.org}.

SDSS-IV is managed by the Astrophysical Research Consortium for the Participating Institutions of the SDSS Collaboration including the Brazilian Participation Group, the Carnegie Institution for Science, Carnegie Mellon University, Center for Astrophysics | Harvard \& Smithsonian, the Chilean Participation Group, the French Participation Group, Instituto de Astrof\'isica de Canarias, The Johns Hopkins University, Kavli Institute for the Physics and Mathematics of the Universe (IPMU)/ University of Tokyo, the Korean Participation Group, Lawrence Berkeley National Laboratory, Leibniz Institut f\"ur Astrophysik Potsdam (AIP), Max-Planck-Institut f\"ur Astronomie (MPIA Heidelberg), Max-Planck-Institut f\"ur Astrophysik (MPA Garching), Max-Planck-Institut f\"ur Extraterrestrische Physik (MPE), National Astronomical Observatories of China, New Mexico State University, New York University, University of Notre Dame, Observat\'ario Nacional/ MCTI, The Ohio State University, Pennsylvania State University, Shanghai Astronomical Observatory, United Kingdom Participation Group, Universidad Nacional Aut\'onoma de M\'exico, University of Arizona, University of Colorado Boulder, University of Oxford, University of Portsmouth, University of Utah, University of Virginia, University of Washington, University of Wisconsin, Vanderbilt University, and Yale University.

% DECaLS
The Legacy Surveys consist of three individual and complementary projects:
the Dark Energy Camera Legacy Survey (DECaLS; Proposal ID \#2014B-0404; PIs: David Schlegel and Arjun Dey), the Beijing--Arizona Sky Survey (BASS; NOAO Prop.\ ID \#2015A-0801; PIs: Zhou Xu and Xiaohui Fan), and the Mayall $z$-band Legacy Survey (MzLS; Prop.\ ID \#2016A-0453; PI: Arjun Dey).
DECaLS, BASS, and MzLS together include data obtained, respectively, at the Blanco telescope, Cerro Tololo Inter-American Observatory, NSF's NOIRLab; the Bok telescope, Steward Observatory, University of Arizona; and the Mayall telescope, Kitt Peak National Observatory, NOIRLab.
Pipeline processing and analyses of the data were supported by NOIRLab and the Lawrence Berkeley National Laboratory (LBNL)\@.
The Legacy Surveys project is honored to be permitted to conduct astronomical research on Iolkam Du'ag (Kitt Peak), a mountain with particular significance to the Tohono O'odham Nation.

NOIRLab is operated by the Association of Universities for Research in Astronomy (AURA) under a cooperative agreement with the National Science Foundation.
LBNL is managed by the Regents of the University of California under contract to the U.S. Department of Energy.

This project used data obtained with the Dark Energy Camera (DECam), which was constructed by the Dark Energy Survey (DES) collaboration.
Funding for the DES Projects has been provided by the U.S. Department of Energy, the U.S. National Science Foundation, the Ministry of Science and Education of Spain, the Science and Technology Facilities Council of the United Kingdom, the Higher Education Funding Council for England, the National Center for Supercomputing Applications at the University of Illinois at Urbana-Champaign, the Kavli Institute of Cosmological Physics at the University of Chicago, Center for Cosmology and Astro-Particle Physics at the Ohio State University, the Mitchell Institute for Fundamental Physics and Astronomy at Texas A\&M University, Financiadora de Estudos e Projetos, Fundacao Carlos Chagas Filho de Amparo, Financiadora de Estudos e Projetos, Fundacao Carlos Chagas Filho de Amparo a Pesquisa do Estado do Rio de Janeiro, Conselho Nacional de Desenvolvimento Cientifico e Tecnologico and the Ministerio da Ciencia, Tecnologia e Inovacao, the Deutsche Forschungsgemeinschaft, and the collaborating institutions in the Dark Energy Survey.
The collaborating institutions are Argonne National Laboratory, the University of California at Santa Cruz, the University of Cambridge, Centro de Investigaciones Energeticas, Medioambientales y Tecnologicas-Madrid, the University of Chicago, University College London, the DES-Brazil Consortium, the University of Edinburgh, the Eidgenossische Technische Hochschule (ETH) Zurich, Fermi National Accelerator Laboratory, the University of Illinois at Urbana-Champaign, the Institut de Ciencies de l'Espai (IEEC/CSIC), the Institut de Fisica d'Altes Energies, Lawrence Berkeley National Laboratory, the Ludwig Maximilians Universitat Munchen and the associated Excellence Cluster Universe, the University of Michigan, NSF's NOIRLab, the University of Nottingham, the Ohio State University, the University of Pennsylvania, the University of Portsmouth, SLAC National Accelerator Laboratory, Stanford University, the University of Sussex, and Texas A\&M University.

BASS is a key project of the Telescope Access Program (TAP), which has been funded by the National Astronomical Observatories of China, the Chinese Academy of Sciences (the Strategic Priority Research Program ``The Emergence of Cosmological Structures'' grant \# XDB09000000), and the Special Fund for Astronomy from the Ministry of Finance.
The BASS is also supported by the External Cooperation Program of Chinese Academy of Sciences (grant \# 114A11KYSB20160057), and Chinese National Natural Science Foundation (grants \# 12120101003, \# 11433005).

The Legacy Survey team makes use of data products from the Near-Earth Object Wide-field Infrared Survey Explorer (NEOWISE), which is a project of the Jet Propulsion Laboratory/California Institute of Technology.
NEOWISE is funded by the National Aeronautics and Space Administration.

The Legacy Surveys imaging of the DESI footprint is supported by the Director, Office of Science, Office of High Energy Physics of the U.S. Department of Energy under contract No.\ DE-AC02-05CH1123, by the National Energy Research Scientific Computing Center, a DOE Office of Science User Facility under the same contract; and by the U.S. National Science Foundation, Division of Astronomical Sciences under contract No.\ AST-0950945 to NOAO\@.
% \end{acknowledgments}

\facilities{Arecibo, ASKAP, Blanco (DECam), FAST:500m, GALEX, IRSA, NED, Sloan, and WISE.}

\software{astropy 5.1 \citep{2022ApJ...935..167A}, astroquery 0.4.2 \citep{2019AJ....157...98G}, numpy 1.21.4 \citep{2020Natur.585..357H}, photutils 1.2.0 \citep{2021zndo...5525286B}, pingouin 0.5.1 \citep{2018JOSS....3.1026V}, emcee 3.1.2 \citep{2013PASP..125..306F}, Python 3.8.10, scipy 1.8.0 \citep{2020NatMe..17..261V}, SExtractor 2.25.0 \citep{1996A&AS..117..393B}, SWarp 2.41.4 \citep{2002ASPC..281..228B}.}

% \section{} % to fool VS Code for collapsing

\appendix

\section{Escaping Velocity and Interlopers}
\label{sec:app:vesc}

We define the 3D escape velocity \vesd as the minimum velocity required for a galaxy to escape the gravitational potential of a group.
With \desd as the 3D group-centric distance and applying the Navarro--Frenk--White CDM cold dark matter halo model \citep{1996ApJ...462..563N},
\begin{equation}
  \vesd = v_{200} \sqrt{\frac{
      2\ln\left(1+xc\right)}{
      x\left[\ln\left(1+c\right)
        -c/\left(1+c\right)\right]}},
\end{equation}
where $x=\desd/\Rzoo$, $v_{200}=\sqrt{G\Mzoo/\Rzoo}$, and the concentration $c$ is set as $4$.

Assuming equipartition and isotropy, the projected values are
\begin{equation}
  \vepj = \frac{\vesd}{\sqrt{3}},\quad
  \dprj = \frac{\pi}{4}\desd,
\end{equation}
respectively.

To estimate the fraction of interlopers (galaxies out of $2\Rzoo$ in 3D) under our sample selection, we use simulated catalogs generated by IllustrisTNG-100 \citep{2018MNRAS.480.5113M,2018MNRAS.477.1206N,2018MNRAS.475..624N,2018MNRAS.475..648P,2018MNRAS.475..676S}.
For all simulated groups with an \Mzoo within \qty{0.3}{\dex} around that of N4636G, we select their \emph{members} using a similar criteria based on the projected PSD, which is introduced in \autoref{ssec:sample_select}.
We find that the interloper fraction is mainly determined by the projected group-centric distance.
At $1.5\Rzoo$ the fraction is $\num{\sim 40}\%$ and is $\num{\sim 20}\%$ at \Rzoo, consistent with the result of \citet{2016MNRAS.463.3083O}.
We confirm that trends introduced in \autoref{sec:env} remain significant if $1.5\Rzoo$ is used as the boundary to select group members.

We excluded certain interlopers using redshift-independent distances available for some of our galaxies from \citetalias{2016AJ....152...50T}.
Four galaxies are excluded whose distances exceed $3\Rzoo$ from \qty{16.2}{\Mpc} (N4636G's distance) after considering the uncertainty.

N4636G also has a massive neighbor cluster, Virgo, to its northwest, making galaxies and groups around it (including N4636G) experiencing gravitational acceleration and having a complex relation between the redshift and heliocentric distance \citep{2017ApJ...850..207S,2020AJ....159...67K}.
This results in unavoidable uncertainty of our sample selection.
This problem is mitigated by using the equipartitioned escape velocity, instead of the maximum one, when selecting member galaxies.
We further assess the significance of the problem by estimating the tidal truncation radius of N4636G as a satellite halo of the Virgo cluster following \citet{2003MNRAS.341..434T}, taking \qty{2.5e14}{\Msun} as the mass of the Virgo \citep{2006PASP..118..517B}.
Only five out of 119 galaxies possibly have a group-centric distance larger than the truncation radius.
Therefore, we concluded that the influence of Virgo is negligible.

\section{Mock Test for Image-Based \Hi Measurements Dependent on Spatial Resolution}
\label{sec:app:hi_mock}

\begin{figure}
  \myplotone{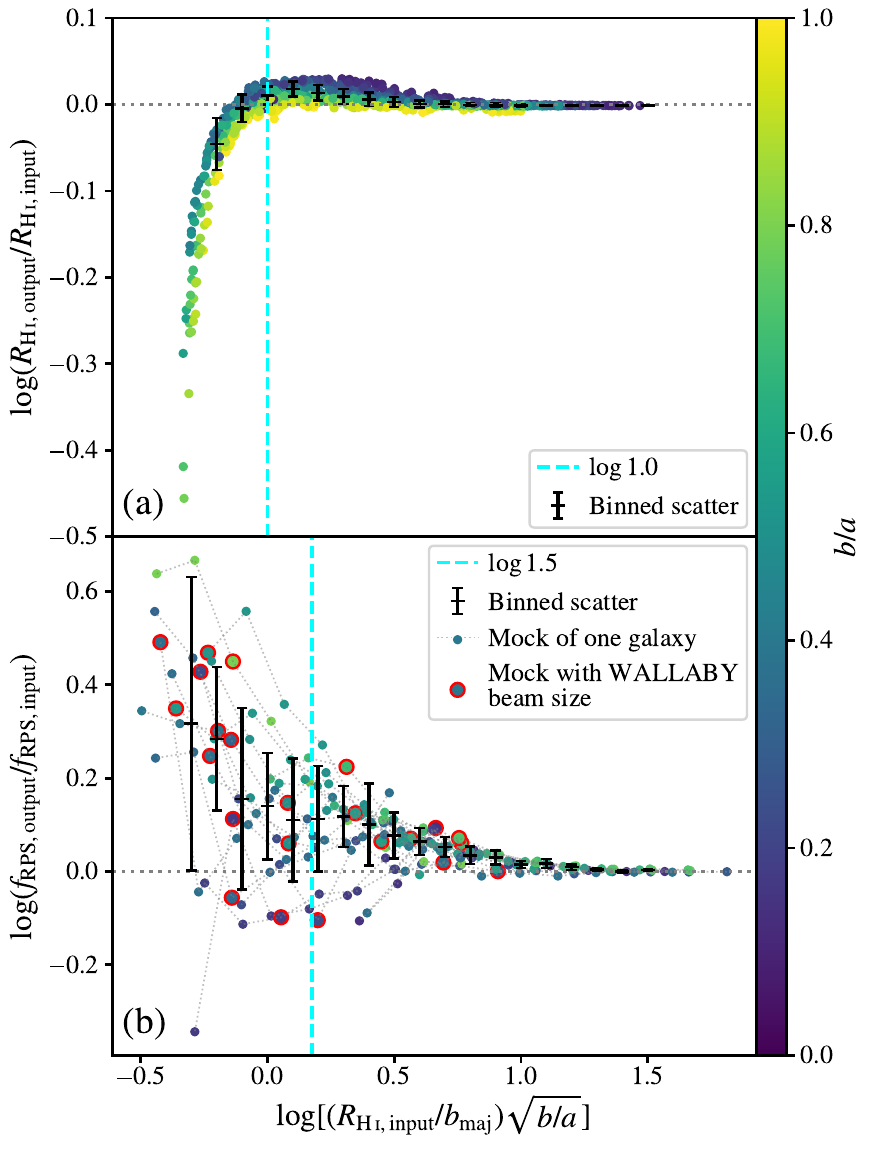}
  \caption{Results of mocking test about the influence of \Hi resolution on moment~(0) map analysis.
    Mock \Hi images are generated assuming a surface density profile of \citet{2016MNRAS.460.2143W} and are convolved using Gaussian kernels of different FWHM, \bmaj.
    The pixel scale is $1/5$ of \bmaj, the same as WALLABY data.
    The ellipticity of the projected disk is expressed as $b/a$, the ratio between minor-axis and major-axis, by which points are color-coded.
    The abscissa is equivalent to the input ratio of projected disk's area to the beam's.
    The criterion of disk's area being $1.5^2$~times the beam's is plotted as vertical dashed cyan line.
    Scatters in \qty{0.1}{\dex} width bins are plotted in black.
    (a)~The ratio of \RHi[output] measured from mock image, after correction introduced in \autoref{ssec:hi_prop}, to the input \RHi[input].
    The images are generated using random geometric parameters.
    (b)~The ratio of \frps measured from mock image, following \citet{2021ApJ...915...70W}, to the \frps directly measured using profiles.
    The ram pressure, disk size, and ellipticity are all true values from our \Hi sample, and mock \bmaj ranges from \ang{;;3.75} to \ang{;;60}.
    Results of one galaxy with different beam sizes are linked with dotted gray lines, and the results using the WALLABY beam size, \ang{;;30}, have red edges.}
  \label{fig:rhi_mock}
\end{figure}

We do simple mock tests to check the influence of spatial resolution on two kinds of measurements from the observed \Hi images, \RHi and~\frps.
Based on results from these tests, we determine the criteria for selecting \Hi disks resolved enough for these image-based measurements.
We select resolved \Hi galaxies to get reliable image-based measurements, using these to assess the consistency with the \Hi-profile-based estimations that is used in the main part of this paper.

For the test related to \RHi, we simulate \Hi images of disks with different axis ratios $b/a$ and at different spatial resolution levels ($\RHi/\bmaj$).
High-resolution, thin \Hi disks are first generated using the \RHi we provide as the ``input'' and the average surface density profile measured by \citet{2020ApJ...890...63W}.
These disks are projected to images with different inclinations, and thus having different $b/a$.
They are then downgraded to different resolution by convolving with Gaussian kernels of different \bmaj and the pixel size is always set to be $\bmaj/5$.
We then directly measure \RHi from the downgraded images, which are denoted as \RHi[obs].
We fix the ellipticity of annuli at the input value, instead of measuring from convolved images.
We empirically correct for the beam smearing effects as $\RHi[output] = \sqrt{\RHi[obs]^2-(\bmaj/2)^2}$, following the formula in \citet{2016MNRAS.460.2143W}, to obtain the final measurements $\RHi[output]$.
We compare the input with output, and the results are shown in \autoref{fig:rhi_mock}(a).

The \RHi[output] generally reproduces the value of \RHi[input] when the projected disk has an area within \RHi[input] larger than $\bmaj^2$ (dashed cyan line), with a small scatter and a systematic offset smaller than 5\%.
There are 11 out of 19 N4636G galaxies detected by WALLABY whose expected \RHi[input] (from \MHi) satisfies this criterion.
We confirm that their image-based \RHi's follow the \Hi size--mass relation with a median offset of \qty{0.03}{\dex} and scatter of \qty{0.03}{\dex}.
The positive offset mainly comes from six galaxies that are close to the \bmaj criterion, consistent with the mock.
We get an offset of \qty{0.00}{\dex} and a scatter of \qty{0.02}{\dex} excluding them.
Because 82\% of the \Hi-sample galaxies are either not resolved enough according to the criterion above or beyond the WALLABY footprint, we use \RHi estimated from the \Hi size--mass relation for scientific analysis in the paper.

For the test related to image-based \frps, we set the input \frps (or equivalently, the ram pressure and stellar mass radial distribution) and \RHi based on the galaxies that have an $\frps>1\%$ from our \Hi sample.
We produce mock images as above for the \RHi-related test, but remove fluxes below the WALLABY depth, \qty{1e20}{\cm^{-2}}.
We derive image-based \frps by comparing the pixel values of the anchoring-force map with the ram pressure, following the procedure of \citet{2021ApJ...915...70W}.
We compare the input and output \frps in \autoref{fig:rhi_mock}(b).
The input \bmaj ranges from \ang{;;3.75} to \ang{;;60} and we highlight the output with the WALLABY beam size, \ang{;;30}, with red circles.

We find that the image-based measurements (output) systematically overestimated the true value (input), especially when the projected disk area is small compared with the beam area.
If the projected disk has an area larger than $1.5^2\bmaj^2$ (dashed cyan line), the median relative error can be less than 25\%.
Only five galaxies from the WALLABY detections are resolved enough by this criterion, and their \rps{}s are weak ($\frps<10\%$), making the image-based method very uncertain.
Due to the small number of reliable image-based measurements, we use the estimation based on median profiles for science analysis in the paper.

\section{Comparison of SDSS and DECaLS Images}
\label{sec:app:sdss_ls}

\begin{figure}
  \myplotone{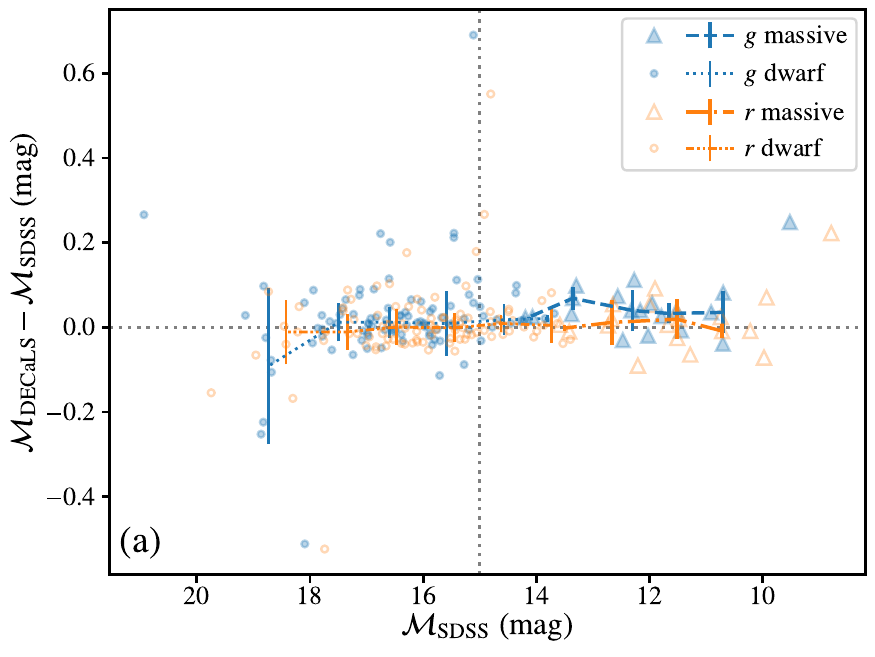}\\
  \myplotone{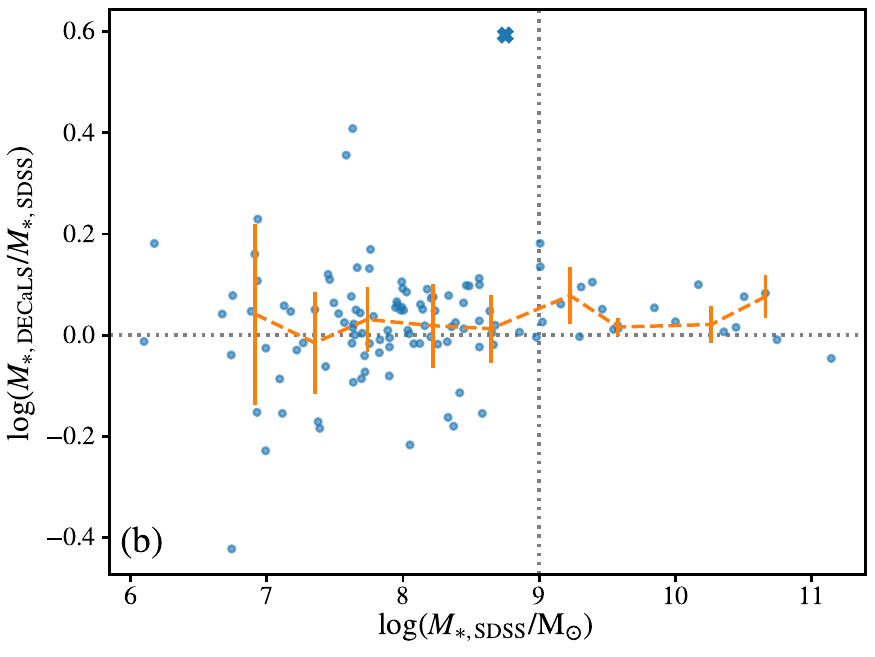}
  \caption{Comparison of the luminosities (and stellar mass $M_*$) from SDSS and DECaLS surveys.
    (a)~The difference of magnitudes using images from two surveys plotted against SDSS magnitudes.
    Measurements are carried out using same apertures for consistency.
    DECaLS magnitudes are corrected following \citet{2019AJ....157..168D}.
    Individual measurements are plotted as translucent symbols.
    Triangles (circles) are high-mass (low-mass) galaxies, and blue (orange) points are $g$-band ($r$-band) measurements.
    Binned, $\sigma$-clipped median values and scatters are plotted for each of these four categories.
    A reference line of \qty{15}{mag} is plotted.
    (b)~The ratio of $M_*$'s from two surveys plotted against SDSS results and corresponding binned, $\sigma$-clipped median values and scatters.
    DECaLS image of \id{75} is of low quality and is plotted as a cross.
    The separation between low- and high-mass galaxies, \qty{1e9}{\Msun}, is plotted.}
  \label{fig:sdss_ls}
\end{figure}

We compare the photometric products from SDSS and DECaLS to justify their preferred usage for high- and low-mass galaxies respectively.

\autoref{fig:sdss_ls}(a) compares CG-based total fluxes from the two surveys.
We have applied the correction reported by \citet{2019AJ....157..168D} for the DECaLS fluxes to be calibrated against the SDSS ones.
We find the following:
\begin{enumerate}
  \item The fluxes from the two surveys are highly consistent when (SDSS) magnitudes are between \qtylist{18;14}{\mag}.
  \item Around $\mathcal{M}=\qty{15}{\mag}$, there are a few galaxies for which DECaLS detects much lower flux than SDSS does.
        We found that these galaxies have a similar diameter to the size of background-estimating box used by the DECaLS data processing pipeline, \ang{;;\sim 250}.
        So, the large difference could be the result of oversubtracting background when the DECaLS pipeline produces the stacked images.
        For these galaxies that have $\mathcal{M}_\text{DECaLS}<\qty{14.8}{\mag}$ and $\mathcal{M}_\text{DECaLS}-\mathcal{M}_\text{SDSS}>\qty{0.15}{\mag}$, we choose to use the SDSS images for measurements.
  \item For the galaxies whose $\mathcal{M}>\qty{18}{\mag}$, DECaLS detects more flux thanks to its better depth.
  \item For the galaxies with magnitudes between \qtylist{14;10}{\mag}, DECaLS clearly underestimates the fluxes, especially in the $g$ band.
  \item For the brightest (and largest) galaxies with $\mathcal{M}<\qty{10}{\mag}$, their fluxes are significantly underestimated by DECaLS\@.
        It is possible that this is similarly due to background oversubtraction of the DECaLS pipeline.
\end{enumerate}

\autoref{fig:sdss_ls}(b) compares the stellar mass estimated using these two surveys.
We find the following:
\begin{enumerate}
  \item The scatter of the difference increases below \qty{1e9}{\Msun} from \num{\sim 0.07} to \qty{\sim 0.19}{\dex}.
  \item For high-mass galaxies, although DECaLS underestimates their fluxes, DECaLS gives a higher \gr value, which leads to a higher stellar mass-to-light ratio and thus a higher stellar mass.
  \item Low-mass galaxies have less systematic difference but a larger scatter.
        Interestingly, the large difference of fluxes between the two surveys for galaxies around $\mathcal{M}=\qty{15}{\mag}$ does not lead to an unacceptably high level of difference in stellar mass.
  \item One galaxy, \id{75} (the cross), has a mosaicking issue in its DECaLS image, which introduces a dramatic overestimation of stellar mass.
        The SDSS image of this galaxy is used for photometric measurements.
\end{enumerate}

Putting these together and after properly treating the few specific cases as described above, we conclude that it is reasonable to set the dividing line between using SDSS and DECaLS photometric products at \qty{1e9}{\Msun}.

\section{The Projective Uncertainties of \texorpdfstring{$S$}{S} and \texorpdfstring{$f$}{f}}
\label{sec:app:proj}
\begin{deluxetable}{cCcC}
  \tablecaption{The Systematic Offset and Uncertainty from Projection.}
  \label{tab:proj}
  \tabletypesize{\footnotesize}
  \tablehead{
    & \multicolumn3c{$\log(x\tsb{3D}/x\tsb{proj})$} \\
    \cline{2-4}
    Quantity $x$ & \colhead{Median} & \multicolumn1{p{5.5em}}{\centering 16th and 84th Percentiles} & \colhead{Scatter} \\
    & \colhead{(\unit{\dex})} & \colhead{(\unit{\dex})} & \colhead{(\unit{\dex})}
  }
  \startdata
  % \begin{tabular}{cS[table-format=+1.2]cS[table-format=1.2]}
%   \toprule
%                                  & \multicolumn{3}{c}{$\log(x\tsb{3D}/x\tsb{proj})$} \\
%   \cmidrule{2-4}
%   Quantity $x$                   & {Median}        &
%   {16th and 84th Percentiles}    & {Scatter} \\
%                                  & {(\unit{\dex})} & {(\unit{\dex})}    & {(\unit{\dex})} \\
%   \midrule
  $|\inc v\tsb{3D}|$             & 0.32            & $^{+0.49}_{-0.24}$ & 0.37 \\
  $d$                            & 0.10            & $^{+0.41}_{-0.10}$ & 0.25 \\
  \Pram                          & 0.41            & $^{+0.99}_{-0.58}$ & 0.78 \\
  \Pram ($|\inc v|>0.5\sigma_v$) & 0.13            & $^{+0.44}_{-0.52}$ & 0.48 \\
  \Stidp[i]                      & -0.71           & $^{+0.44}_{-0.92}$ & 0.68 \\
  \Stidp[\perp,i]                & -0.55           & $^{+0.47}_{-0.96}$ & 0.71
%   \\\bottomrule
% \end{tabular}

  \\\enddata
  \tablecomments{The 16th and 84th percentiles are given as the value relative to the median.
    The scatter given in the rightmost column is half the difference between these two percentiles.
    The 3D and the line-of-sight velocities are denoted as $|\inc v\tsb{3D}|$ and $|\inc v|$ respectively.
    Please refer to the text in \autoref{sec:app:proj} for more details.}
\end{deluxetable}

Assuming that the projective effects of $|\inc v\tsb{3D}|$ and $d$ are the only source of uncertainties in \Pram and \Stid, we use the same data as used in \autoref{sec:app:vesc}, and calculate the logarithmic difference between the 3D and projected values of $|\inc v\tsb{3D}|$, $d$, \Pram, and \Stid.
The median value and scatter, i.e., the systematic offset and random uncertainty for these parameters, are listed in \autoref{tab:proj}.
Interlopers are included in the calculation because they are included in the observations.
The systematic offset of $d$ is consistent with those reported by \citet{2020ApJ...903..103W}.

With our ICM model, $\Pram\propto{|\inc v\tsb{3D}|}^2/d^{1.32}$.
We find that the underestimation and projective scatter of \Pram increase significantly when $|\inc v|<0.5v_\sigma$.
This introduces a bias against galaxies with a small $|\inc v|$, making our \rps sample incomplete but rather clean.
We confirm that all but one of our \rps galaxies satisfy $|\inc v|>0.5v_\sigma$, and thus the scatter of \Pram in this region (\qty{0.48}{\dex}) is used in the main text.

Although \Stid is calculated for galaxy pairs, we estimate its projective effect using $1/|\inc v\tsb{3D}|d^2$.
The scatter of the latter is \qty{0.68}{\dex}, almost independent of $|\inc v|$ or \dprj.
We emphasize that the observed \Stid is the sum of all possible pairs, which would reduce the final logarithmic scatter.
Following \citet[see their Figure~2]{2022ApJ...927...66W}, we find that the three strongest perturbers have a median contribution of 90\% in total \Stid.
We thus use $0.68/\sqrt{3}=\qty{0.41}{\dex}$ as the scatter.
According to the theoretical definition of \Stid, only the components of $ v\tsb{3D}$ that are perpendicular to the 3D $d$ need consideration.
This, however, does not change the scatter much (\Stidp[\perp,i] in \autoref{tab:proj}).

We note that, although \Srps and \Stid are systematically underestimated and overestimated respectively, our results are not altered by it, because what matters is their relative values\footnote{%
  In particular, the deviation of \Scri{tid} makes \ftid insensitive to the systematic bias in \Stid.
  We acknowledge that the absolute value of \Srps (based on the \citealp{1972ApJ...176....1G} model) is used to determine \frps, but using \Scri{\rps} to determine \frps does not change our major results.
}.
We do emphasize, however, that their large scatters demand that our methods be used statistically.

We further propagate the uncertainty in $S$ and get the uncertainty of the radius of stripping boundary ($0.1\RHi$ for \rps, and \qty{0.3}{\dex} for TS).
Then, the uncertainty in $\log f$ could be calculated, which decreases strongly with an increasing $f$.
For typical $\log\frps\approx\qty{-1.25}{\dex}$ and $\log\ftid\approx\qty{-0.5}{\dex}$, the uncertainties are \qtylist{0.25;0.7}{\dex} respectively.

\section{Measurement of Strippable-gas Fraction}
\label{sec:app:fstr}
We provide details about estimating the strippable \Hi fraction here.
The estimated anchoring force as a function of the radius, which is used by \citet{2021ApJ...915...70W}, is given as \autoeqref{eq:fanc}.
The stellar density profile measurement is reported in \autoref{ssec:mass} and $\Sig{\text{gas}}=1.4\Sig{\tHi}$, where the factor $1.4$ accounts for helium, and \Sig{\tHi} is the \RHi-normalized median \Hi-surface-density profile of galaxies \citep{2020ApJ...890...63W} scaled with \RHi.
The \Hi~profile is truncated at $1.5\RHi$ and if the \Sig{*} profile is not extended enough, it is extrapolated as an exponential disk.

With \autoeqref{eq:fanc}, the radius where $\Fanc=\Pram$ can be derived.
The fraction of \Hi beyond this radius is taken as~\frps, calculated by integrating the \Sig{\tHi} profile.
If an \frps is smaller than $1\%$, we take this as unreliable and set it to $0$.

\citet{2021ApJ...915...70W} also calculated \frps using the moment~(0) map for galaxies with an $\RHi>1.5\bmaj$ and found a correction factor of $1.4$ for the ``predicted'' \frps.
In this study, we do not apply any correction factor:
all five galaxies resolved enough for the moment~(0) map method have \frps less than 10\%, making the determination of a correction factor difficult.
Similarly, only two resolved galaxies have \ftid larger than 10\%, and no correction factor is applied to~\ftid.

\section{Substructures in N4636G}
\label{sec:app:substructure}
\begin{figure}
  \myplotone{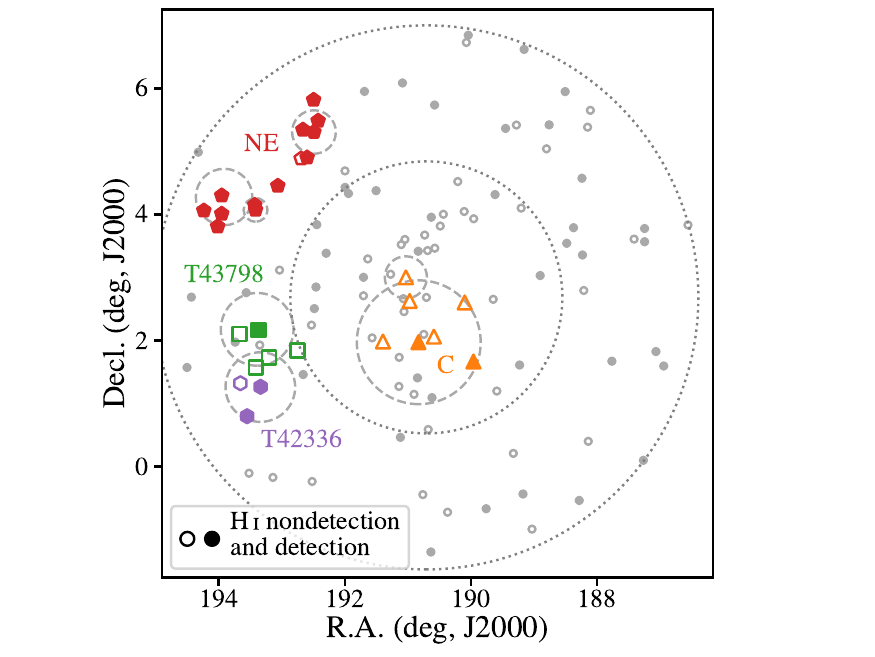}
  \caption{The substructure map.
    Four substructures (C, T43798, NE, T42336) are plotted in different colors and symbols and are labeled.
    Several groups listed in \citetalias{2017ApJ...843...16K} that share galaxies with these substructures are plotted as dashed gray circles, radii of which are corresponding \Rzoo's.}
  \label{fig:substructure}
\end{figure}

\begin{figure*}
  \myplotone{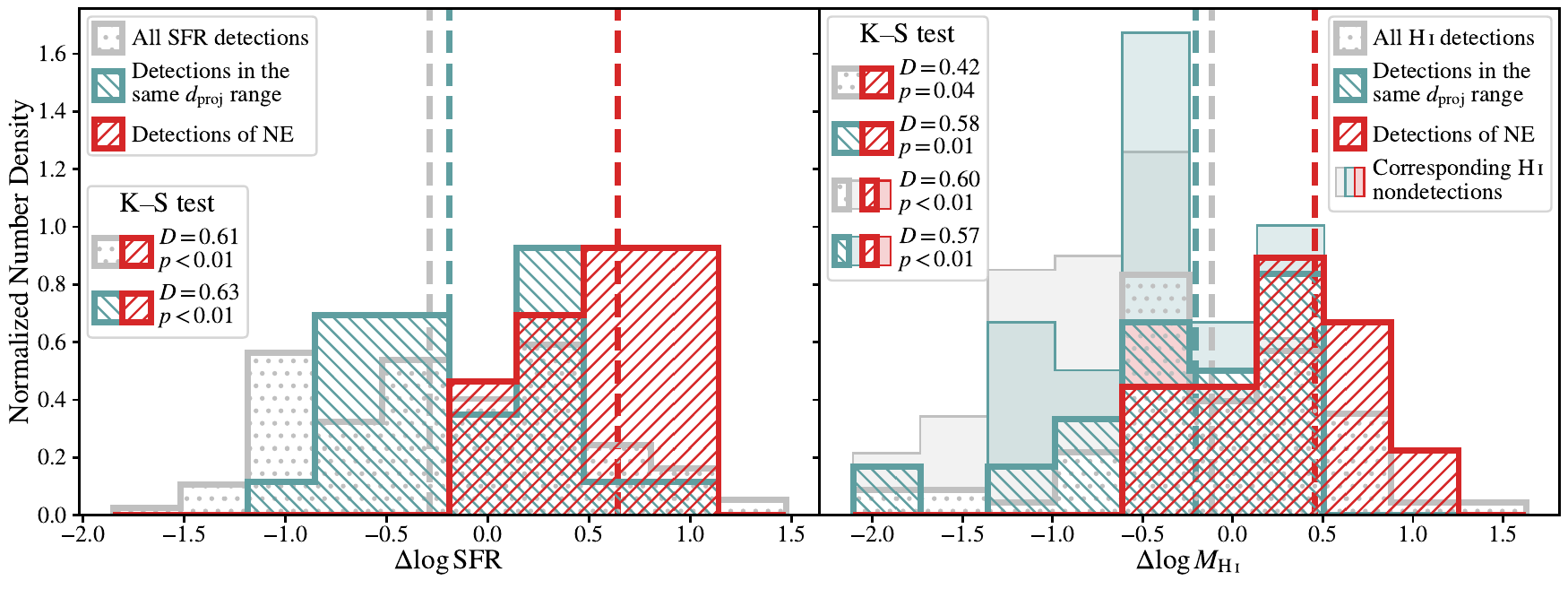}
  \caption{The galaxies in NE (red slashes) compared with galaxies in the same range of~\dprj (dark-cyan backslashes) or in the whole sample (gray dots).
    The distributions of the deviations from SFMS (left) and the mean \MHi relation (right) are plotted.
    In the left panel, two galaxies that have only lower limits of \sfr are included, which makes no apparent difference to the results.
    In the right panel, \Hi~nondetections are plotted as translucent patches in corresponding colors using the upper limits of~\MHi.
    The median values of \delSFR and \Hi-detected \delMHi in each sample are indicated with vertical dashed lines.
    The results of the K--S tests between the NE sample and the whole sample and between the NE sample and the same-\dprj sample are listed.
    For \delMHi, K--S tests are also performed with \Hi~nondetections.
    All distributions are normalized.}
  \label{fig:ne_dist}
\end{figure*}

We identified four possible substructures in N4636G by crossmatching the sample with the \citetalias{2017ApJ...843...16K} catalog, and by inspecting the clustering of galaxies in the redshift map (\autoref{fig:sample}) and the stripping map [\autoref{fig:spatial_dist}(c)].
The results are shown in \autoref{fig:substructure}.

Substructure~C consists of two groups, PGC1--42797 and PGC1--1242969, identified in \citetalias{2017ApJ...843...16K}.
Substructure~NE comprises three groups (PGC1--43413, PGC1--44086, and PGC1--1263098) from \citetalias{2017ApJ...843...16K} plus seven other galaxies, all of which have similar and negative heliocentric velocities relative to the center of N4636G\@.

We note that these identifications are of large uncertainty due to the gravitational influence of the Virgo cluster and the lack of redshift-independent distance measurements.

All substructures, apart from NE, have few \Hi-detected galaxies, and thus do not \emph{contaminate} much our analysis sample in the main part of paper.
\autoref{fig:ne_dist} compares the galaxies in NE with those in the same range  of~\dprj or in the whole group.
The galaxies of NE have both their median \delMHi (\Hi-detections only) and \delSFR significantly larger than those of the two comparison samples.
The differences are significant according to the K--S test probabilities.
If we include \Hi-nondetections in the \delMHi comparison, the $p$-values are even lower.
NE, with such high \sfr and \Hi~content, could be a young and small group falling into N4636G recently.

One question is whether galaxies in these substructures (especially NE), which might have experienced significant preprocessing, follow the same relations discussed in \autoref{sec:env}.
\autoref{fig:spatial_dist}(a) and~(b) reveal that the galaxies in NE undergo both RPS and TS\@.
We confirm that in Figures \ref{fig:s_tid} and~\ref{fig:p_mid} the NE galaxies do not stand out.
It is worth future investigation to determine whether such relations apply to more systems.

\section{Models of \Hi Stripping}

\subsection{MCMC Fitting of the Model}
\label{sec:app:evol_fit}

\begin{figure*}
  \myplotone{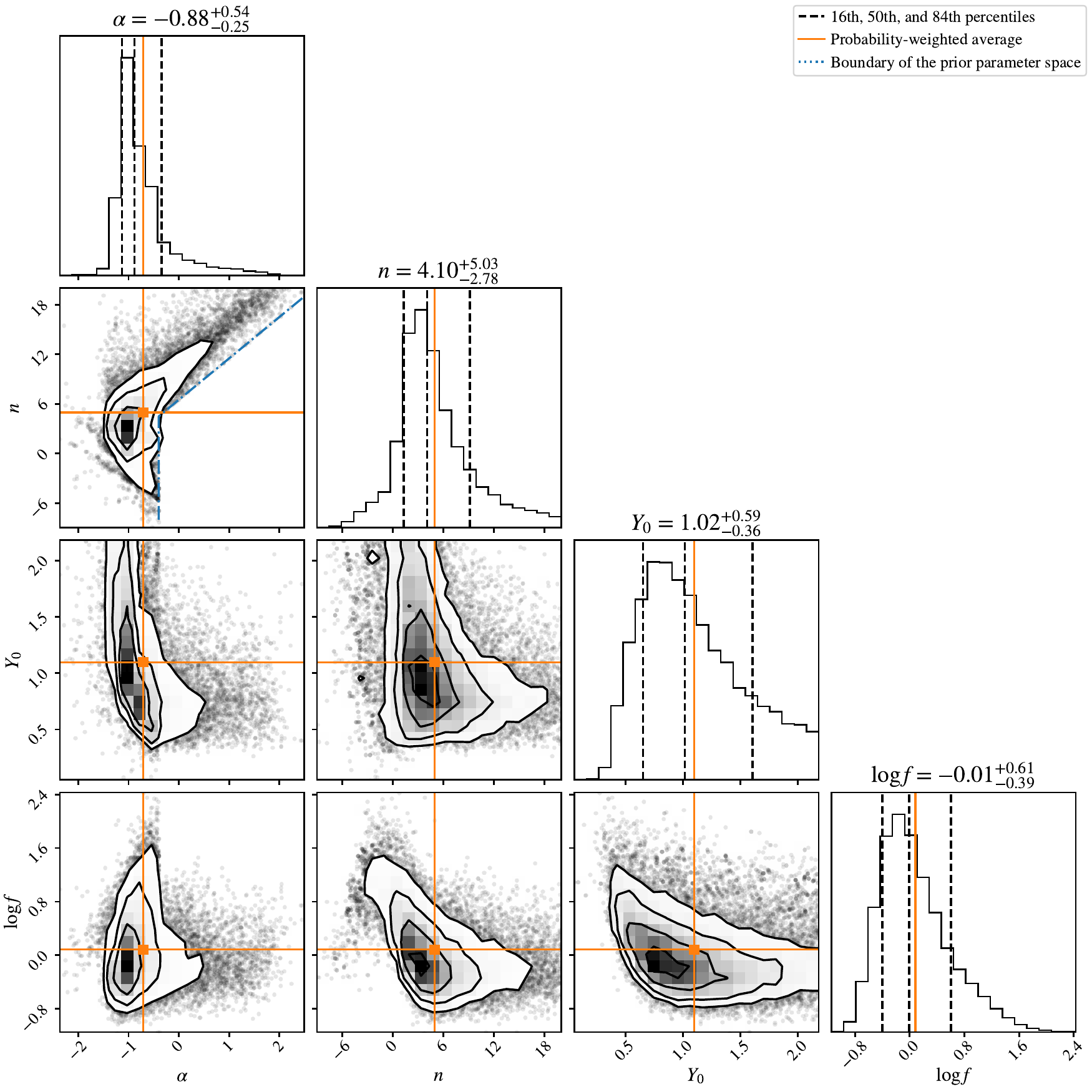}
  \caption{The posterior probability distributions of four parameters of evolutionary path model.
    The probability-weighted average is plotted in orange, and the 16th, 50th, and 84th percentiles are given as black dashed lines.
    Above each histogram, the values of three percentiles are listed.
    The region on the $\alpha$--$n$ plane where the prior probability is set to $0$ is bordered by blue dashed--dotted line.}
  \label{fig:mcmc}
\end{figure*}

While fitting the evolutionary path described by \autoeqref{eq:ivp}, besides $\alpha$ and $n$, we also introduced parameters $Y_0$, the initial value of \delMHi at $\dprj=2\Rzoo$, and $\log f$, the factor of variance underestimation.
The posterior probability distributions of all four parameters are shown in \autoref{fig:mcmc} and the median values are reported with $1\sigma$ uncertainties.
The probability-weighted values (orange lines) are similar to the corresponding median values that we used in science analyses.

We adopt such a prior probability distribution that $Y_0$ cannot result in a strippable gas fraction \fstr larger than 100\%.
We also found that there is a second peak of posterior probability on $n$--$\alpha$ corresponding to an unphysically large $\alpha$, which would result in a \tstr much larger than the Hubble time out of \Rzoo and an abrupt stripping at $\num{\sim 0.5}\Rzoo$.
We set the prior probability within this region as $0$.
The region is bordered by a blue dashed--dotted line in \autoref{fig:mcmc}.

\subsection{The Uncertainty of the Fitted Evolution Path}
\label{sec:app:evol}

\begin{figure}
  \myplotone{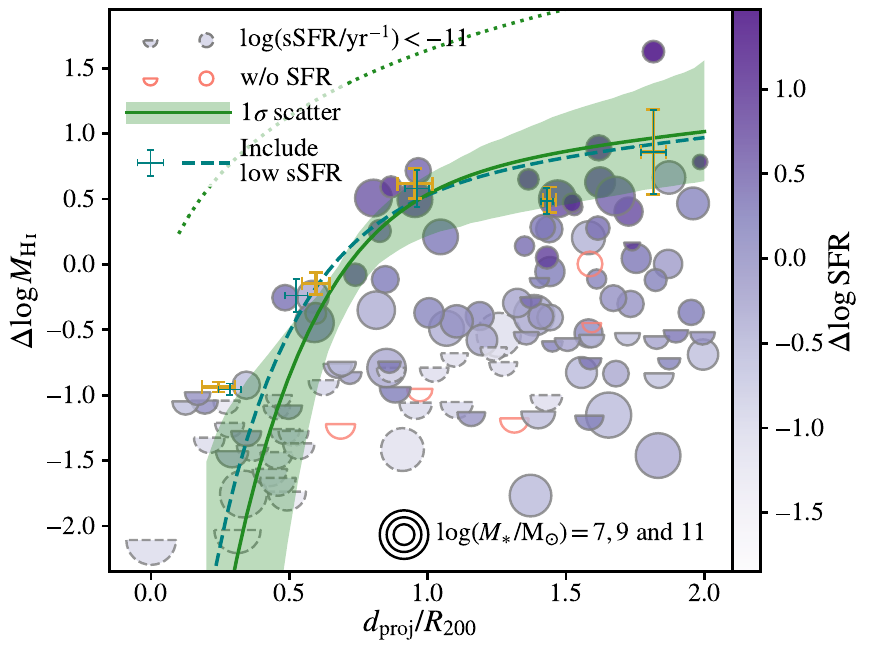}
  \caption{The uncertainty of the fitted evolutionary path, shown as green shading.
    The meanings of circles and semicircles are the same as those in \autoref{fig:logmhi_drop} and symbols with dashed edges represent galaxies with $\log\left(\ssfr/\unit{\per\yr}\right)\le -11$.
    Galaxies without \sfr measurements are plotted as empty salmon symbols.
    The evolutionary path with the median values of $\alpha$ and $n$ is given as solid line.
    The $1\sigma$ uncertainty of the path is estimated using bootstrap method and is plotted as green shading.
    The 95th percentiles and evolutionary path taking into account low-\ssfr galaxies are plotted as narrow blue error bars and dashed blue line respectively.}
  \label{fig:logmhi_drop_app}
\end{figure}

The uncertainty of fitted~$\alpha$ and~$n$ would be propagated into the uncertainty of the evolutionary path itself.
We show the bootstrapped $1\sigma$ uncertainty as the green shading in \autoref{fig:logmhi_drop_app}.
The 95th percentiles of \delMHi (with $\ssfr>\qty{1e-11}{\per\yr}$) and the path of median parameters fitted from them are shown as  orange data points and the solid green line respectively, as references.
The green shading covers the percentile points well.

\subsection{Modified Models with Different Assumptions}
\label{ssec:app:strip_model}

\begin{figure*}
  \myplotone{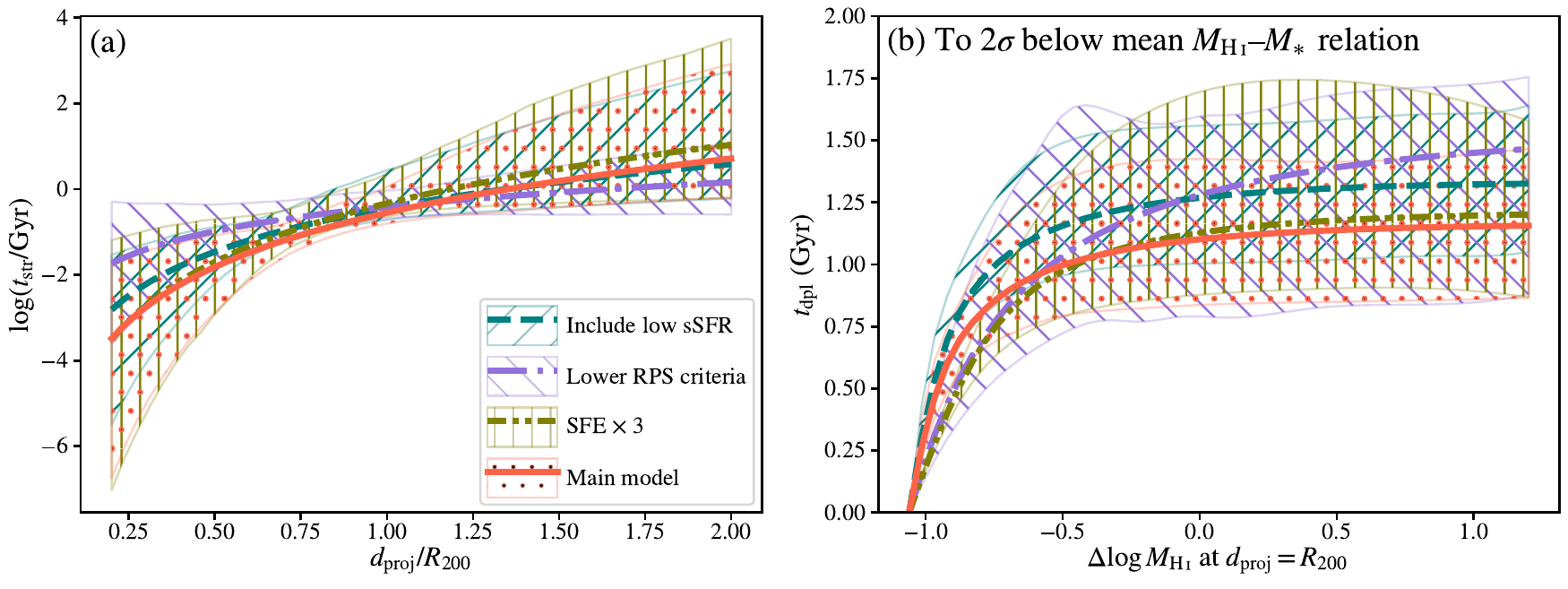}
  \caption{The influence of model specification on the values of~\tstr and~\tdpl.
    Panels are similar to \autoref{fig:str_time}(a) and~(b).
    Four models are plotted with different line-styles and patches.
    The main model we used is the solid red line.
    For details of other three models, please refer to the text.
    In~(b) the criterion of gas depletion is reaching \qty{1.06}{\dex} below the mean \MHi--$M_*$ relation.}
  \label{fig:str_time_app}
\end{figure*}

We test three different modifications to our fiducial gas-stripping empirical model described in \autoref{ssec:strip_model}, and present the changes in~\tstr and~\tdpl they introduce.

One modification is to use the 95th percentiles of all galaxies for MCMC fitting, instead of only those with high \ssfr.
In \autoref{fig:logmhi_drop_app} we show all galaxies, and their 95th percentiles are shown as narrow blue error bars.
They are similar to the percentiles of high-\ssfr galaxies (orange error bars), and the fitted result (dashed line) does not deviate much, except that the plummet happens nearer to the group center.

Shown in \autoref{fig:str_time_app} as dashed green lines, in this model, the \tstr around group center is a bit longer, and the \tdpl increases by \qty{\sim 150}{\Myr} because of the delay of stripping.

Another modification is lowering the criterion of \rps from $\Pram=\Fanc(r)$ to $\Pram/\Fanc(r)=\Scri{\rps}$, in the fashion of defining \ftid.
From \autoref{fig:size_ratio} we have $\Scri{\rps}=10^\numScrirpsshort$.

The change due to this modification (dashed--dotted purple lines) is shown in \autoref{fig:str_time_app}.
All \frps's consistently increase to 3~times the original values.
Correspondingly, the stripping timescale \tstr increases by~\qty{\sim2}{\dex} toward the group center, where \rps is strongest, since the extent of \Hi disk for stripping is larger.
At the outskirts, however, \tstr decreases a bit.
Because of the flatter relation of \tstr with \dprj under this model, \tdpl shows a larger slope with respect to the initial \delMHi.

The third modification considers the combined effect of SFE decreasing toward low-mass galaxies \citep{2012ApJ...756..113H} and the increase of stellar mass as a function of time because of star formation.
These two effects lead to the lower and higher decrease rates in \delMHi respectively than the fiducial model would imply.
The combined effect can be summarized into multiplying the slope term related to SFE by a factor of \num{\sim 3}.

This modified model (dashed--dotted--dotted brown lines in \autoref{fig:str_time_app}) does not change \tstr much, but decreases the \tdpl of galaxies with a low \delMHi at \Rzoo.
The reason is that, for these initially gas-poor galaxies, enhanced star-formation consumption quickly depletes the \Hi before the stripping takes over.

\section{Colors of High-mass TS Galaxies}
\label{sec:app:highmass_ts}

\begin{figure*}
  \myplotone{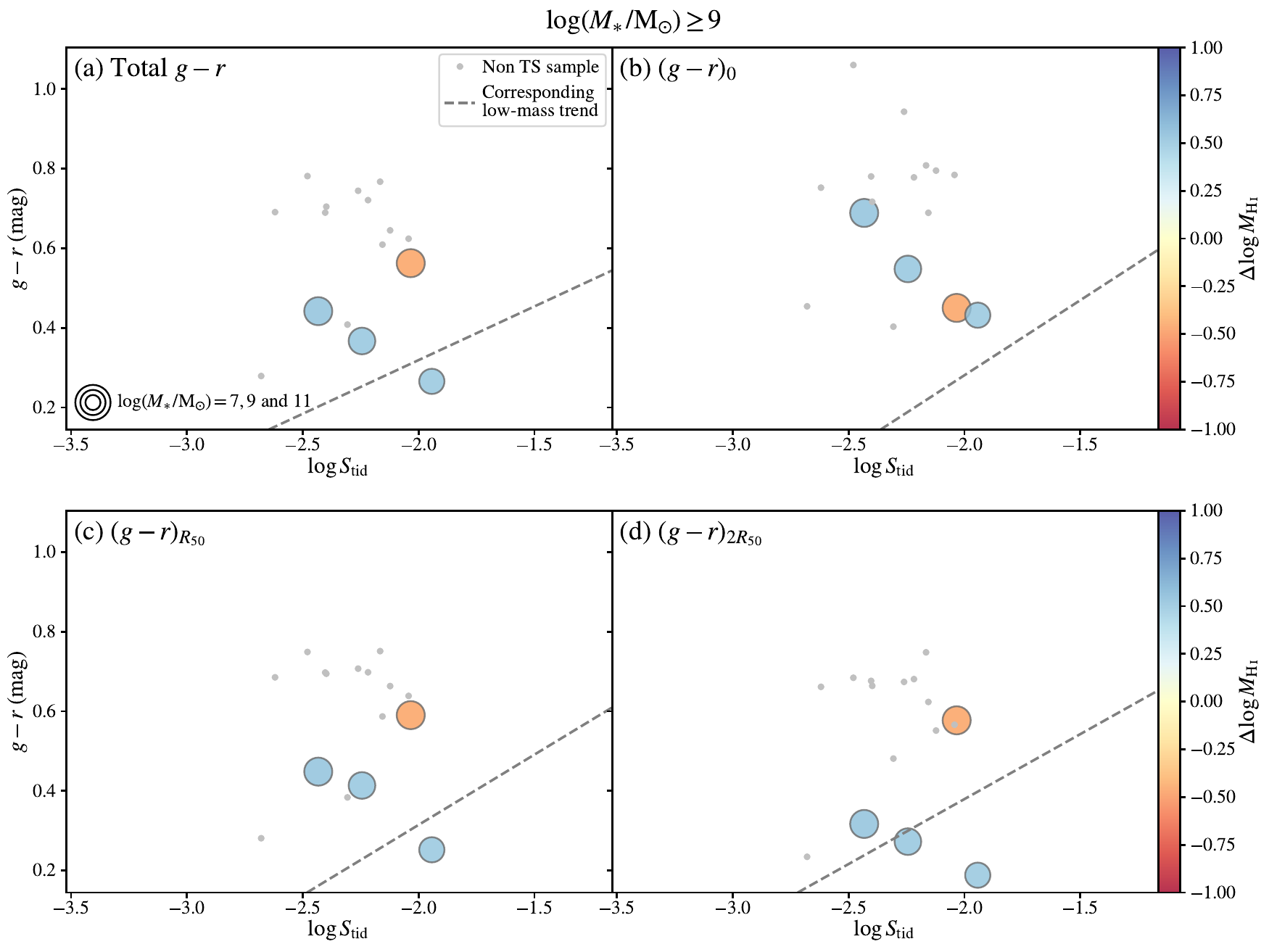}
  \caption{The same as \autoref{fig:s_tid}, but only \Hi-detected high-mass galaxies are plotted.
    The bisector fitting results of \autoref{fig:s_tid} are plotted as references.
    The $x$-axis range is the same as those in \autoref{fig:s_tid}.}
  \label{fig:s_tid_highmass}
\end{figure*}

We plotted the four high-mass TS galaxies in \autoref{fig:s_tid_highmass}.
Compared with low-mass ones, they do not have a very high \Stid.
They scatter around the linear fittings we get in \autoref{ssec:color_tidal}.

\section{Atlas of \Hi-resolved Galaxies}
\label{sec:app:overlay}

\begin{figure*}
  \myplotone{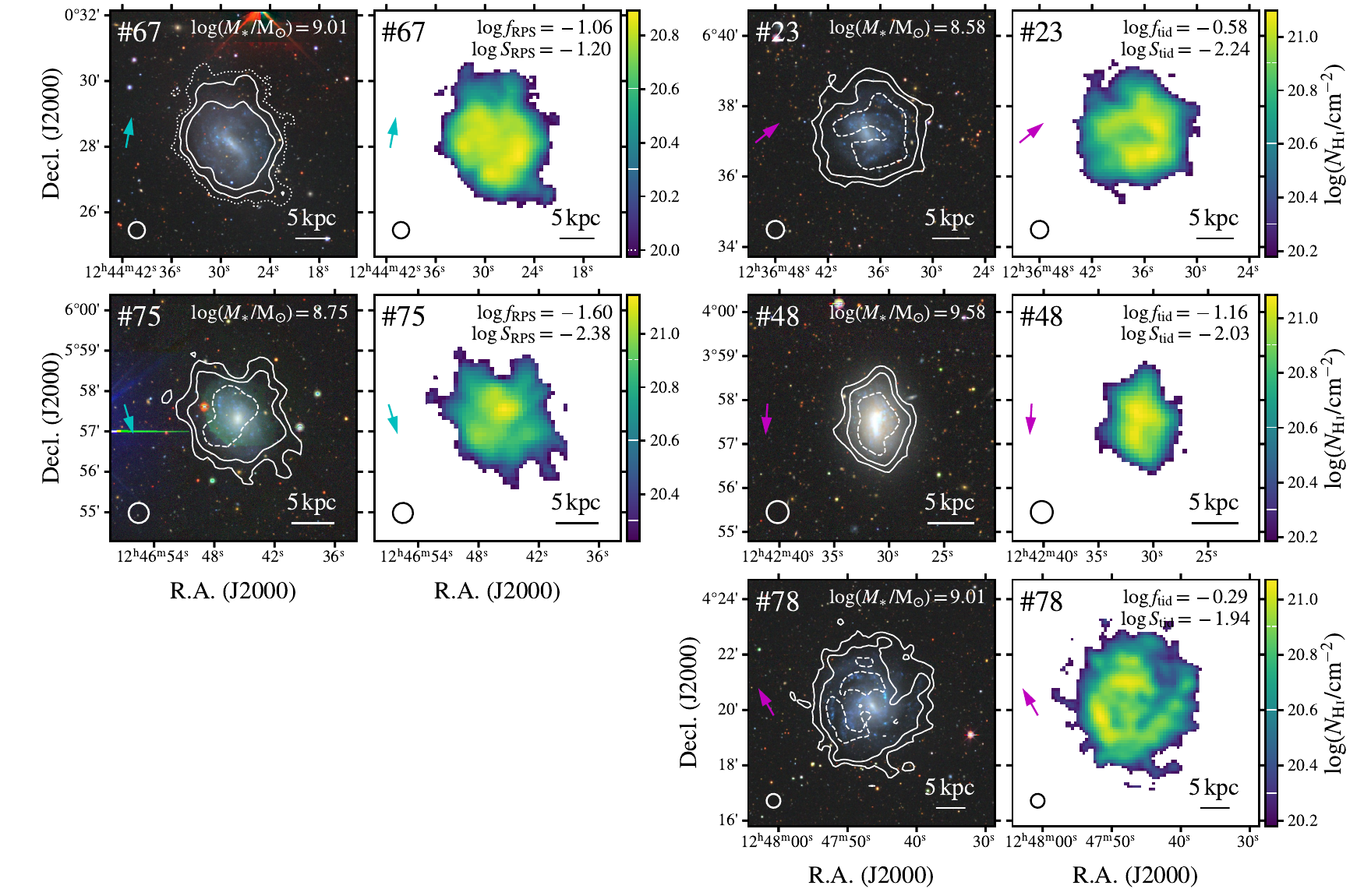}
  \caption{The atlas of five \Hi-resolved galaxies (see also \autoref{sec:app:hi_mock}).
    RPS and TS galaxies are plotted on the left and right respectively.
    For each galaxy, the DECaLS DR9 optical image and the WALLABY \Hi-column-density map are placed adjacently.
    Of the \Hi intensity map, only pixels with values $\mathord{>}5\sigma$ are shown.
    The optical image is overlaid with the contour of $N\sbHi$, with levels of \qtylist[list-exponents=combine,print-unity-mantissa=true,list-units=single]{1e20;2e20;4e20;8e20}{\cm^{-2}}.
    The lowest and the highest level are plotted as dotted and dashed lines respectively for clarity.
    The ID, beam size (\ang{;;\sim 30}), \qty{5}{kpc} scale bar, stellar mass, and the relevant values of $f$ and $S$ are given at each panel.
    The direction of the group center is indicated by a cyan arrow for RPS galaxies, with the direction of the major tidal perturber by a magenta arrow for TS ones.}
  \label{fig:overlay}
\end{figure*}

We provide the optical images and \Hi-column-density maps of five galaxies resolved by WALLABY as \autoref{fig:overlay}.
Two of them (left column) are among the \rps sample, and the other three (right) belong to the TS sample.
All of them have a stellar mass larger than \qty[parse-numbers=false]{10^{8.5}}{\Msun}, and thus the irregularity of the \Hi disk, if any, is less likely due to internal supernovae feedback than to external perturbations.

Both galaxies undergoing \rps have a low \frps (\qty[quantity-product=]{<10}{\%}), and do not show strong directional irregularities.
Galaxy \id{67} has the largest \frps, and shows a compressed feature in the southeast, \emph{backward} to the group center (cyan arrow).
It is possible that \id{67}, with $\dprj=1.05\Rzoo$ and $\inc v=0.93\sigma_v$, is at the backsplash stage, moving outward from the group.

Two of three TS galaxies (\id{23} and \id{78}) have high values of \ftid.
Compared with the other three galaxies with weaker tidal interaction, a striking feature is that both galaxies have a half-ring-like \Hi structure in the direction of the major tidal perturber, which provides the largest portion of \Stid.
It is a tentative result worth future investigation when more resolved \Hi images are available.

\section{Galaxy Properties}
\label{sec:app:catalogue}

Basic properties and the \sfr and optical measurements of our sample are listed in \autoref{tab:Optical_table}.
\Hi-related measurements of our \Hi sample (and the merging pair) are given in \autoref{tab:HI_table}.

\begin{longrotatetable}
  \ifxetex\else%
    \global\pdfpageattr\expandafter{/Rotate 90}
  \fi
  \renewcommand{\thefootnote}{\alph{footnote}}
  \setlength{\tabcolsep}{1.7pt}
\begin{deluxetable*}{ccCCCcCCccCCCCCCCCCCC}
    \tablecaption{Galaxy Properties and Non-\Hi Measurements\label{tab:Optical_table}}
    \tabletypesize{\scriptsize}
    \tablehead{
      \ifxetex\fi
      &&&&&&&&&&&&&& \multicolumn3{c}{\gr Color At} &&&&\\
      \cline{15-17}
      ID & Name & \colhead{R.A.} & \colhead{Decl.} & \colhead{$v\tsb{hel}$} & Refs & \colhead{$\log\sfr$} & \colhead{\delSFR} & \multicolumn1{p{2em}}{\centering SFR Note} & \multicolumn1{p{3em}}{\centering Image Source} & \colhead{$\log\Pram$} & \colhead{$\log\Stid$} & \colhead{$g$} & \colhead{$r$} & \colhead{Center} & \colhead{\Rso} & \colhead{2\Rso} & \colhead{$\log M_*$} & \colhead{\Rzsg} & \colhead{\Rso} & \colhead{\Rqo}\\
      & & \colhead{(\unit{\deg})} & \colhead{(\unit{\deg})} & \colhead{(\unit{\km\per\s})} & & \colhead{[\unit{\Msun\per\yr}]} & \colhead{(\unit{\dex})} & & & \colhead{[\unit{\Pa}]} & \colhead{(\unit{\dex})} & \colhead{(\unit{\mag})} & \colhead{(\unit{\mag})} & \colhead{(\unit{\mag})} & \colhead{(\unit{\mag})} & \colhead{(\unit{\mag})} & \colhead{[\unit{\Msun}]} & \colhead{(\unit{\kpc})} & \colhead{(\unit{\kpc})} & \colhead{(\unit{\kpc})}
    }
    \colnumbers
    \startdata
    % \begin{tabular}{ccS[table-format=3.4]S[table-format=+1.4]S[table-format=4.0]cS[table-format=+1.2(2)]S[table-format=+1.2]ccS[table-format=+2.1]S[table-format=+1.2]*{2}{S[table-format=2.3(2)]}S[table-format=+1.3]*{2}{S[table-format=1.3]}S[table-format=2.2(2), table-align-text-after=false]S[table-format=2.3(2)]S[table-format=1.3(2)]S[table-format=1.2(2)]}
%   \toprule
%   ID  & Name     & {R.A.}          & {Decl.}         & {$v\tsb{hel}$}       & Refs & {$\log\sfr$}            & {\delSFR}       & SFR Note & Image Source & {$\log\Pram$}  & {$\log\Stid$}   & {$g$}           & {$r$}           & {\gr[0]}        & {\gr[\Rso]}     & {\gr[2\Rso]}    & {$\log M_*$}                 & {\Rzsg}         & {\Rso}          & {\Rqo} \\
%       &          & {(\unit{\deg})} & {(\unit{\deg})} & {(\unit{\km\per\s})} &      & {[\unit{\Msun\per\yr}]} & {(\unit{\dex})} &          &              & {[\unit{\Pa}]} & {(\unit{\dex})} & {(\unit{\mag})} & {(\unit{\mag})} & {(\unit{\mag})} & {(\unit{\mag})} & {(\unit{\mag})} & {[\unit{\Msun}]}             & {(\unit{\kpc})} & {(\unit{\kpc})} & {(\unit{\kpc})} \\
%   \midrule
  1   & {\ldots} & 186.5579        & 3.8371          & 809                  & S    & -2.68\pm0.08            & -0.07           & cn       & D            & -15.6          & -2.55           & 17.467\pm0.007  & 16.929\pm0.007  & 0.449           & 0.589           & 0.555           & 7.57\pm0.15                  & 0.726\pm0.006   & 0.438\pm0.003   & 1.11\pm0.02 \\
  2   & {\ldots} & 186.9430        & 1.5996          & 1298                 & A    & -1.57\pm0.08            & 1.47            & uf       & S            & -14.5          & -3.40           & 16.358\pm0.040  & 16.257\pm0.027  & -0.017          & 0.258           & 0.093           & 7.11\pm0.17                  & 1.324\pm0.050   & 1.815\pm0.055   & 4.01\pm0.25 \\
\enddata
    \tablecomments{Column~(1): ID\@.
      Column~(2): NGC/UGC/IC ID\@.
      Columns~(3)--(4): R.A. and decl.\ in decimal degrees (J2000).
      Column~(5): heliocentric radial velocity.
      Column~(6): source of $v\tsb{hel}$; FAST (F), ALFALFA (A), \citetalias{2017ApJ...843...16K} (K), SDSS DR16 \texttt{SpecObj} table (S), and H = HIPASS Bright Galaxy Catalog (BGC; H)\@.
      Column~(7): star formation rate.
      Column~(8): offset from SFMS\@.
      Column~(9): note on SFR measurement; estimated solely from UV data (u), from W4-band data (w), and UV and W4 data combined (c); if UV data are used, FUV (f, preferred), and NUV (n, when FUV is unavailable).
      Column~(10): source of optical images; SDSS (S), and DECaLS (D)\@.
      Column~(11): ram pressure estimated using projected parameters.
      Column~(12): tidal strength parameter.
      Columns~(13)--(14): total $g$- and $r$-bands AB magnitude using GC method.
      Columns~(15)--(17): the \gr color at the galaxy center, \Rso, and $2\Rso$.
      Column~(18): stellar mass; the uncertainty from distance is not included.
      Column~(19): the radius at which the $g$-band SB drops to \qty{25}{\mag.arcsec^{-2}}, assuming a precise distance of \qty{16.2}{\Mpc}.
      Columns~(20)--(21): the radii enclosing $50\%$ and $90\%$ of $r$-band fluxes.}
    \tablecomments{This table is also available in machine-readable form.}
    \tablenotetext{a}{This \Hi-nondetected galaxy (\id{20}) has a distinctly red color ($\gr=\qty{1.767}{\mag}$) and we assigned it the 95th percentile of \gr among our samples (\qty{0.744}{\mag}) for stellar mass calculation instead.
      Since most of our discussion is focused on \Hi-detections, this color assignment does not change our major results.}
  \end{deluxetable*}
\end{longrotatetable}
\ifxetex\else%
  \global\pdfpageattr\expandafter{}
\fi

\startlongtable
\begin{deluxetable*}{cccccCcCCCCCc}
  \renewcommand{\thefootnote}{\alph{footnote}}
  \setlength{\tabcolsep}{3pt}
  \tablecaption{\Hi-related Properties and Measurements\label{tab:HI_table}}
  \tabletypesize{\footnotesize}
  \tablehead{
    ID & FAST & ALFALFA & HIPASS & WALLABY & \colhead{$\log\MHi$} & Refs & \colhead{\delMHi} & \colhead{$\log\Srps$} & \colhead{$\log\frps$} & \colhead{$\log\ftid$} & \colhead{$\log\fstr$} & \multicolumn1{p{4em}}{\centering Stripping Status} \\
    & & & & & \colhead{[\unit{\Msun}]} & & \colhead{(\unit{\dex})} & \colhead{(\unit{\dex})} & \colhead{(\unit{\dex})} & \colhead{(\unit{\dex})} & \colhead{(\unit{\dex})} &
  }
  \colnumbers
  \startdata
  % \begin{tabular}{*{5}{c}S[table-format=1.2(2)]c*{5}{S[table-format=+1.2]}c}
%   \toprule
%   ID  & FAST     & ALFALFA  & HIPASS   & WALLABY        & {$\log\MHi$}                & Refs & {\delMHi}       & {$\log\Srps$}   & {$\log\frps$}   & {$\log\ftid$}   & {$\log\fstr$}   & Status \\
%       &          &          &          &                & {[\unit{\Msun}]}            &      & {(\unit{\dex})} & {(\unit{\dex})} & {(\unit{\dex})} & {(\unit{\dex})} & {(\unit{\dex})} & \\
%   \midrule
  2   & {\ldots} & 221639   & {\ldots} & {\ldots}       & 9.14\pm0.04                 & A    & 1.63            & -1.55           & -0.95           & {\ldots}        & -0.95           & R \\
  3   & {\ldots} & 224278   & {\ldots} & {\ldots}       & 7.70\pm0.05                 & A    & -0.30           & -3.86           & {\ldots}        & {\ldots}        & {\ldots}        & N \\
  4   & {\ldots} & 7609     & {\ldots} & {\ldots}       & 8.40\pm0.04                 & A    & -1.15           & {\ldots}        & {\ldots}        & {\ldots}        & {\ldots}        & N \\
  5   & 32       & {\ldots} & {\ldots} & {\ldots}       & 7.11\pm0.05                 & F    & -0.83           & -0.84           & {\ldots}        & -0.21           & -0.21           & T \\
  6   & 24       & 225760   & {\ldots} & {\ldots}       & 7.71\pm0.04                 & F    & -0.69           & -1.31           & -1.64           & {\ldots}        & -1.64           & R \\
  8   & 27       & 221570   & {\ldots} & J123103+014032 & 8.23\pm0.04                 & F    & -0.05           & -1.98           & -1.37           & {\ldots}        & -1.37           & R \\
  13  & 33       & 220761   & {\ldots} & {\ldots}       & 7.78\pm0.04                 & F    & -0.40           & -1.32           & -1.39           & {\ldots}        & -1.39           & R \\
\enddata
  \tablecomments{Column~(1): ID\@.
    Column~(2): ID in \citet{2022RAA....22i5016Z}.
    Column~(3): ALFALFA catalog number.
    Column~(4): HIPASS BGC name.
    Column~(5): WALLABY identifier.
    Column~(6): \Hi mass; the uncertainties in \Hi flux and the calibration are considered here.
    Column~(7): source of \Hi flux; FAST (F), ALFALFA (A), HIPASS BGC (H), and WALLABY Pilot Survey (W).
    Column~(8): offset from the mean \MHi relation.
    Column~(9): \rps strength parameter; galaxies with too little gas for the calculation of \Srps are omitted.
    Columns~(10)--(12): fraction of gas under \rps, TS, and total stripping; values smaller than $1\%$ are omitted.
    Column~(13): stripping status; RPS sample (R), TS sample (T), non-stripping galaxies (N), and merging pair (M).
  }
  \tablecomments{This table is also available in machine-readable form.}
  \tablenotetext{a}{Galaxy~\id{110} and~\id{111} are merging, and their \Hi fluxes are inseparable.
    The total \MHi of them is reported here and should be considered as the upper limit.}
\end{deluxetable*}

% \section{} % to fool VS Code for collapsing
\onecolumngrid
\phantomsection
\addcontentsline{toc}{section}{\refname}
\bibliography{N4636}

\end{document}